\documentclass[pre,floatfix,longbibliography,superscriptaddress,twocolumn,final,notitlepage,nofootinbib]{revtex4-1}

\usepackage{tabularx}
\usepackage[caption=false]{subfig}
\usepackage{graphicx}
\usepackage{amsfonts,amsmath,amssymb}
\usepackage{comment}
\usepackage{lineno}
\usepackage{hyperref}
\usepackage[explicit]{titlesec}
\usepackage{footmisc}

\usepackage[usenames,dvipsnames,svgnames,table]{xcolor}
\hypersetup{
 bookmarks=true, 
 unicode=false, 
 pdftoolbar=true, 
 pdfmenubar=true, 
 pdffitwindow=false, 
 pdfstartview={FitH}, 
 pdftitle={}, 
 pdfauthor={}, 
 pdfcreator={}, 
 pdfproducer={}, 
 pdfkeywords={} {} {}, 
 pdfnewwindow=true, 
 colorlinks=true, 
 linkcolor=red, 
 citecolor=ForestGreen, 
 filecolor=magenta, 
 urlcolor=cyan 
}

\newcommand{\reportcaption}{(top-left) A linear hierarchy diagram showing inferred SpringRank scores. Circles correspond to nodes; blue edges point down the hierarchy and red edges point up. (top-middle) A histogram shows the empirical distribution of ranks: the vertical axis is the rank $s_{i}$ (binned) and the horizontal axis is the count of nodes having a rank in that bin. (top-right) A sparsity plot of rank-ordered adjacency matrix; blue and red dots represent non-zero entries going down and up the hierarchy, respectively. (middle-right) Results of statistical significance test with randomized edge directions. The histogram represents the energies obtained in the randomized samples: the dotted line is the ground state energy obtained on the observed real network.} 

\newcommand{\reportcaptionb}{(bottom) Nodes' ranks are plotted, ordered by rank, from top rank (left) to bottom rank (right), and shaded by tier. The tiers are calculated by the $k$-means algorithm.}

\newcommand{\be}{\begin{equation}}
\newcommand{\ee}{\end{equation}} 
\newcommand{\bea}{\begin{eqnarray}}
\newcommand{\eea}{\end{eqnarray}}

\newcommand{\e}{\mathrm{e}}

\newcommand{\f}[2]{\frac{#1}{#2}}

\newcommand{\bup}[1]{\left(#1\right)}
\newcommand{\rup}[1]{\left[#1\right]}

\renewcommand{\b}[1]{\bar{#1}}

\newcommand{\id}{\mathbb{I}}
\newcommand{\Exp}{\mathbb{E}}
\renewcommand{\mid}{\,|\,}
\newcommand{\eps}{\sigma} 
\newcommand{\ones}{\mathbf{1}}

\renewcommand{\ref}[1]{[\ref{#1}]}

\newcommand{\calN}{\mathcal{N}}

\begin{document}

\title{A physical model for efficient ranking in networks}

\author{Caterina De Bacco}
	\email{cdebacco@santafe.edu}
	\thanks{Contributed equally.}
	\affiliation{Data Science Institute, Columbia University, New York, NY 10027, USA}
	\affiliation{Santa Fe Institute, Santa Fe, NM 87501, USA}

\author{Daniel B. Larremore}
	\email{daniel.larremore@colorado.edu}
	\thanks{Contributed equally.}
	\affiliation{Department of Computer Science, University of Colorado, Boulder, CO 80309, USA}
	\affiliation{BioFrontiers Institute, University of Colorado, Boulder, CO 80303, USA}
	\affiliation{Santa Fe Institute, Santa Fe, NM 87501, USA}

\author{Cristopher Moore} 
	\email{moore@santafe.edu}
	\affiliation{Santa Fe Institute, Santa Fe, NM 87501, USA}



\begin{abstract}
We present a physically-inspired model and an efficient algorithm to infer hierarchical rankings of nodes in directed networks. It assigns real-valued ranks to nodes rather than simply ordinal ranks, and it formalizes the assumption that interactions are more likely to occur between individuals with similar ranks. It provides a natural statistical significance test for the inferred hierarchy, and it can be used to perform inference tasks such as predicting the existence or direction of edges. The ranking is obtained by solving a linear system of equations, which is sparse if the network is; thus the resulting algorithm is extremely efficient and scalable. We illustrate these findings by analyzing real and synthetic data, including datasets from animal behavior, faculty hiring, social support networks, and sports tournaments. We show that our method often outperforms a variety of others, in both speed and accuracy, in recovering the underlying ranks and predicting edge directions.
\end{abstract}

\maketitle

\section*{Introduction}

In systems of many individual entities, interactions and their outcomes are often correlated with these entities' ranks or positions in a hierarchy. While in most cases these rankings are hidden from us, their presence is nevertheless revealed in the asymmetric patterns of interactions that we observe. For example, some social groups of birds, primates, and elephants are organized according to dominance hierarchies, reflected in patterns of repeated interactions in which dominant animals tend to assert themselves over less powerful subordinates~\cite{drews1993concept}. Social positions are not directly visible to researchers, but we can infer each animal's position in the hierarchy by observing the network of pairwise interactions. Similar latent hierarchies have been hypothesized in systems of endorsement in which status is due to prestige, reputation, or social position \cite{power2017,clauset2015systematic}. For example, in academia, universities may be more likely to hire faculty candidates from equally or more prestigious universities~\cite{clauset2015systematic}.

In all these cases, the direction of the interactions is affected by the status, prestige, or social position of the entities involved. But it is often the case that even the \emph{existence} of an interaction, rather than its direction, contains some information about those entities' relative prestige. For example, in some species, animals are more likely to interact with others who are close in dominance rank~\cite{cote2001reproductive,hobson2015social,dey2014individual,dey2013network,cant2006individual}; human beings tend to claim friendships with others of similar or slightly higher status~\cite{ball2013friendship}; and sports tournaments and league structures are often designed to match players or teams based on similar skill levels \cite{szymanski2003economic,baumann2010anomalies}. This suggests that we can infer the ranks of individuals in a social hierarchy using both the existence and the direction of their pairwise interactions. It also suggests assigning real-valued ranks to entities rather than simply ordinal rankings, for instance in order to infer clusters of entities with roughly equal status with gaps between them.

In this work we introduce a physically-inspired model that addresses the problems of hierarchy inference, edge prediction, and significance testing. The model, which we call SpringRank, maps each directed edge to a directed spring between the nodes that it connects, and finds real-valued positions of the nodes that minimizes the total energy of these springs. Because this optimization problem requires only linear algebra, it can be solved for networks of millions of nodes and edges in seconds. 

We also introduce a generative model for hierarchical networks in which the existence and direction of edges depend on the relative ranks of the nodes.  This model formalizes the assumption that individuals tend to interact with others of similar rank, and it can be used to create synthetic benchmark networks with tunable levels of hierarchy and noise.  It can also predict unobserved edges, allowing us to use cross-validation as a test of accuracy and statistical significance. Moreover, the maximum likelihood estimates of the ranks coincides with SpringRank asymptotically. 

We test SpringRank and its generative model version on both synthetic and real datasets, including data from animal behavior, faculty hiring, social support networks, and sports tournaments. We find that it infers accurate rankings, provides a simple significance test for hierarchical structure, and can predict the existence and direction of as-yet unobserved edges.  In particular, we find that SpringRank often predicts the direction of unobserved edges more accurately than a variety of existing methods, including popular spectral techniques, Minimum Violation Ranking, and the Bradley-Terry-Luce method.

\subsection*{Related work}

Ranking entities in a system from pairwise comparisons or interactions is a fundamental problem in many contexts, and many  methods have been proposed.  One family consists of spectral methods like Eigenvector Centrality~\cite{bonacich1987}, PageRank~\cite{page1999}, Rank Centrality \cite{negahban2016rank}, and the method of Callaghan et al.~\cite{callaghan2007random}. These methods propose various types of random walks on the directed network and therefore produce real-valued scores. However, by design these methods tend to give high ranks to a small number of important nodes, giving us little information about the lower-ranked nodes. In addition, they often require explicit regularization, adding a small term to every element of the adjacency matrix if the graph of comparisons is not strongly connected.

A second family focuses on ordinal rankings, i.e., permutations, that minimize various penalty functions. This family includes Minimum Violation Rank~\cite{ali1986minimum,slater1961inconsistencies,gupte2011finding} and SerialRank \cite{fogel2014serialrank} and SyncRank \cite{cucuringu2016sync}.  Minimum Violation Rank (MVR) imposes a uniform penalty for every violation or ``upset,'' defined as an edge that has a direction opposite to the one expected by the rank difference between the two nodes. Non-uniform penalties and other generalizations are often referred to as \emph{agony} methods \cite{letizia2018resolution}. For common choices of the penalty function, minimization can be computationally difficult~\cite{slater1961inconsistencies,tatti2017tiers}, forcing us to use simple heuristics that find local minima.

SerialRank constructs a matrix of similarity scores between each pair of nodes by examining whether they produce similar outcomes when compared with the other nodes, thereby relating the ranking problem to a more general ordering problem called seriation. SyncRank is a hybrid method which first solves a spectral problem based on synchronization, embeds node positions on a half-circle in the complex plane, and then chooses among the circular permutations of those ranks by minimizing the number of violations as in MVR.

Random Utility Models~\cite{train2009discrete}, such as the Bradley-Terry-Luce (BTL) model~\cite{bradley1952,luce1959}, are designed to infer real-valued ranks from data on pairwise preferences. These models assign a probability to the direction of an edge conditioned on its existence, but they do not assign a probability to the existence of an edge. They are appropriate, for instance, when an experimenter presents subjects with choices between pairs of items, and asks them which they prefer. 

Methods like David's Score~\cite{david1987ranking} and the Colley matrix~\cite{colley2002colley} compute rankings from proportions of wins and losses. The latter, which was originally developed by making mathematical adjustments to winning percentages, is equivalent to a particular case of the general method we introduce below. Elo score~\cite{elo1978rating}, Go Rank~\cite{coulom2008whole}, and TrueSkill~\cite{herbrich2007trueskill} are also widely used win-loss methods, but these schemes update the ranks after each match rather than taking all previous interactions into account. This specialization makes them useful when ranks evolve over sequential matches, but less useful otherwise.

Finally, there are fully generative models such the Probabilistic Niche Model of ecology~\cite{williams2010probabilistic,williams2011probabilistic,jacobs2015untangling}, models of friendship based on social status~\cite{ball2013friendship}, and more generally latent space models~\cite{hoff01latentspace} which assign probabilities to the existence and direction of edges based on real-valued positions in social space.  However, inference of these models tends to be difficult, with many local optima.  Our generative model can be viewed as a special case of these models for which inference is especially easy.

In the absence of ground-truth rankings, we can compare the accuracy of these methods using cross-validation, computing the ranks using a subset of the edges in the network and then using those ranks to predict the direction of the remaining edges. Equivalently, we can ask them to predict unobserved edges, such as which of two sports teams will win a game. However, these methods do not all make the same kinds of predictions, requiring us to use different kinds of cross-validation. Methods such as BTL produce probabilistic predictions about the direction of an edge, i.e., they estimate the probability one item will be preferred to another. Fully generative models also predict the probability that an edge exists, i.e., that a given pair of nodes in the network interact. On the other hand, ordinal ranking methods such as MVR do not make probabilistic predictions, but we can interpret their ranking as a coarse prediction that an edge is more likely to point in one direction than another.

\section*{The SpringRank model}
\label{sec:SRmodel}

We represent interactions between $N$ entities as a weighted directed network, where $A_{ij}$ is the number of interactions $i\!\to\!j$ suggesting that $i$ is ranked above $j$.  This allows both ordinal and cardinal input, including where pairs interact multiple times. For instance, $A_{ij}$ could be the number of fights between $i$ and $j$ that $i$ has won, or the number of times that $j$ has endorsed $i$.

Given the adjacency matrix $A$, our goal is to find a ranking of the nodes. To do so, the SpringRank model computes the optimal location of nodes in a hierarchy by imagining the network as a physical system. Specifically, each node $i$ is embedded at a real-valued position or rank $s_i$, and each directed edge $i\!\to\!j$ becomes an oriented spring with a nonzero resting length and displacement $s_i - s_j$.  Since we are free to rescale the latent space and the energy scale, we set the spring constant and the resting length to $1$.  Thus, the spring corresponding to an edge $i\!\to\!j$ has energy 
\begin{equation}
\label{eq:hij}
H_{ij} = \frac{1}{2} \bup{s_i - s_j -1}^2 \, , 
\end{equation}
which is minimized when $s_i - s_j = 1$.

This version of the model has no tunable parameters.  Alternately, we could allow each edge to have its own rest length or spring constant, based on the strength of each edge. However, this would create a large number of parameters, which we would have to infer from the data or choose \emph{a priori}.  We do not explore this here.

According to this model, the optimal rankings of the nodes are the ranks $s^* = (s_1^*, \ldots, s_N^*)$ which minimize the total energy of the system given by the Hamiltonian
\begin{equation}
	H(s) = \sum_{i,j=1}^N A_{ij} H_{ij} = \frac{1}{2} \sum_{i,j} A_{ij} \left(s_i -s_j -1 \right)^2  \, . 
	\label{eq:SRhamiltonian}
\end{equation}
Since this Hamiltonian is convex in $s$, we can find $s^*$ by setting $\nabla H(s) = 0$, yielding the linear system
\begin{equation}
	\left[ D^{\text{out}}+D^{\text{in}} - \left(A+ A^{T} \right) \right ] s^* 
	=\left[D^{\text{out}}-D^{\text{in}}  \right ] \ones \, ,
	\label{eq:linearsystSR}
\end{equation}
where $\ones$ is the all-ones vector and $D^{\text{out}}$ and $D^{\text{in}}$ are diagonal matrices whose entries are the weighted in- and out-degrees, $D^{\text{out}}_{ii} = \sum_j {A}_{ij}$ and  $D^{\text{in}}_{ii} = \sum_j {A}_{ji}$.  See SI Text~\ref{apx:gradient} for detailed derivations.

The matrix on the left side of Eq.~\eqref{eq:linearsystSR} is not invertible. This is because $H$ is translation-invariant: it depends only on the relative ranks $s_i-s_j$, so that if $s^* = \{s_i\} $ minimizes $H(s)$ then so does $\{ s_i+a \}$ for any constant $a$.  One way to break this symmetry is to invert the matrix in the subspace orthogonal to its nullspace by computing a Moore-Penrose pseudoinverse. If the network consists of a single component, the nullspace is spanned by the eigenvector $\ones$, in which case this method finds the $s^*$ where the average rank $(1/N) \sum_i s_i = (1/N) s^* \cdot \ones$ is zero. This is related to the random walk method of~\cite{callaghan2007random}: if a random walk moves along each directed edge with rate $\frac{1}{2}+\varepsilon$ and against each one with rate $\frac{1}{2}-\varepsilon$, then $s^*$ is proportional to the perturbation to the stationary distribution to first order in $\varepsilon$.

In practice, it is more efficient and accurate to fix the rank of one of the nodes and solve the resulting equation using a sparse iterative solver (see SI Text \ref{apx:gradient}). Faster still, because this matrix is a Laplacian, recent results~\cite{spielman2014nearly,koutis2011nearly} allow us to solve Eq.~\eqref{eq:linearsystSR} in nearly linear time in $M$, the number of non-zero edges in $A$. 

Another way to break translation invariance is to introduce an ``external field'' $H_0(s_i) = \frac{1}{2} \alpha s_i^2$ affecting each node, so that the combined Hamiltonian is
\begin{equation}
	H_\alpha(s)= H(s) + \frac{\alpha}{2} \sum_{i=1}^N s_i^2 \, . 
	\label{eq:fullH}
\end{equation}
The field $H_0$ corresponds to a spring that attracts every node to the origin.  We can think of this as imposing a Gaussian prior on the ranks, or as a regularization term that quadratically penalizes ranks with large absolute values.  This version of the model has a single tunable parameter, namely the spring constant $\alpha$.  Since $H(s)$ scales with the total edge weight $M = \sum_{i,j} {A}_{ij}$ while $H_0(s)$ scales with $N$, for a fixed value of $\alpha$ this regularization becomes less relevant as networks become more dense and the average (weighted) degree $M/N$ increases.  

For $\alpha > 0$ there is a unique $s^*$ that minimizes $H_\alpha$, given by
\begin{equation}
	\left[ D^{\text{out}}+D^{\text{in}}-\left({A} +{A}^T \right)+\alpha \id \right ] s^* = \left[  D^{\text{out}}-D^{\text{in}} \right] \ones \, , 
	\label{eq:FullLinearSystem}
\end{equation}
where $\id$ is the identity matrix.  The matrix on the left side is now invertible, since the eigenvector $\ones$ has eigenvalues $\alpha$ instead of $0$.  In the limit $\alpha \to 0$, we recover Eq.~\eqref{eq:linearsystSR};  the value $\alpha=2$ corresponds to the Colley matrix method~\cite{colley2002colley}. 

Minimizing $H(s)$, or the regularized version $H_\alpha(s)$, corresponds to finding the ``ground state'' $s^*$ of the model. In the next section we show that this corresponds to a maximum-likelihood estimate of the ranks in a generative model. However, we can use SpringRank not just to maximize the likelihood, but to compute a joint distribution of the ranks as a Boltzmann distribution with Hamiltonian Eq.~\eqref{eq:fullH}, and thus estimate the uncertainty and correlations between the ranks.  In particular, the ranks $s_i$ are random variables following an $N$-dimensional Gaussian distribution with mean $s^*$ and covariance matrix (SI Text~\ref{SI:multivariatenormal})
\begin{equation}
\label{eq:posterior}
\Sigma = \frac{1}{\beta} \left[ D^{\text{out}} + D^{\text{in}} - \left({A}+ {A}^{T} + \alpha \id \right)\right]^{-1} \, .
\end{equation}
Here $\beta$ is an inverse temperature controlling the amount of noise in the model. In the limit $\beta\!\to\!\infty$, the rankings are sharply peaked around the ground state $s^*$, while for $\beta\!\to\!0$ they are noisy.  As we discuss below, we can estimate $\beta$ from the observed data in various ways.

The rankings given by SpringRank Eq.~\eqref{eq:linearsystSR} and its regularized form Eq.~\eqref{eq:FullLinearSystem} are easily and rapidly computed by standard linear solvers. In particular, iterative solvers that take advantage of the sparsity of the system can find $s^*$ for networks with millions of nodes and edges in seconds.  However, as defined above, SpringRank is not a fully generative model that assigns probabilities to the data and allows for Bayesian inference. In the next section we introduce a generative model for hierarchical networks and show that it converges to SpringRank in the limit of strong hierarchy.
 
\subsection*{A generative model}
\label{sec:gm}

In this section we propose a probabilistic generative model that takes as its input a set of node ranks $s_1,\ldots,s_N$ and produces a weighted directed network. The model also has a temperature or noise parameter $\beta$ and a density parameter $c$. Edges between each pair of nodes $i,j$ are generated independently of other pairs, conditioned on the ranks. The expected number of edges from $i$ to $j$ is proportional to the Boltzmann weight of the corresponding term in the Hamiltonian Eq.~\eqref{eq:SRhamiltonian},
\begin{equation}
	\Exp[A_{ij}] = c \exp(-\beta H_{ij}) = c \exp{\left[- \frac{\beta}{2} \bup{s_i-s_j-1}^2 \right]} \, , \nonumber
\end{equation}
where the actual edge weight $A_{ij}$ is drawn from a Poisson distribution with this mean. The parameter $c$ controls the overall density of the network, giving an expected number of edges
\begin{equation}
	\Exp[M] = \sum_{i,j} \Exp[A_{ij}] = c\sum_{i,j} \exp{\left[- \frac{\beta}{2} \bup{s_i-s_j-1}^2 \right]} \, , \nonumber 
\end{equation}
while the inverse temperature $\beta$ controls the extent to which edges respect (or defy) the ranks $s$. For smaller $\beta$, edges are more likely to violate the hierarchy or to connect distant nodes, decreasing the correlation between the ranks and the directions of the interactions: for $\beta = 0$ the model generates a directed Erd\H{o}s-R\'enyi graph, while in the limit $\beta \to \infty$ edges only exist between nodes $i, j$ with $s_i - s_j = 1$, and only in the direction $i\!\to\!j$. 
 
The Poisson distribution may generate multiple edges between a pair of nodes, so this model generates directed multigraphs.  This is consistent with the interpretation that $A_{ij}$ is the number, or total weight, of edges from $i$ to $j$. However, in the limit as $\Exp[A_{ij}] \to 0$, the Poisson distribution approaches a Bernoulli distribution, generating binary networks with $A_{ij} \in \{0,1\}$. 

The likelihood of observing a network $A$ given ranks $s$, inverse temperature $\beta$, and density $c$ is 
\begin{equation}
	P(A \mid s,\beta,c) = \prod_{i,j} \frac{\left[c\e^{-\frac{\beta}{2}(s_i-s_j-1)^2 } \right ]^{\!A_{ij}} } {A_{ij}!} \exp\!\left[-c\e^{-\frac{\beta}{2}(s_i-s_j-1)^2 } \right]. 
	\label{eq:generativemodel}
\end{equation}
Taking logs, substituting the maximum-likelihood value of $c$, and discarding constants that do not depend on $s$ or $\beta$ yields a log-likelihood (see Supplemental Text~\ref{poisson})
\begin{equation}
	\mathcal{L}(A \mid s,\beta) = -\beta H(s)  - M  \log \bigg[ \sum_{i,j} \e^{-\frac{\beta}{2} (s_i - s_j - 1)^2} \bigg ] \, ,
	\label{eq:MLgen}
\end{equation}
where $H(s)$ is the SpringRank energy defined in Eq.~\eqref{eq:SRhamiltonian}. In the limit of large $\beta$ where the hierarchical structure is strong, the $\hat{s}$ that maximizes Eq.~\eqref{eq:MLgen}, approaches the solution $s^*$ of Eq.~\eqref{eq:linearsystSR} that minimizes $H(s)$. Thus the maximum likelihood estimate $\hat{s}$ of the rankings in this model approaches the SpringRank ground state.

As discussed above, we can break translational symmetry by adding a field $H_0$ that attracts the ranks to the origin.  This is is equivalent to imposing a prior $P(s) \propto
\prod_{i=1}^N \e^{-\frac{\alpha\beta}{2}  (s_i -1 )^2 }$.  The maximum a posteriori estimate $\hat{s}$ then approaches the ground state $s^*$ of the  Hamiltonian in Eq.~\eqref{eq:fullH}, given by Eq.~\eqref{eq:FullLinearSystem}.  

This model belongs to a larger family of generative models considered in ecology and network theory~\cite{williams2010probabilistic,williams2011probabilistic,ball2013friendship}, and more generally the class of latent space models~\cite{hoff01latentspace}, where an edge points from $i$ to $j$ with probability $f(s_i-s_j)$ for some function $f$.  These models typically have complicated posterior distributions with many local optima, requiring Monte Carlo methods (e.g.~\cite{jacobs2015untangling}) that do not scale efficiently to large networks.  In our case, $f(s_i-s_j)$ is a Gaussian centered at $1$, and the posterior converges to the multivariate Gaussian Eq.~\eqref{eq:posterior} in the limit of strong structure.

\subsection*{Predicting edge directions}
\label{sec:ed}
\label{sec:edlnkpr}

If hierarchical structure plays an important role in a system, it should allow us to predict the direction of previously unobserved interactions, such as the winner of an upcoming match, or which direction social support will flow between two individuals. This is a kind of cross-validation, which lets us test the statistical significance of hierarchical structure.  It is also a principled way of comparing the accuracy of various ranking methods for datasets where no ground-truth ranks are known.

We formulate the edge prediction question as follows: given a set of known interactions, and given that there is an edge between $i$ and $j$, in which direction does it point? In one sense, any ranking method provides an answer to this question, since we can predict the direction according to which of $i$ or $j$ is ranked higher based on the known interactions. When comparing SpringRank to methods such as SyncRank, SerialRank, and MVR, we use these ``bitwise'' predictions, and define the accuracy $\eps_b$ as the fraction of edges whose direction is consistent with the inferred ranking.

But we want to know the odds on each game, not just the likely winner---that is, we want to estimate the probability that an edge goes in each direction. A priori, a ranking algorithm does not provide these probabilities unless we make further assumptions about how they depend on the relative ranks. Such assumptions yield  generative models like the one defined above, where the conditional probability of an edge $i \to j$ is 
\begin{equation}
	P_{ij}(\beta) = \frac{\e^{-{\beta} H_{ij}}}{\e^{-{\beta} H_{ij}}+\e^{-{\beta} H_{ji}}} = \frac{1}{1+\e^{-2\beta\bup{s_i -s_j }}}\, .
	\label{eq:pofd} 
\end{equation}
The density parameter $c$ affects the probability that an edge exists, but not its direction. Thus our probabilistic prediction method has a single tunable parameter $\beta$.

Note that $P_{ij}$ is a logistic curve, is monotonic in the rank difference $s_i -s_j $, and has width determined by the inverse temperature $\beta$.  SpringRank has this in common with two other ranking methods: setting $\gamma_i = \e^{2\beta s_i}$ recovers the Bradley-Terry-Luce model~\cite{bradley1952,luce1959} for which $P_{ij} = \gamma_i / (\gamma_i + \gamma_j)$, and setting $k=2\beta$ recovers the probability that $i$ beats $j$ in the Go rank~\cite{coulom2008whole}, where $P_{ij} = 1/(1+\e^{-k\bup{s_i - s_j }})$.  However, SpringRank differs from these methods in how it infers the ranks from observed interactions, so SpringRank and BTL make different probabilistic predictions.

In our experiments below, we test various ranking methods for edge prediction by giving them access to $80\%$ of the edges in the network (the training data) and then asking them to predict the direction of the remaining edges (the test data). We consider two measures of accuracy: $\eps_a$ is the average probability assigned to the correct direction of an edge, and $\eps_L$ is the log-likelihood of generating the directed edges given their existence.  For simple directed graphs where $A_{ij}+A_{ji} \in \{0,1\}$, these are
\begin{equation}
	\eps_a = \sum_{i,j} A_{ij} P_{ij} 
	\quad \text{and} \quad
	\eps_L = \sum_{i,j} A_{ij} \log P_{ij} \, .
\end{equation}
In the multigraph case, we ask how well $P_{ij}$ approximates the fraction of interactions between $i$ and $j$ that point from $i$ to $j$ [see Eqs.~\eqref{eq:localaccuracy} and~\eqref{eq:loglikelihood}].
For a discussion of other performance measures, see Supplemental Text \ref{SI:perfmetrics}.

We perform our probabilistic prediction experiments as follows.  Given the training data, we infer the ranks using Eq.~\eqref{eq:FullLinearSystem}. We then choose the temperature parameter $\beta$ by maximizing either $\eps_a$ or $\eps_L$ on the training data while holding the ranks fixed.  The resulting values of $\beta$, which we denote $\hat{\beta}_a$ and $\hat{\beta}_L$ respectively, are generally distinct (Supplemental Table~\ref{tableresults} and  Text~\ref{SI:bestbeta}).  This is intuitive, since a single severe mistake where $A_{ij}=1$ but $P_{ij} \approx 0$ reduces the likelihood by a large amount, while only reducing the accuracy by one edge.  As a result, predictions using $\hat{\beta}_a$ produce fewer incorrectly oriented edges, achieving a higher $\eps_a$ on the test set, while predictions using $\hat{\beta}_L$ will produce fewer dramatically incorrect predictions where $P_{ij}$ is very low, and thus achieve higher $\eps_L$ on the test set. 

\subsection*{Statistical significance using the ground state energy}

We can measure statistical significance using any test statistic, by asking whether its value on a given dataset would be highly improbable in a null model.  One such statistic is the accuracy of edge prediction using a method such as the one described above. However, this may become computationally expensive for cross-validation studies with many replicates, since each fold of each replicate requires inference of the parameter $\hat{\beta}_{a}$.  Here we propose a test statistic which is very easy to compute, inspired by the physical model behind SpringRank: namely, the ground state energy.  For the unregularized version~ Eq.~\eqref{eq:SRhamiltonian}, the energy per edge is (see SI Text~\ref{SI:rewriteE}) 
\begin{equation}
	\frac{H(s^*)}{M} = \frac{1}{2M} \sum_{i} (d_{i}^{\text{in}}-d_{i}^{\text{out}})\, s^*_{i} + \frac{1}{2} \, .
	\label{eq:groundstateE}
\end{equation}
Since the ground state energy depends on many aspects of the network structure, and since hierarchical structure is statistically significant if it helps us predict edge directions, like~\cite{elephant} we focus on the following null model: we randomize the direction of each edge while preserving the total number $\bar{A}_{ij} = A_{ij} + A_{ji}$ of edges between each pair of vertices.  If the real network has a ground state energy which is much lower than typical networks drawn from this null model, we can conclude that the hierarchical structure is statistically significant.

This test correctly concludes that directed Erd\H{o}s-R\'enyi graphs have no significant structure.  It also finds no significant structure for networks created using the generative model Eq.~\eqref{eq:generativemodel} with $\beta=0.1$, i.e., when the temperature or noise level $1/\beta$ is sufficiently large the ranks are no longer relevant to edge existence or direction (Fig.~\ref{SIfig:nullmodel}).  However, we see in the next section that it shows statistically significant hierarchy for a variety of real-world datasets, showing that $H(s^*)$ is both useful and computationally efficient as a test statistic.

\begin{figure}[h]
	\includegraphics[width=1.0\linewidth]{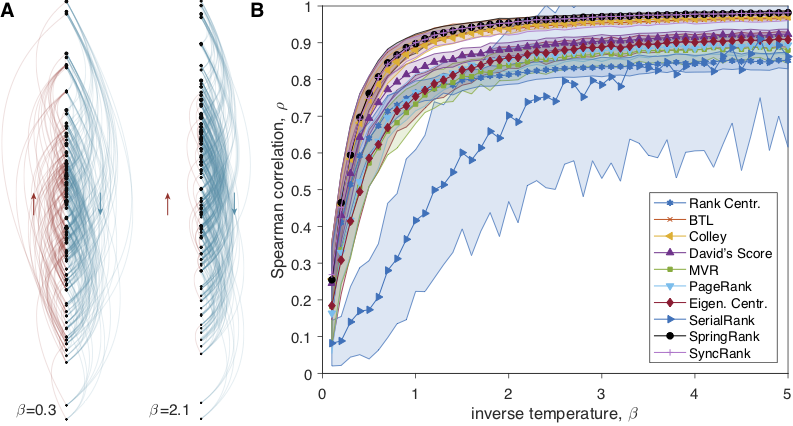} 
	\includegraphics[width=1.0\linewidth]{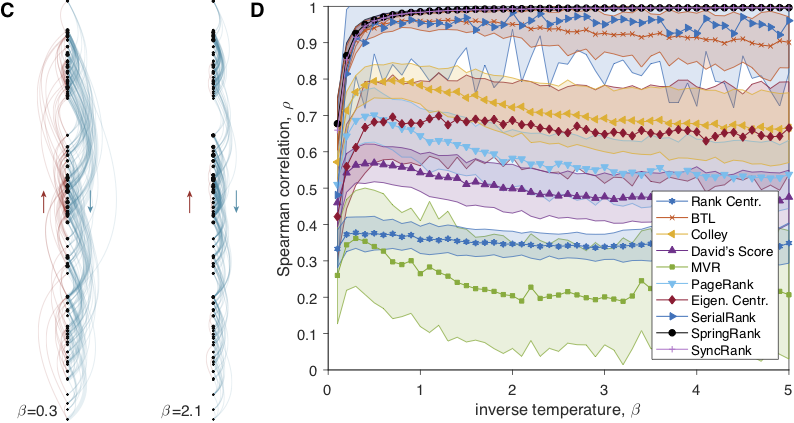}
	\caption{\textbf{Performance on synthetic data.}  (A) A synthetic network of $N=100$ nodes, with ranks drawn from a standard Gaussian and edges drawn via the generative model Eq.~\eqref{eq:generativemodel} for two different values of $\beta$ and average degree $5$. Blue edges point down the hierarchy and red edges point up, indicated by arrows. (B) The accuracy of the inferred ordering defined as the Spearman correlation averaged over $100$ indendepently generated networks; error bars indicate one standard deviation.  (C, D) Identical to A and B but with ranks drawn from a mixture of three Gaussians so that the nodes cluster into three tiers (Materials and Methods). See Fig.~\ref{sfig:pearson} for performance curves for Pearson correlation $r$.}
	\label{fig:synthetic}
\end{figure}
   
\section*{Results on real and synthetic data}

Having introduced SpringRank, an efficient procedure for inferring real-valued ranks, a corresponding generative model, a method for edge prediction, and a test for the statistical significance of hierarchical structure, we now demonstrate it by applying it to both real and synthetic data.  
For synthetic datasets where the ground-truth ranks are known, our goal is to see to what extent SpringRank and other algorithms can recover the actual ranks. For real-world datasets, in most cases we have no ground-truth ranking, so we apply the statistical significance test defined above, and compare the ability of SpringRank and other algorithms to predict edge directions given a subset of the interactions.

We compare SpringRank to other widely used methods: the spectral methods \mbox{PageRank}~\cite{page1999}, Eigenvector Centrality~\cite{bonacich1987} and Rank Centrality \cite{negahban2016rank};  Minimum Violation Ranking (MVR)~\cite{ali1986minimum,slater1961inconsistencies}, SerialRank \cite{fogel2014serialrank} and SyncRank \cite{cucuringu2016sync}, which produce ordinal rankings;
 David's score~\cite{david1987ranking}; and the BTL random utility model~\cite{bradley1952,luce1959} using the algorithm proposed in~\cite{hunter2004mm}, which like our generative model makes probabilistic predictions.  We also compare unregularized SpringRank with the regularized version $\alpha = 2$, corresponding to the Colley matrix method~\cite{colley2002colley}.  Unfortunately, Eigenvector Centrality, Rank Centrality, David's score, and BTL are undefined when the network is not strongly connected, e.g. when there are nodes with zero in- or out-degree.  In such cases we follow the common regularization procedure of adding low-weight edges between every pair of nodes (see Supplemental Text~\ref{apx:algoparameters}). 

\begin{figure}[t]
	\centering
		\includegraphics[width=0.9\linewidth]{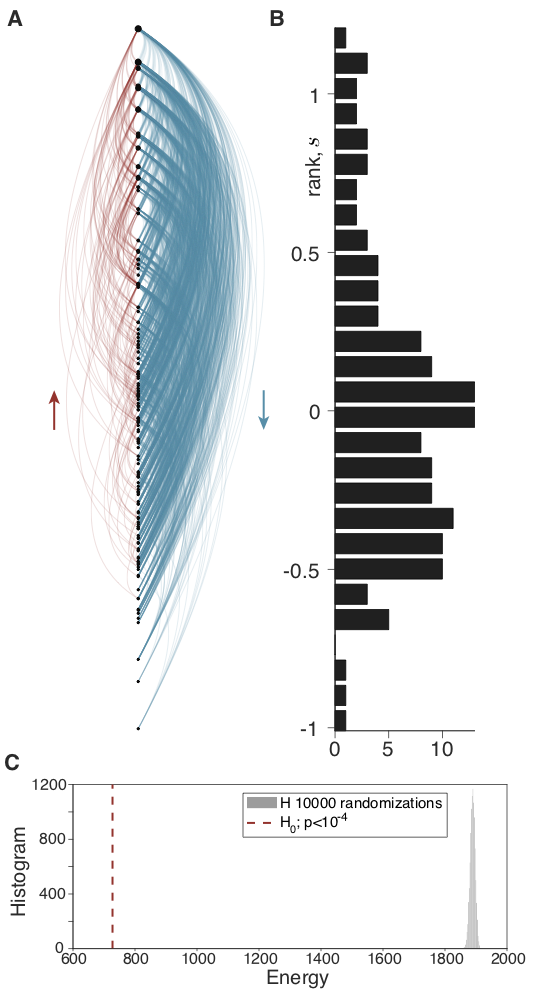}
	\caption{\textbf{Ranking the History faculty hiring network \cite{clauset2015systematic}.} 
	(A) Linear hierarchy diagram with nodes embedded at their inferred SpringRank scores. Blue edges point down the hierarchy and red edges point up. (B) Histogram of the empirical distribution of ranks, with a vertical axis of ranks matched to panel A. (C) Histogram of ground-state energies from $10,000$ randomizations of the network according to the null model where edge directions are random; the dashed red line shows the ground state energy of the empirical network depicted in panels A and B.  The fact that the ground state energy is so far below the tail of the null model is overwhelming evidence that the hierarchical structure is statistically significant, with a $p$-value $<10^{-4}$}.
	\label{fig:history}
\end{figure}

\subsection*{Performance for synthetic networks} 

We study two types of synthetic networks, generated by the model described above. Of course, since the log-likelihood in this model corresponds to the SpringRank energy in the limit of large $\beta$, we expect SpringRank to do well on these networks, and its performance should be viewed largely as a consistency check.  But by varying the distribution of ranks and the noise level, we can illustrate types of structure that may exist in real-world data, and test each algorithm's ability to identify them.

In the first type, the ranks are normally distributed with mean zero and variance one (Fig.~\ref{fig:synthetic}A).  In the second type, the ranks are drawn from an equal mixture of three Gaussians with different means and variances, so that nodes cluster into high, middle, and low tiers (Fig.~\ref{fig:synthetic}C).  This second type is intended to focus on the importance of real-valued ranks, and to measure the performance of algorithms that (implicitly or explicitly) impose strong priors on the ranks when the data defy their expectations. In both cases, we vary the amount of noise by changing $\beta$ while keeping the total number of edges constant (see Materials and Methods).  

Since we wish to compare SpringRank both to methods such as MVR that only produce ordinal rankings, and to those like PageRank and David's Score that produce real-valued ranks, we measure the accuracy of each algorithm according to the Spearman correlation $\rho$ between its inferred rank order and the true one.  Results for the Pearson correlation, where we measure the algorithms' ability to infer the real-valued ranks as opposed to just their ordering, are shown in Fig.~\ref{sfig:pearson}.

We find that all the algorithms do well on the first type of synthetic network.  As $\beta$ increases so that the network becomes more structured, with fewer edges (shown in red in Fig.~\ref{fig:synthetic}A) pointing in the ``wrong'' direction, all algorithms infer ranks that are more correlated with the ground truth.  SpringRank and SyncRank have the highest accuracy, followed closely by the Colley matrix method and BTL (Fig.~\ref{fig:synthetic}B).  Presumably the Colley matrix works well here because the ranks are in fact drawn from a Gaussian prior, as it implicitly assumes.  

\begin{figure*}[t]%
	\centering
	\includegraphics[width=0.48\linewidth]{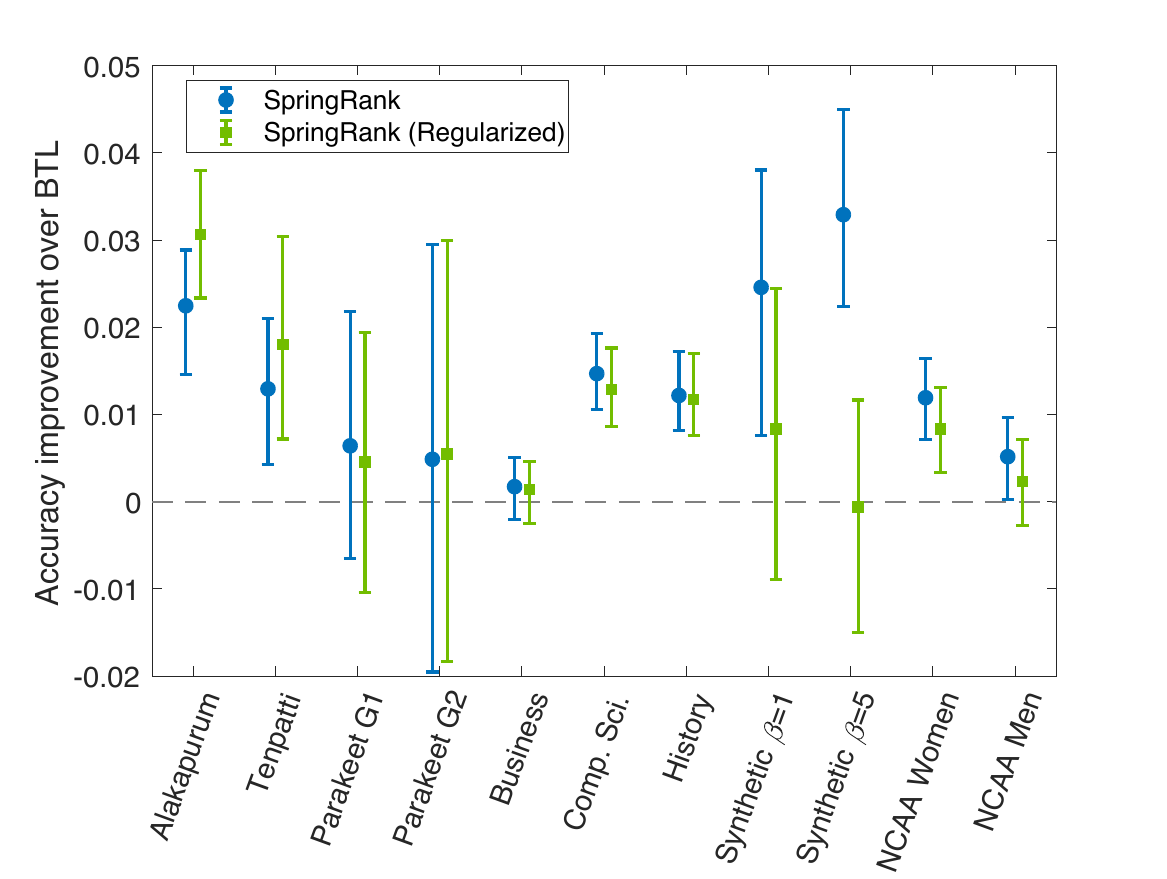}
	\includegraphics[width=0.48\linewidth]{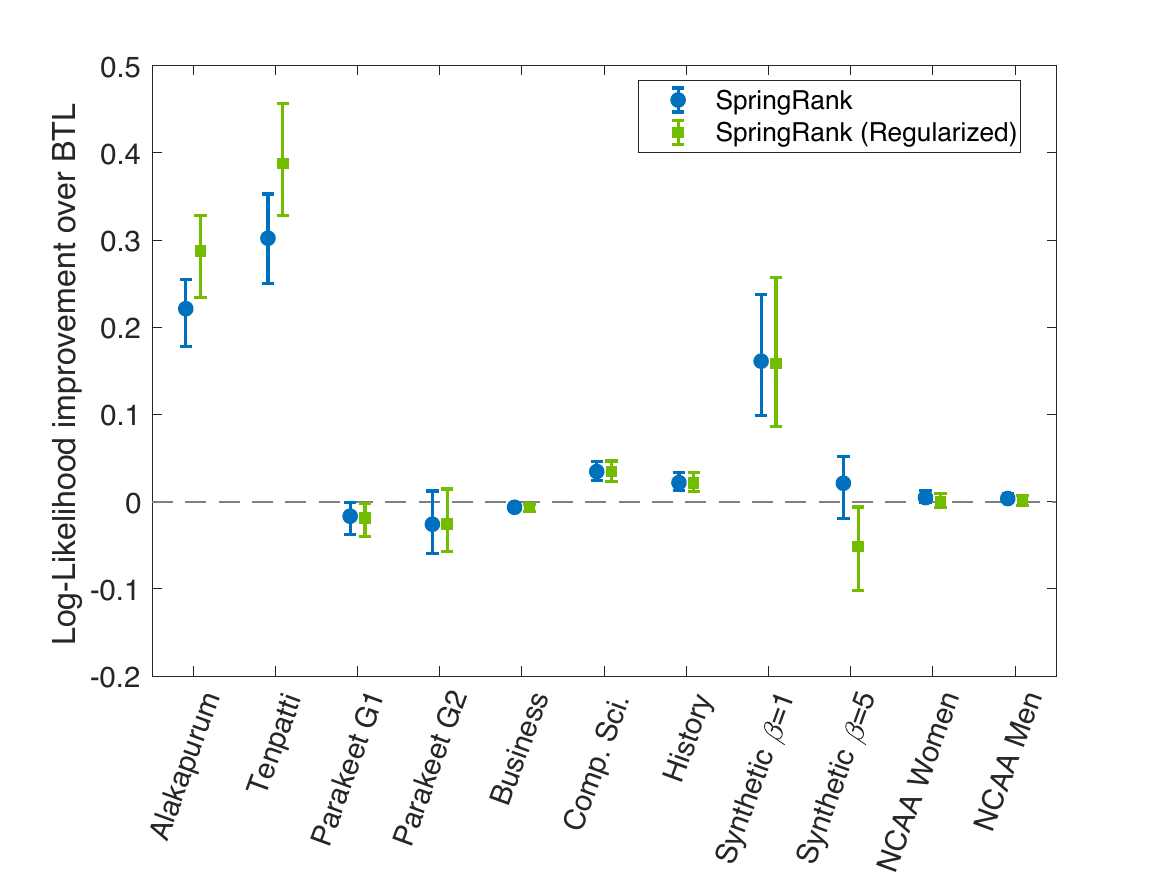}
	\caption{\textbf{Edge prediction accuracy over BTL}. Distribution of differences in performance of edge prediction of SpringRank compared to BTL on real and synthetic networks defined as (A) edge-prediction accuracy $\eps_a$ Eq.~\eqref{eq:localaccuracy} and (B) the conditional log-likelihood $\eps_L$ Eq.~\eqref{eq:loglikelihood}.  Error bars indicate quartiles and markers show medians, corresponding to $50$ independent trials of 5-fold cross-validation, for a total of 250 test sets for each network. The two synthetic networks are generated with $N=100$, average degree $5$, and Gaussian-distributed ranks as in Fig.~\ref{fig:synthetic}A, with inverse temperatures $\beta=1$ and $\beta=5$. For each experiment shown, the fractions of trials in which each method performed equal to or better than BTL are shown in Table~\ref{table:performance}. These differences correspond to prediction of an additional 1 to 12 more correct edge directions, on average.}
	\label{fig:ED}
\end{figure*}

Results for the second type of network are more nuanced.  The accuracy of SpringRank and SyncRank increases rapidly with $\beta$ with exact recovery around $\beta=1$. SerialRank also performs quite well on average. Interestingly, the other methods do not improve as $\beta$ increases, and many of them decrease beyond a certain point (Fig.~\ref{fig:synthetic}D). This suggests that these algorithms become confused when the nodes are clustered into tiers, even when the noise is small enough that most edges have directions consistent with the hierarchy.  SpringRank takes advantage of the fact that edges are more likely between nodes in the same tier (Fig.~\ref{fig:synthetic}C), so the mere existence of edges helps it cluster the ranks. 

These synthetic tests suggest that real-valued ranks capture information that ordinal ranks do not, and that many ranking methods perform poorly when there are substructures in the data such as tiered groups.  Of course, in most real-world scenarios, the ground-truth ranks are not known, and thus edge prediction and other forms of cross-validation should be used instead. We turn to edge prediction in the next section.

\subsection*{Performance for real-world networks}

As discussed above, in most real-world networks, we have no ground truth for the ranks.  Thus we focus on our ability to predict edge directions from a subset of the data, and measure the statistical significance of the inferred hierarchy.  

We apply our methods to datasets from a diverse set of fields, with sizes ranging up to $N=415$ nodes and up to $7000$ edges (see Table~\ref{tableresults}): three North American academic hiring networks where $A_{ij}$ is the number of faculty at university $j$ who received their doctorate from university $i$, for History (illustrated in Figs.~\ref{fig:history}A and B), Business, and Computer Science departments~\cite{clauset2015systematic}; two networks of animal dominance among captive monk parakeets~\cite{hobson2015social} and one among Asian elephants~\cite{elephant} where $A_{ij}$ is the number of dominating acts by animal $i$ toward animal $j$; and social support networks from two villages in Tamil Nadu referred to (for privacy reasons) by the pseudonyms ``Te\underbar npa\d t\d ti'' and ``A\underbar lak\= apuram,'' 
where $A_{ij}$ is the number of distinct social relationships (up to five) through which person $i$ supports person $j$~\cite{power2017}; and 53 networks of NCAA Women's and Men's college basketball matches during the regular season, spanning 1985-2017 (Men) and 1998-2017 (Women), where $A_{ij}=1$ if team $i$ beat team $j$. Each year's network comprises a different number of matches, ranging from 747 to 1079 \cite{ncaa}.

\begin{figure}[t]
	\centering
	\includegraphics[width=1.0\linewidth]{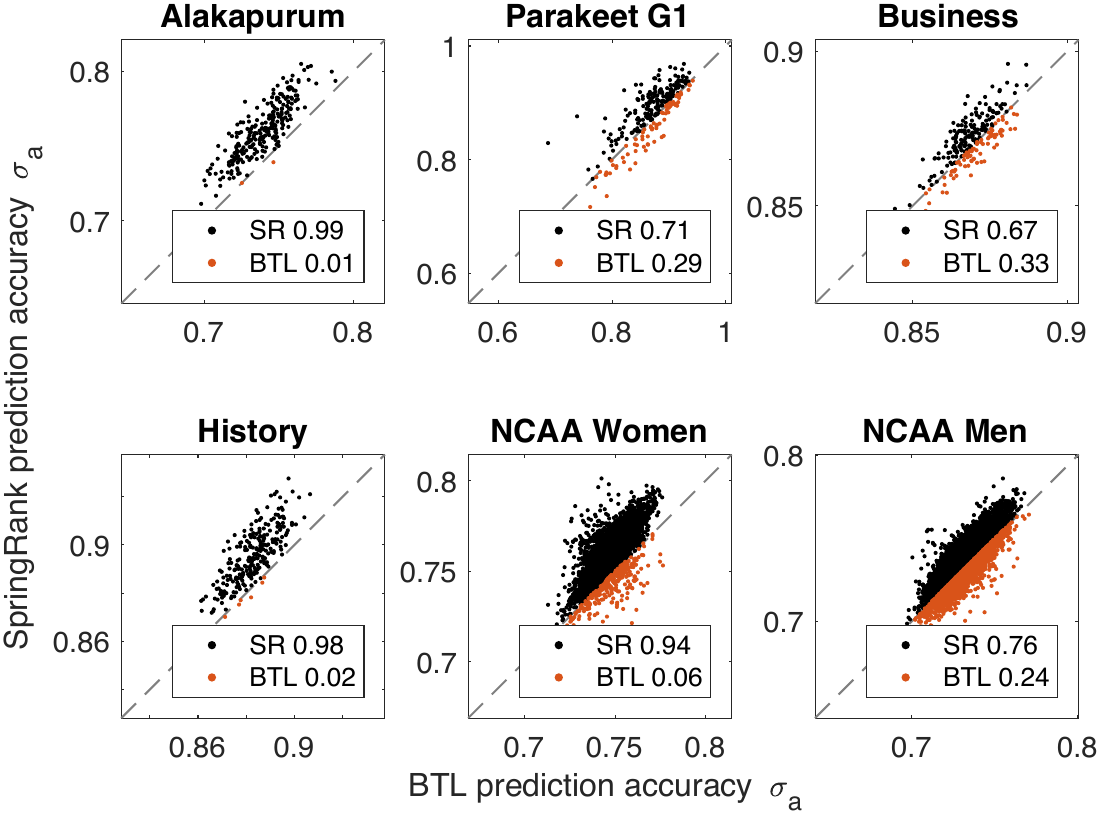}
	\caption{\textbf{Probabilistic edge prediction accuracy $\eps_a$ of SpringRank vs.\ BTL}. For 50 independent trials of 5-fold cross-validation (250 total folds per network), the values of $\eps_a$ for SpringRank and BTL are shown on the vertical axis and the horizontal axis respectively. Points above the diagonal, shown in blue, are trials where SpringRank is more accurate than BTL. The fractions for which each method is superior are shown in plot legends, matching Table~\ref{table:performance}.}
	\label{fig:prediction_scatter}
\end{figure}

Together, these examples cover prestige, dominance, and social hierarchies. In each of these domains, inferring ranks from interactions is key to further analysis. Prestige hierarchies play an unequivocal role in the dynamics of academic labor markets~\cite{way2016gender}; in behavioral ecology, higher-ranked individuals in dominance hierarchies are believed to have higher fitness~\cite{drews1993concept,majolo2012fitness}; and patterns of aggression are believed to reveal animal strategies and cognitive capacities~\cite{cote2001reproductive,hobson2015social,dey2014individual,dey2013network,cant2006individual}.  Finally, in social support networks, higher ranked individuals have greater social capital and reputational standing~\cite{lin2002social,cook2009whom}, particularly in settings in which social support is a primary way to express and gain respect and prestige~\cite{mines1994public}.

We first applied our ground state energy test for the presence of statistically significant hierarchy, rejecting the null hypothesis with $p < 10^{-4}$ in almost all cases (e.g., for History faculty hiring, see Fig.~\ref{fig:history}C). The one exception is the Asian Elephants network for which $p > 0.4$. This corroborates the original study of this network~\cite{elephant}, which found that counting triad motifs shows no significant hierarchy~\cite{shizuka2012social}.  This is despite the fact that one can find an appealing ordering of the elephants using the Minimum Violation Rank method, with just a few violating edges (SI Fig.~\ref{SFelephants}).  Thus the hierarchy found by MVR may well be illusory.

As described above, we performed edge prediction experiments using 5-fold cross-validation, where $80\%$ of the edges are available to the algorithm as training data, and a test set consisting of $20\%$ of the edges is held out (see Materials and Methods). To test SpringRank's ability to make probabilistic predictions, we compare it to BTL.

We found that SpringRank outperforms BTL, both in terms of the accuracy $\eps_a$ (Fig.~\ref{fig:ED}A) and, for most networks, the log-likelihood $\eps_L$ (Fig.~\ref{fig:ED}B).  The accuracy of both methods has a fairly broad distribution over the trials of cross-validation, since in each network some subsets of the edges are harder to predict than others when they are held out. However, as shown in Fig.~\ref{fig:prediction_scatter}, in most trials SpringRank was more accurate than BTL.  Fig.~\ref{fig:ED}A and Table~\ref{table:performance} show that SpringRank predicts edge directions more accurately in the majority of trials of cross-validation for all nine real-world networks, where this majority ranges from $62\%$ for the parakeet networks to $100\%$ for the Computer Science hiring network. 

\begin{figure}[t]
	\centering
	\includegraphics[width=1.0\linewidth]{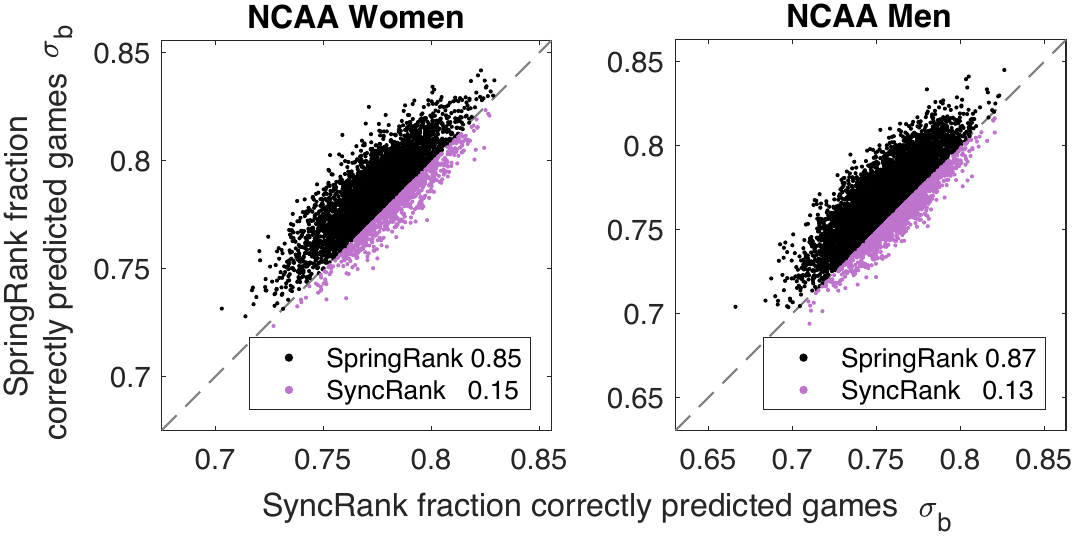}
	\caption{\textbf{Bitwise prediction accuracy $\eps_b$ of SpringRank vs.\ SyncRank}. For 50 independent trials of 5-fold cross-validation (250 total folds per NCAA season), the fraction of correctly predicted game outcomes $\eps_b$ for SpringRank and Syncrank are shown on the vertical axis and the horizontal axis respectively. Points above the equal performance line, shown in blue, are trials where SpringRank is more accurate than SyncRank; the fractions for which each method is superior are shown in plot legends.}
	\label{fig:prediction_scatter_b}
\end{figure}

Table~\ref{table:performance} shows that SpringRank also obtained a higher log-likelihood $\eps_L$ than BTL for $6$ of the $9$ real-world networks. Regularizing SpringRank with $\alpha=2$ does not appear to significantly improve either measure of accuracy (Fig.~\ref{fig:ED}). We did not attempt to tune the regularization parameter $\alpha$.

To compare SpringRank with methods that do not make probabilistic predictions, including those that produce ordinal rankings, we measured the accuracy $\eps_b$ of ``bitwise'' predictions, i.e., the fraction of edges consistent with the infered ranking. We found that spectral methods perform poorly here, as does SerialRank. Interestingly, BTL does better on the NCAA networks in terms of bitwise prediction than it does for probabilistic predictions, suggesting that it is better at rank-ordering teams than determining their real-valued position.

\begin{table*}
	\centering
	\begin{tabular}{|r|r|c|c|c|c|}
	\hline
	& & \multicolumn{2}{c|}{\% trials higher $\eps_a$ vs BTL} & \multicolumn{2}{c|}{\% trials higher $\eps_L$ vs BTL} \\
	Dataset & Type & SpringRank & +regularization & SpringRank & +regularization  \\
	 \hline
	Comp. Sci.~\cite{clauset2015systematic} & Faculty Hiring	& 100.0$\ \ \ $ & 97.2 & 100.0$\ \ $  & 99.6\\	  
	A\underbar lak\= apuram~\cite{power2017} & Social Support	& 99.2$^\dagger$ & 99.6 & 100.0$\ \ $  & 100.0$\ \ $  \\
	Synthetic $\beta=5$ & Synthetic & 98.4$\ $ & 63.2 & 76.4 &  \emph{46.4}\\
	History~\cite{clauset2015systematic}	 & Faculty Hiring	& 97.6$^\dagger$ & 96.8 & 98.8 & 98.8\\
	NCAA Women (1998-2017)~\cite{ncaa}& Basketball	& 94.4$^\dagger$ &	87.0 & 69.1 &	51.0 \\
	Te\underbar npa\d t\d ti~\cite{power2017}	& Social Support	& 88.8$\ $ & 93.6 & 100.0$\ \ $  & 100.0$\ \ $ \\
	Synthetic $\beta=1$ & Synthetic & 83.2$\ $ & 65.2 & 98.4 & 98.4\\
	NCAA Men (1985-2017)~\cite{ncaa}& Basketball	& 76.0$^\dagger$ &	62.3 & 68.5 &	52.4 \\
	Parakeet G1~\cite{hobson2015social}	& Animal Dominance	& 71.2$^\dagger$ & 56.8 & \emph{41.2} & \emph{37.2}\\
	Business~\cite{clauset2015systematic}	& Faculty Hiring	& 66.8$^\dagger$ & 59.2 & \emph{39.2} & \emph{36.8}\\
	Parakeet G2~\cite{hobson2015social}	& Animal Dominance	& 62.0$\ $ & 51.6 & \emph{47.6} & \emph{47.2}\\
	\hline
	\end{tabular}
	\caption{\textbf{Edge prediction with BTL as a benchmark.} During 50 independent trials of 5-fold cross-validation (250 total folds per network), columns show the the percentages of instances in which SpringRank Eq.~\eqref{eq:linearsystSR} and regularized SpringRank Eq.~\eqref{eq:FullLinearSystem} with $\alpha = 2$ produced probabilistic predictions with equal or higher accuracy than BTL. Distributions of accuracy improvements are shown in Fig.~\ref{fig:ED}. Center columns show accuracy $\eps_a$ and right columns show $\eps_L$ (Materials and Methods). Italics indicate where BTL outperformed SpringRank for more than 50\% of tests. $^\dagger$ Dagger symbols indicate tests that are shown in detail in Fig.~\ref{fig:prediction_scatter}. NCAA Basketball datasets were analyzed one year at a time}.
	\label{table:performance}
\end{table*}

We found that SyncRank is the strongest of the ordinal methods, matching SpringRank's accuracy on the parakeet and business school networks, but SpringRank outperforms SyncRank on  the social support and NCAA networks (see Fig.~\ref{SIfig:ED}). We show a trial-by-trial comparison of SpringRank and SyncRank in Fig.~\ref{fig:prediction_scatter_b}, showing that in most trials of cross-validation SpringRank makes more accurate predictions for the NCAA networks.

To check whether our results were dependent on the choice of holding out $20\%$ of the data, we repeated our experiments using 2-fold cross-validation, i.e., using 50\% of network edges as training data and trying to predict the other 50\%. We show these results in Fig.~\ref{fig:ED2}. While all algorithms are less accurate in this setting, the comparison between algorithms is similar to that for 5-fold cross-validation.

Finally, the real-valued ranks found by SpringRank shed light on the organization and assembly of real-world networks  (see Figs.~\ref{SI:CS},~\ref{SI:HS},~\ref{SI:BS},~\ref{SI:Ten}, and~\ref{SI:Ala}). For example, we found that ranks in the faculty hiring networks have a long tail at the top, suggesting that the most prestigious universities are more separated from those below them than an ordinal ranking would reveal. In contrast, ranks in the social support networks have a long tail at the bottom, suggesting a set of people who do not have sufficient social status to provide support to others.  SpringRank's ability to find real-valued ranks makes these distributions amenable to statistical analysis, and we suggest this as a direction for future work.

\subsection*{Conclusions}
\label{sec:conclusion}

SpringRank is a mathematically principled, physics-inspired model for hierarchical structure in networks of directed interactions.  It yields a simple and highly scalable algorithm, requiring only sparse linear algebra, which enables analysis of networks with millions of nodes and edges in seconds. Its ground state energy provides a natural test statistic for the statistical significance of hierarchical structure.

While the basic SpringRank algorithm is nonparametric, a parameterized regularization term can be included as well, corresponding to a Gaussian prior. While regularization is often required for BTL, Eigenvector Centrality, and other commonly used methods (Supplemental Text~\ref{apx:algoparameters}) it is not necessary for SpringRank and our tests indicate that its effects are mixed. 

We also presented a generative model that allows one to create synthetic networks with tunable levels of hierarchy and noise, whose posterior coincides with SpringRank in the limit where the effect of the hierarchy is strong. By tuning a single temperature parameter, we can use this model to make probabilistic predictions of edge directions, generalizing from observed to unobserved interactions. Therefore, after confirming its ability to infer ranks in synthetic networks where ground truth ranks are known, we measured SpringRank's ability to to predict edge directions in real networks. We found that in networks of faculty hiring, animal interactions, social support, and NCAA basketball, SpringRank often makes better probabilistic predictions of edge predictions than the popular Bradley-Terry-Luce model, and performs as well or better than SyncRank and a variety of other methods that produce ordinal rankings.

SpringRank is based on springs with quadratic potentials, but other potentials may be of interest.  For instance, to make the system more tolerant to outliers while remaining convex, one might consider a piecewise potential that is quadratic for small displacements and linear otherwise. We leave this investigation of alternative potentials to future work.

Given its simplicity, speed, and high performance, we believe that SpringRank will be useful in a wide variety of fields where hierarchical structure appears due to dominance, social status, or prestige. 

\section*{Materials and methods}
\vspace{-0.1in}

\subsection*{Synthetic network generation}
\vspace{-0.1in}

Networks were generated in three steps. First, node ranks $s_\text{planted}$ were drawn from a chosen distribution. For Test 1, $N=100$ ranks were drawn from a standard normal distribution, while for Test 2, $34$ ranks were drawn from each of three Gaussians, $\calN(-4,2)$, $\calN(0,\frac{1}{2})$, and $\calN(4,1)$ for a total of $N=102$. Second, an average degree $\langle k \rangle$ and a value of the inverse temperature $\beta$ were chosen. Third, edges were drawn generated to Eq.~\eqref{eq:generativemodel} with $c = \langle k \rangle N / \sum_{i,j} \exp{[-(\beta/2)(s_i - s_j - 1)^2]}$ so that the expected mean degree is $\langle k \rangle$ (see SI Text~\ref{SI:cforsparsity}). 

This procedure resulted in directed networks with the desired hierarchical structure, mean degree, and noise level. Tests were conducted for $\langle k \rangle \in [5,15]$, $\beta \in [0.1,5]$, and all performance plots show mean and standard deviations for $100$ replicates.

\subsection*{Performance measures for edge prediction} 
\vspace{-0.1in}
For multigraphs, we define the accuracy of probabilistic edge prediction as the extent to which $P_{ij}$ is a good estimate of the fraction of interactions between $i$ and $j$ that point from $i$ to $j$, given that there are any edges to predict at all, i.e., assuming $\bar{A}_{ij} =A_{ij} + A_{ji}>0$.  If this prediction were perfect, we would have $A_{ij} = \bar{A}_{ij} P_{ij}$.  We define $\eps_A$ as $1$ minus the sum of the absolute values of the difference between $A_{ij}$ and this estimate,
\begin{equation}
	\eps_a = 1 - \frac{1}{2M} \, \sum_{i,j} \left \lvert A_{ij} - \bar{A}_{ij} \,P_{ij} \right \rvert \, ,
	\label{eq:localaccuracy}
\end{equation}
where $M$ is the number of directed edges in the subset of the network under consideration, e.g., the training or test set.  
Then $\eps_a = 1$ if $P_{ij} = A_{ij} / \bar{A}_{ij}$ for all $i,j$, and $\eps_a=0$ if for each $i,j$ all the edges go from $i$ to $j$ (say) but $P_{ij}=0$ and $P_{ji}=1$.

To measure accuracy via the conditional log-likelihood, we ask with what probability we would get the directed network $A$ from the undirected network $\bar{A}$ if each edge between $i$ and $j$ points from $i \to j$ with probability $P_{ij}$ and from $j \to i$ with probability $P_{ji} = 1-P_{ij}$.  This gives
\begin{align}
	\eps_L &= \log \Pr[A \mid \bar{A}] \nonumber \\
	&= \sum_{i,j}  \log \binom{A_{ij}+A_{ji}}{A_{ij}} + \log \left[ P_{ij}^{A_{ij}} \, \rup{1-P_{ij}}^{A_{ji }} \right ] \, ,
	\label{eq:loglikelihood}
\end{align}
where $\binom{x}{y}$ is the binomial coefficient. We disregard the first term of this sum since it does not depend on $P$. If we wish to compare networks of different sizes as in Fig.~\ref{fig:ED}, we can normalize $\eps_L$ by the number of edges. For an extensive discussion of performance metrics see Supplementary Text~\ref{SI:perfmetrics}.

\subsection*{Statistical significance of ranks} 
\vspace{-0.1in}
We compute a standard left-tailed $p$-value for the statistical significance of the ranks $s^*$ by comparing the ground state energy Eq.~\eqref{eq:groundstateE} of the real network $A$ with the null distribution of ground state energies of an ensemble of networks $\tilde A$ drawn from the null model where $\bar{A}_{ij}$ is kept fixed, but the direction of each edge is randomized.
\begin{equation}
	p\text{-value} = \Pr[H(s^*;A) \leq H(\tilde s^*; \tilde A)] \, .
	\label{eq:pvalue}
\end{equation}
In practice, this $p$-value is estimated by drawing many samples from the null distribution by randomizing the edge directions of $A$ to produce $\tilde A$, computing the ranks $\tilde{s}^*$ from Eq.~\eqref{eq:linearsystSR}, and then computing the ground state energy Eq.~\eqref{eq:groundstateE} of each.

\subsection*{Cross-validation tests}
\vspace{-0.1in}
We performed edge prediction using 5-fold cross-validation. In each realization, we divide the interacting pairs $i,j$, i.e., those with nonzero $\bar{A}_{ij} = A_{ij}+A_{ji}$, into five equal groups.  We use four of these groups as a training set, inferring the ranks and setting $\beta$ to maximize $\eps_a$ or $\eps_L$ (on the left and right of Fig.~\ref{fig:ED} respectively).  We then use the fifth group as a test set, asking the algorithm for $P_{ij}$ for each pair $i,j$ in that group, and report $\eps_a$ or $\eps_L$ on that test set.  By varying which group we use as the test set, we get 5 trials per realization: for instance, 50 realizations give us 250 trials of cross-validation.
Results for 2-fold cross-validation are reported in SI.

\section*{Acknowledgements}
\vspace{-0.1in}
{CDB and CM were supported by the John Templeton Foundation. CM was also supported by the Army Research Office under grant W911NF-12-R-0012. DBL was supported by NSF award SMA-1633747 and the Santa Fe Institute Omidyar Fellowship.  We thank Aaron Clauset and Johan Ugander for helpful comments. Competing Interests: The authors declare that they have no competing interests. All authors derived the model, analyzed results, and wrote the manuscript. CDB wrote Python implementations and DBL wrote MATLAB implementations. Open-source code in Python, MATLAB, and SAS/IML available at \href{https://github.com/cdebacco/SpringRank}{https://github.com/cdebacco/SpringRank}.} 
\bibliographystyle{Science_inctitles}
\bibliography{bibliography}


\newcommand{\beginsupplement}{%
        \setcounter{table}{0}
        \renewcommand{\thetable}{S\arabic{table}}%
        \setcounter{figure}{0}
        \renewcommand{\thefigure}{S\arabic{figure}}%
        \setcounter{equation}{0}
        \renewcommand{\theequation}{S\arabic{equation}}
         \setcounter{section}{0}
        \renewcommand{\thesection}{S\arabic{section}}
 }
     
\clearpage
\beginsupplement
\begin{widetext}

\section*{{Supporting Information (SI)}}

\section{Deriving the linear system minimizing the Hamiltonian}
\label{apx:gradient}

The SpringRank Hamiltonian Eq.~\eqref{eq:SRhamiltonian} is convex in $s$ and we set its gradient $\nabla H(s)=0$ to obtain the global minimum:
\begin{equation}
	\f{\partial H}{\partial s_i }= \sum_{j } \rup{ A_{ij}\bup{s_i  -s_j  -1} -  A_{ji}\bup{s_j  -s_i -1 } }=0 \, .
	\label{eq:hamsolution1}
\end{equation}
Let the weighted out-degree and in-degree be $d_{i}^{\text{out}}=\sum_j A_{ij}$ and $d_{i}^{\text{in}}=\sum_{j}A_{ji}$, respectively. Then Eq.~\eqref{eq:hamsolution1} can be written as
\begin{equation}
	\bup{d_{i}^{\text{out}}+d_{i}^{\text{in}}} s_i  -  \bup{d_{i}^{\text{out}}-d_{i}^{\text{in}}} - \sum_{j}\rup{A_{ij}+A_{ji} } s_j  = 0 \, .
	\label{eq:derivative}
\end{equation}
We now write the system of $N$ equations together by introducing the following matrix notation. Let $D^{\text{out}}=\text{diag}(d_{1}^{\text{out}}, \dots,d_{N}^{\text{out}} )$ and $D^{\text{in}}=\text{diag}(d_{1}^{\text{in}}, \dots,d_{N}^{\text{in}} )$ be diagonal matrices, let $\ones$ be the $N$-dimensional vector of all ones. Then Eq.~\eqref{eq:derivative} becomes
\begin{equation}
	\rup{ D^{\text{out}}+D^{\text{in}} - \bup{{A}+ {A}^{T}}} s = \rup{D^{\text{out}}-D^{\text{in}}  } \,\ones \, .
\end{equation}
This is a linear system of the type $B \, s = b$, where $B=\rup{ D^{\text{out}}+D^{\text{in}} - \bup{ {A}+ {A}^{T}}}$ and $b=\rup{D^{\text{out}}-D^{\text{in}}  } \,\ones$. The rank of $B$ is at most $N-1$ and more generally, if the network represented by $A$ consists of $C$ disconnected components, $B$ will have rank $N-C$. In fact, $B$ has an eigenvalue $0$ with multiplicity $C$, and the eigenvector $\ones$ is in the nullspace. $B$ is not invertible, but we can only invert in the $N-C$-dimensional subspace orthogonal to the nullspace of $B$. The family of translation-invariant solutions $s^*$ is therefore defined by 
\begin{equation}
	s^* =\rup{ D^{\text{out}}+D^{\text{in}} - \bup{ {A}+ {A}^{T}}}^{-1} \rup{D^{\text{out}}-D^{\text{in}}  } \,\ones \, ,
\end{equation}
in which the notation $[\cdot]^{-1}$ should be taken as the Moore-Penrose pseudoinverse.

In practice, rather than constructing the pseudo-inverse, it will be more computationally efficient (and for large systems, more accurate) to solve the linear system in an iterative fashion. Since we know that solutions may be translated up or down by an arbitrary constant, the system can be made full-rank by fixing the position of an arbitrary node $0$. Without loss of generality, let $s_N=0$. In this case, terms that involve $s_N$ can be dropped from Eq.~\eqref{eq:derivative}, yielding
\begin{align}
	\bup{d_{i}^{\text{out}}+d_{i}^{\text{in}}} s_i  -  \bup{d_{i}^{\text{out}}-d_{i}^{\text{in}}} - \sum_{j=1}^{N-1} \rup{A_{ij}+A_{ji} } s_j  &= 0 \, , \qquad i\neq N \label{eq:derivative2i} \\
	 - \bup{d_{N}^{\text{out}}-d_{N}^{\text{in}}} - \sum_{j=1}^{N-1}\rup{A_{Nj} +A_{jN}  } s_j  &= 0	\label{eq:derivative2N} \, .
\end{align}
Adding Eq.~\eqref{eq:derivative2i} to Eq.~\eqref{eq:derivative2N} yields
\begin{equation}
	\bup{d_{i}^{\text{out}}+d_{i}^{\text{in}}} s_i  
	- \bup{d_{i}^{\text{out}}+ d_{N}^{\text{out}} - d_{i}^{\text{in}} -d_{N}^{\text{in}}} 
	- \sum_{j=1}^{N-1} \rup{A_{ij}+ A_{Nj} + A_{ji}  +A_{jN} } s_j  = 0 \, ,\\	
\end{equation}
which can be written in matrix notation as
\begin{equation}
	\left[ D^{\text{out}}+D^{\text{in}} - \mathring{A} \right ] s = \rup{D^{\text{out}}-D^{\text{in}}  } \,\ones +  \left( d_N^{\text{out}} - d_N^{\text{in}} \right) \ones\, ,
	\label{eq:nonsingular}
\end{equation}
where
\begin{equation}
	\mathring{A}_{ij} =A_{ij} + A_{Nj} + A_{ji} +A_{jN} \, .
\end{equation}
In this formulation, Eq.~\eqref{eq:nonsingular} can be solved to arbitrary precision using iterative methods that take advantage of the sparsity of $\mathring{A}$. The resulting solution may then be translated by an arbitrary amount as desired. 

\section{Poisson generative model}\label{poisson}

The expected number of edges from node $i$ to node $j$ is $c \exp{\left[ -\frac{\beta}{2}(s_i-s_j-1)^2\right]}$ and therefore the likelihood of observing a network $A$, given parameters $\beta$, $s$, and $c$ is
\begin{equation}
	P(A \mid s,\beta,c)\! =\! \prod_{i,j} \! \frac{\left[c\,\,\e^{-\frac{\beta}{2}\bup{s_i-s_j-1}^2 } \right ]^{\!A_{ij}} } {A_{ij}!} 
	\,\exp\!\left[-c\,\e^{-\frac{\beta}{2}\bup{s_i-s_j-1}^2 } \right] \, .
	\label{poissonloglikelihood}
\end{equation}
Taking logs yields
\begin{equation}
	\log P(A \mid s,\beta,c)\! =\! \sum_{i,j} A_{ij} \log c - \frac{\beta}{2} A_{ij}\bup{s_i - s_j - 1}^2 - \log \left[ A_{ij} ! \right ] - c\e^{- \frac{\beta}{2} \bup{s_i - s_j - 1}^2} \, .
\end{equation}
Discarding the constant term $\log \left[ A_{ij} ! \right]$, and recognizing the appearance of the SpringRank Hamiltonian $H(s)$, yields
\begin{equation}
	\mathcal{L}(A \mid s,\beta,c)\! =\! -\beta H(s) +  \sum_{i,j} A_{ij} \log c  - \sum_{i,j} c\e^{- \frac{\beta}{2} \bup{s_i - s_j -1}^2} \, .
\end{equation}
Taking $\partial \mathcal{L}/ \partial c$ and setting it equal to zero yields
\begin{equation}
	\hat{c} = \frac{\sum_{i,j} A_{ij}}{\sum_{i,j} \e^{- \frac{\beta}{2} \bup{s_i - s_j -1}^2}} \, ,
\end{equation}
which has the straightforward interpretation of being the ratio between the number of observed edges and the expected number of edges created in the generative process for $c=1$. Substituting in this solution and letting $M = \sum_{i,j} A_{ij}$ yields
\begin{align}
	\mathcal{L}(A \mid s,\beta)\! &=\! -\beta H(s) + M  \log \hat{c}  - \sum_{i,j} \hat{c} \,\e^{-\frac{\beta}{2} (s_i - s_j -1 )^2} \nonumber \\
	&=\! -\beta H(s) +M  \log M  - M  \log \left[ \sum_{i,j} \e^{-\frac{\beta}{2} \bup{s_i - s_j -1}^2} \right ]  - M \, .
\end{align}
The terms $M\log M $ and $M $ may be neglected since they do not depend on the parameters, and we divide by $\beta$, yielding a log-likelihood of 
\begin{equation}
	\mathcal{L}(A \mid s,\beta)\! =\! - H(s) - \frac{M}{\beta} \log \left[ \sum_{i,j} \e^{- \frac{\beta}{2} (s_i - s_j - 1)^2} \right ] \, .
\end{equation}
Note that the SpringRank Hamiltonian may be rewritten as  $H(s) = M \left[\frac{1}{2} \langle A_{ij}(s_i - s_j - 1)^{2} \rangle_{E} \right ]$ where $\langle \cdot \rangle_{E}$ denotes the average over elements in the edge set $E$. In other words, $H(s)$ scales with $M$ and the square of the average spring length. This substitution for $H(s)$ allows us to analyze the behavior of the log-likelihood
\begin{equation}\label{SIeqn:limit}
	\mathcal{L}(A \mid s,\beta)\! =\! - M  \left \{\frac{1}{2} \left \langle \, A_{ij} \, \bup{s_i - s_j - 1}^{2} \right \rangle_{E}  +\frac{1}{\beta} \log \left[ \sum_{i,j} \e^{- \frac{\beta}{2} (s_i - s_j - 1)^2} \right ] \right \} \, .
\end{equation}
Inside the logarithm there are $N^2$ terms of finite value, so that the logarithm term is of order $O(\frac{\log N}{\beta})$. 
Thus, for well-resolved hierarchies, i.e. when $\beta$ is large enough that the sampled edges consistently agree with the score difference between nodes, the maximum likelihood ranks $\hat{s}$ approach the ranks $s^*$ found by minimizing the Hamiltonian. In practice, exactly maximizing the likelihood would require extensive computation, e.g. by using local search heuristic or Markov chain Monte Carlo sampling.

\section{Rewriting the energy}
\label{SI:rewriteE}

The Hamiltonian Eq.~\eqref{eq:SRhamiltonian} can be rewritten as
 \bea
 2 \,H(s)&=&\sum_{i,j=1}^N  A_{ij}  \, \bup{s_i -s_j -1}^2  \nonumber \\
 &=& \sum_{i,j=1}^N  A_{ij} \, \bup{s_i ^2 +s_j ^2  -2 s_i s_j  +1 -2 s_i  +2 s_j } \nonumber \\
 &=&  \sum_{i}^N  s_i ^2  \sum_{j=1}^N  A_{ij} + \sum_{j}^N  s_j ^2  \sum_{i=1}^N  A_{ij}  -2 \sum_{i} s_i \sum_{j} A_{ij} s_j  + M \nonumber \\
 && -2\sum_{i=1}^N  s_i  \sum_{j=1}^N A_{ij}  +2\sum_{j=1}^N  s_j  \sum_{i=1}^N A_{ij}  \nonumber \\
 &=&  \sum_{i}^N  s_i ^2  \bup{ d_{i}^{\text{out}} + d_{i}^{\text{in}} } -2 \sum_{i} s_i  \bup{ d_{i}^{\text{out}}-d_{i}^{\text{in}}} +M   -2 \sum_{i} s_i \sum_{j} A_{ij} s_j  \, .
 \label{eq:rewritingH}
 \eea
 From Eq.~\eqref{eq:derivative} we have
 \begin{equation}
	 \sum_{j} s_i ^2  \bup{ d_{i}^{\text{out}} + d_{i}^{\text{in}} } - \sum_{i} s_i  \bup{ d_{i}^{\text{out}}-d_{i}^{\text{in}}} = \sum_{i} s_{i } \sum_{j}\rup{A_{ij}+A_{ji}  } s_j \, .
\end{equation}
 We can substitute this into Eq.~\eqref{eq:rewritingH}
 \bea
	 2H(s) &=& \sum_{i} s_{i } \sum_{j}\rup{A_{ij}+A_{ji}  } s_j  - \sum_{i} s_i  \bup{ d_{i}^{\text{out}}-d_{i}^{\text{in}}} + M  -2 \sum_{i} s_i \sum_{j} A_{ij} s_j \nonumber \\
	 &=&\sum_{i} s_i  \bup{ d_{i}^{\text{in}}-d_{i}^{\text{out}}} + M  \nonumber \\ 
	 &=&\sum_{i} h_{i} s_i  + M\, ,  \label{quenched}  
 \eea
 where $h_{i}\equiv d_{i}^{\text{in}}-d_{i}^{\text{out}}$.

 \section{Ranks distributed as a multivariate Gaussian distribution}\label{SI:multivariatenormal}
 
 Assuming that the ranks are random variables distributed as a multivariate Gaussian distribution of average $\b s$ and covariance matrix $\Sigma$, we have:
\begin{equation}
	P(s) \propto \exp{\left(-\frac{1}{2} (s-\b s)^{\intercal} \Sigma^{-1} (s-\b s)\right)} \, .
	\label{eq:MN} 
\end{equation}
 We can obtain this formulation by considering a Boltzman distribution with the Hamiltonian Eq.~\eqref{eq:SRhamiltonian} as the energy term and inverse temperature $\beta$  so that
\begin{equation}
	P(s) \propto \exp{\bup{ -\frac{\beta}{2} \sum_{i, j=1}^N   A_{ij} \left(s_i -s_j -1 \right)^2 } }\, . 
	\label{BoltzmanSR}
\end{equation}
Manipulating the exponent of Eq.~\eqref{eq:MN} yields
\begin{equation}
	\frac{1}{2}(s-\bar s)^{T}  \Sigma^{-1} (s-\bar s) = \frac{1}{2} \left(s^{T} \Sigma^{-1} s - 2s^{T}\Sigma^{-1}\bar s + \bar s^{T} \Sigma^{-1} \bar s \right) \, ,
\end{equation} 
whereas the parallel manipulation of Eq.~\eqref{BoltzmanSR} yields
\begin{equation}
	\frac{\beta}{2} \sum_{i, j=1}^N   A_{ij} \left(s_i -s_j -1 \right)^2 =\frac{\beta}{2} \rup{ s^T \bup{D^{\text{out}} + D^\text{in} - {A}^T - {A}}s + 2\, \,s^T(D^\text{in} - D^\text{out})\ones  + M  }\, ,
\end{equation}
where $\ones$ is a vector of ones and $D^{\text{in}}$ are diagonal matrices whose entries are the in- and out-degrees, $D^{\text{out}}_{ii} = \sum_j {A}_{ij}$ and  $D^{\text{in}}_{ii} = \sum_j {A}_{ji}$ and $M=\sum_{i,j}{A}_{ij}$. Comparing these last two expressions and removing terms that do not depend on $s$ because irrelevant when accounting for normalization, we obtain:
\begin{equation}
	\Sigma =\tfrac{1}{\beta} \left( D^\text{out}+D^\text{in} - {A}^{T} - {A} \right)^{-1} \quad \text{and} \quad \bar s =  \, \beta \,\Sigma \bup{D^\text{out} - D^\text{in}} \ones = s^* \, .
\end{equation}

\section{Bayesian SpringRank}\label{SI:bayesianSR}
 
Adopting a Bayesian approach with a factorized Gaussian prior for the ranks,  we obtain that the $s$ that maximizes the posterior distribution is the one that minimizes the regularized SpringRank Hamiltonian Eq.~\eqref{eq:fullH}, i.e. the $s$ that solves the linear system Eq.~\eqref{eq:FullLinearSystem}. In fact, defining:

\begin{equation}
	P( s)=Z^{-1}(\beta,\alpha) \prod_{i \in V}\e^{-\beta \, \f{\alpha}{2}  (s_i -1 )^2 }= Z^{-1}(\beta,\alpha) \prod_{i \in V}\e^{-\beta \, \alpha H_0(s_i )}\, ,
\end{equation}
where $Z(\beta,\alpha)=\rup{\f{2 \pi}{\beta \, \alpha} }^{N/2}$ is a normalization constant that depends on $\alpha$ and $\beta$,
and following the same steps as before we get:
\begin{align}
	\log P( s \mid A) =&\sum_{i,j}\log P(A_{ij}\mid  s) -\beta \alpha \sum_{i \in V} H_0(s_i ) +\log\bup{Z\bup{\beta,\alpha}\,} \nonumber  \\
	 = & -\beta \rup{ H( s)  + \alpha \sum_{i \in V} H_0(s_i )} + C\, ,
\end{align}
where $C$ is a constant that does not depend on the parameters, and thus may be ignored when maximizing $\log P(s \mid A)$.

\section{Fixing $c$ to control for sparsity}\label{SI:cforsparsity}
The parameter $c$ included in the generative model \eqref{eq:generativemodel} controls for network's sparsity. We can indeed fix it so to obtain a network with a desired expected number of edges $\langle M \rangle$ as follows:

\begin{equation}
	\langle M \rangle \equiv \sum_{i,j} \langle A_{ij} \rangle= c\sum_{i,j} \e^{-\f{\beta}{2} \bup{s_i-s_j-1}^2 } \, .\nonumber
\end{equation}
For a given  vector of ranks $s$ and inverse temperature $\beta$, the $c$ realizing the desired sparsity will then be:
\begin{equation}
	c=\f{\langle M \rangle }{\sum_{i,j} \e^{-\f{\beta}{2} \bup{s_i -s_j -1 }^2 }} =\f{\langle k \rangle N }{\sum_{i,j} \e^{-\f{\beta}{2} \,\bup{s_i-s_j-1}^2 }} \, ,
\end{equation}
where $\langle k \rangle$ is the expected node degree $\langle k \rangle=\sum_{i=1}^N  \rup{d_{i}^{\text{in}}+d_{i}^{\text{out}}}$. Similar arguments apply when considering a generative model with Bernoulli distribution.

\section{Comparing optimal $\beta$ for predicting edge directions}
\label{SI:bestbeta}

In the main text, \eqref{eq:localaccuracy} and Eq.~\eqref{eq:loglikelihood} define the accuracy of edge prediction, in terms of the number of edges predicted correctly in each direction and the log-likelihood conditioned on the undirected graph. Here we compute the optimal values of $\beta$ for both notions of accuracy. In both computations that follow, the following two facts will be used:
\begin{align}
	P'_{ij}(\beta) &=  2\bup{s_i -s_j } \, \e^{-2 \beta \bup{s_i -s_j }} \, P^2 _{ij}(\beta)\, ,
	\label{prelim1}
\end{align}
and
\begin{align}
	1 &= \f{P_{ij}(\beta)}{1-P_{ij}(\beta)}\, \e^{-2 \beta \bup{s_i -s_j }}\, .
	\label{prelim2}
\end{align}

\subsection{Choosing $\beta$ to optimize edge direction accuracy}

We take the derivative of Eq.~\eqref{eq:localaccuracy} with respect to $\beta$, set it equal to zero, and solve as follows.
\begin{equation}
	0 \equiv \frac{\partial \eps_a(\beta)}{\partial \beta} = \frac{\partial}{\partial \beta} \left[1- \frac{1}{2m} \, \sum_{i,j} \left| A_{ij} -\bup{A_{ij} +A_{ji}  } \,P_{ij}(\beta) \right|  \right ]\, .
	\label{setup}
\end{equation}
In preparation to take the derivatives above, note that $P'_{ij}(\beta) = - P'_{ji}(\beta)$ and that whenever the $(i,j)$ term of $ \eps_a(\beta)$ takes one sign, the $(j,i)$ term takes the opposite sign,
\begin{equation}
	A_{ij} - \bup{A_{ij} +A_{ji}  }P_{ij}(\beta) = - \left[ A_{ji} - \bup{A_{ij} +A_{ji}  }P_{ji}(\beta) \right]\, .
\end{equation}
Without loss of generality, assume that the $(i,j)$ term is positive and the $(j,i)$ term is negative. This implies that 
\begin{align}
	\frac{\partial}{\partial \beta} \left | A_{ij} - \bup{A_{ij} +A_{ji}  }P_{ij}(\beta) \right | &= -\bup{A_{ij} +A_{ji}  }P'_{ij}(\beta)\, ,
\end{align}
and 
\begin{align}
	\frac{\partial}{\partial \beta} \left | A_{ji} - \bup{A_{ij} +A_{ji}  }P_{ji}(\beta) \right | &= -\bup{A_{ij} +A_{ji}  }P'_{ij}(\beta)\, .
\end{align}
In other words, the derivatives of the $(i,j)$ and $(j,i)$ terms are identical, and the sign of both depends on whether the quantity $\left[ A_{ij} - (A_{ij} + A_{ji})P_{ij}(\beta) \right]$ is positive or negative. We can make this more precise by directly including the sign of the $(i,j)$ term, and by using Eq.~\eqref{prelim1}, to find that
\begin{equation}
	\frac{\partial}{\partial \beta} \left | A_{ij} - \bup{A_{ij} +A_{ji}  }P_{ij}(\beta) \right | = - 2 \bup{A_{ij} +A_{ji}  }\bup{s_i -s_j } \, \e^{-2 \beta \bup{s_i -s_j }} \, P^2 _{ij} \times \text{sign} \big \{A_{ij} - \bup{A_{ij} +A_{ji}  }P_{ij}(\beta) \big \}\, .
\end{equation}
Expanding $P_{ij}^2$ and reorganizing yields
\begin{equation}
	\frac{\partial}{\partial \beta} \left | A_{ij} - \bup{A_{ij} +A_{ji}  }P_{ij}(\beta) \right | = - 2 \frac{\bup{A_{ij} +A_{ji}  }\bup{s_i - s_j}}{2 \cosh\left[2\beta\bup{s_i - s_j}\right] + 2 } \times \text{sign} \big \{A_{ij} - \bup{A_{ij} +A_{ji}  }P_{ij}(\beta) \big \}\, .
\end{equation}
Combining terms $(i,j)$ and $(j,i)$, the optimal inverse temperature for local accuracy $\hat{\beta}_a$ is that which satisfies
\begin{equation}
	0 = \sum_{(i,j) \in U(E)}^N \frac{\bup{A_{ij} +A_{ji}  }\bup{s_i - s_j}}{\cosh\left[2\hat{\beta}_a\bup{s_i - s_j}\right] + 1 } \times \text{sign} \big \{A_{ij} - \bup{A_{ij} +A_{ji}  }P_{ij}(\hat{\beta}_a) \big \} \, ,
	\label{deltaepsilon}
\end{equation}
which may be found using standard root-finding methods.

\subsection{Choosing $\beta$ to optimize the conditional log likelihood}

We take the derivative of Eq.~\eqref{eq:loglikelihood} with respect to $\beta$, set it equal to zero, and partially solve as follows.
\begin{equation}
	0 \equiv \frac{\partial \eps_L(\beta)}{\partial \beta} = \frac{\partial}{\partial \beta} \left[\sum_{i,j}  \log \binom{A_{ij}+A_{ji}}{A_{ij}} + \log \left[ P_{ij}(\beta)^{A_{ij}} \, \rup{1-P_{ij}(\beta)}^{A_{ji }} \right ] \,\right ]\, .
	\label{setup2}
\end{equation}
Combining the $(i,j)$ and $(j,i)$ terms, we get 
\begin{align}
	0 \equiv &\f{\partial}{\partial \beta} \sum_{(i,j) \in U(E)} \log \binom{A_{ij}+A_{ji}}{A_{ij}} + \log \binom{A_{ij}+A_{ji}}{A_{ji}} + \rup{\, A_{ij} \log P_{ij}(\beta) \, + A_{ji} \log \rup{1-P_{ij}(\beta)}\,} \nonumber \\
	=& 	\sum_{(i,j) \in U(E)}  \rup{\f{ A_{ij}}{P_{ij}(\beta)} \,  -\f{ A_{ji}}{1-P_{ij}(\beta)} 			 } \f{\partial  P_{ij}(\beta)}{\partial \beta} \nonumber \\
	=& 		 \sum_{(i,j) \in U(E)} 2 \bup{s_i -s_j } \, \rup{ A_{ij} -\bup{A_{ij}+A_{ji}} \, P_{ij}(\beta) }  \f{P_{ij}(\beta)}{1-P_{ij}(\beta)}\, \e^{-2 \beta \bup{s_i -s_j }}\, .
\end{align}
Applying both Eq.~\eqref{prelim1} and Eq.~\eqref{prelim2}, the optimal inverse temperature for the conditional log likelihood $\hat{\beta}_{L}$ is that which satisfies
\begin{equation}
	0= \sum_{(i,j) \in U(E)} 2 \bup{s_i -s_j } \, \rup{ A_{ij} -\bup{A_{ij}+A_{ji}} \, P_{ij}(\hat{\beta}_{L}) }\, ,
	\label{deltalogL}
\end{equation}
which, like Eq.~\eqref{deltaepsilon} may be found using standard root-finding methods. Comparing equations Eq.~\eqref{deltaepsilon} and Eq.~\eqref{deltalogL}, we can see that the values of $\beta$ that maximize the two measures may, in general, be different. Table \ref{tableresults} shows for optimal values for $\hat{\beta}_L$ and $\hat{\beta}_a$ for various real-world datasets.

\section{Bitwise accuracy $\eps_b$}
\label{apx:maxfraction}

Some methods provide rankings but do not provide a model to estimate $P_{ij}$, meaning that Eq.~\eqref{eq:localaccuracy} and Eq.~\eqref{eq:loglikelihood} cannot be used. Nevertheless, such methods still estimate one bit of information about each pair $(i,j)$: whether the majority of the edges are from $i$ to $j$ or vice versa. This motivates the use of a bitwise version of $\eps_a$, which we call $\eps_b$,
\begin{equation}
	\eps_b =1- \frac{1}{N^2 - t} \sum_{i,j} \Theta\bup{s_i -s_j }\Theta(A_{ji}-A_{ij}) \, ,
	\label{maxfractionED}
\end{equation}
where $\Theta(x)=1$ if $x>0$ and $\Theta(x) = 0$ otherwise, and $N$ is the number of nodes and $t$ is the number of instances in which $A_{ij}=A_{ji}$; there are $N^2-t$ total bits to predict. Results in terms of this measure on the networks considered in the main text are shown in Figure~\ref{SIfig:ED}.
In the special case that the network is unweighted ($A$ is a binary adjacency matrix) and there are no bi-directional edges (if $A_{ij}=1$, then $A_{ji}=0$), then $1-\eps_b$ is the fraction of edges that violate the rankings in $s$. In other words, for this particular type of network, $1-\eps_b$ is the minimum violations rank penalty normalized by the total number of edges in the network, i.e.,  $\frac{1}{M} \sum_{i,j} \Theta(s_{i}-s_{j})\, A_{ji}$.

\section{Performance metrics}\label{SI:perfmetrics}
When evaluating the performance of a ranking algorithm in general one could consider a variety of different measures. 
One possibility is to focus on the ranks themselves, rather than the outcomes of pairwise interactions, and calculate correlation coefficients as in Fig. \ref{fig:synthetic}; this is a valid strategy when using synthetic data thanks to the presence of ground truth ranks, but can only assess the performance with respect of the specific generative process used to generate the pairwise comparisons, as we point out in the main text. This strategy can also be applied for comparisons with \emph{observed} real world ranks, as we did in Table~\ref{SItbl:USCSpearson} and it has been done for instance in \cite{cucuringu2016sync,fogel2014serialrank} to compare the ranks with those observed in real data in sports. However, the observed ranks might have been derived from a  different process than the one implied by the ranking algorithm considered. For instance, in the faculty hiring networks, popular ranking methods proposed by domain experts for evaluating the prestige of universities do not consider interactions between institutions, but instead rely on a combination of performance indicators such as first-year student retention or graduation rates. The correlation between observed and inferred ranks should thus be treated as a qualitative indicator of how well the two capture similar features of the system, such as prestige, but should not be used to evaluate the performance of a ranking algorithm.

Alternatively, one can look at the outcomes of the pairwise comparisons and relate them to the rankings of the nodes involved as in Eqs.~\eqref{eq:localaccuracy} and~\eqref{eq:loglikelihood} for testing prediction performance. A popular metric of this type is the number of violations (also called upsets), i.e., outcomes where a higher ranked node is defeated by a lower ranked one. This is very similar to the bitwise accuracy defined in (\ref{maxfractionED}), indeed when there are no ties and two nodes are compared only once, then they are equivalent. These can be seen as low-resolution or coarse-grained measures of performance: for each comparison predict a winner, but do not distinguish between cases where the winner is easy to predict and cases where there is almost a tie. In particular, an upset between two nodes ranked nearby counts as much as an upset between two nodes	that are far away in the ranking. The latter case signals a much less likely scenario. In order to distinguish these two situations, one can penalize each upset by the nodes' rank difference elevated to a certain power $d$. This is what the \emph{agony} function does~\cite{gupte2011finding} with the exponent $d$ treated as a parameter to tune based on the application. When $d=0$ we recover the standard number of unweighted upsets.

Note that optimization of agony is often used as a non-parametric approach to detect hierarchies \cite{letizia2018resolution}, in particular for ordinal ranks. For ordinal ranks, rank differences are integer-valued and equal to one for adjacent-ranked nodes, yet for real-valued scores this is not the case. Therefore the result of the agony minimization problem can vary widely between ordinal and real valued ranking algorithms.
(We note that the SpringRank objective function, i.e., the Hamiltonian in Eq.~\eqref{eq:SRhamiltonian}, can be considered a kind of agony. However, since we assume that nearby pairs are more likely to interact, it is large for a edge from $i$ to $j$ if $i$ is ranked far above or far below $j$, and more specifically whenever $s_i$ is far from $s_j+1$.)

In contrast to the coarse prediction above---which competitor is more likely to win?---we require, when possible, more precise predictions in Eqs.~\eqref{eq:localaccuracy} and~\eqref{eq:loglikelihood}, which ask {\it how much more likely} is one competitor to win? This, however, requires the ranking algorithm to provide an estimate of $P_{ij}$, the probability that $i$ wins over $j$, which is provided only by BTL and SpringRank; all other methods compared in this study provide orderings or embeddings without probabilistic predictions.

The conditional log-likelihood $\eps_L$ as defined in Eq.~\eqref{eq:loglikelihood} can be seen as a \emph{Log Loss} often used as a classification loss function \cite{rosasco2004loss} in statistical learning. This type of function heavily penalizes ranking algorithms that are very confident about an incorrect outcome, e.g. when the predicted $P_{ij}$ is close to 1, $i$ very likely to win over $j$, but the observed outcome is that $j$ wins over $i$. For this reason, this metric is more sensitive to outliers, as when in sports a very strong team loses against one at the bottom of the league. The accuracy $\eps_b$ defined in Eq.~\eqref{eq:localaccuracy} focuses instead in predicting the correct proportions of wins/losses between two nodes that are matched in several comparisons. This is less sensitive to outliers, and in fact if $P_{ij}$ is close but not exactly equal to 1, for a large number of comparisons between $i$ and $j$, we would expect that $j$ should indeed win few times, e.g. if $P_{ij}=0.99$ and $i$, $j$ are compared 100 times, $\eps_a$ is maximized when $i$ wins 99 times and $j$ wins once.

\section{Parameters used for regularizing ranking methods}\label{apx:algoparameters}
When comparing SpringRank to other methods, we need to deal with the fact that certain network structures cause other methods to fail to return any output. Eigenvector Centrality cannot, for example, be applied to directed trees, yet this is precisely the sort of structure that one might expect when hierarchy becomes extreme.

More generally, many spectral techniques fail on networks that are not \emph{strongly connected}, i.e., where it is not the case that one can reach any node from any other by moving along a path consistent with the edge directions, since in that case the adjacency matrix is not irreducible and the Perron-Frobenius theorem does not apply. In particular, nodes with zero out-degree---sometimes called ``dangling nodes'' in the literature~\cite{page1999}---cause issues for many spectral methods since the adjacency matrix annihilates any vector supported on such nodes. In contrast, the SpringRank optimum given by Eq.~\eqref{eq:linearsystSR} is unique up to translation whenever the network is connected in the undirected sense, i.e., whenever we can reach any node from any other by moving with or against directed edges.

A different issue occurs in the case of SyncRank. When edges are reciprocal in the sense that an equal number of edges point in each direction, they effectively cancel out. That is, if $A_{ij}=A_{ji}$, the corresponding entries in the SyncRank comparison matrix will be zero, $C_{ij}=C_{ji}=0$, as if $i$ and $j$ were never compared at all. As a result, there can be nodes $i$ such that $C_{ij} = C_{ji} = 0$ for all $j$. While rare, these pathological cases exist in real data and during cross-validation tests, causing the output of SyncRank to be undefined.
 
In all these cases, regularization is required. Our regularized implementations of five ranking methods are described below:
\begin{itemize}
	\item \textbf{Regularized Bradley-Terry-Luce (BTL).} If there exist dangling nodes, the Minimization-Maximization algorithm to fit the BTL model to real data proposed in~\cite{hunter2004mm} requires a regularization. In this case we set the total number of out-edges $d_{i}^{\text{out}}=10^{-6}$ for nodes that would have $d_{i}=0$ otherwise. This corresponds to $W_{i}$ in Eq.(3) of~\cite{hunter2004mm}. 
	\item \textbf{Regularized PageRank.} If there exist dangling nodes, we add an edge of weight $1/N$ from each dangling node to every other node in the network. For each dataset we tried three different values of the teleportation parameter, $\alpha\in \{0.4,\, 0.6,\, 0.8\}$, and reported the best results of these three.
	\item \textbf{Regularized Rank Centrality.} If there exist dangling nodes, we use the regularized version of the algorithm presented in Eq. (5) of \cite{negahban2016rank} with $\epsilon=1$.
	\item \textbf{Regularized SyncRank.}  If there are nodes whose entries in the comparison matrix $C$ are zero, we add a small constant $\epsilon=0.001$ to the entries of $H$ in Eq.~(13) of Ref.~\cite{cucuringu2016sync}, so that $D$ is invertible.
	\item \textbf{Regularized Eigenvector Centrality.} 
	If the network is not strongly connected, we add a weight of $1/N$ to every entry in $A$ and then diagonalize.
\end{itemize}

\clearpage
\section{Supplemental Tables}

\begin{table}[h]\label{SItbl:USCSpearson}
	\centering
\begin{tabular}{l | rrrrrr}
\hline
\textbf{Comp. Sci.} &   SpringRank &  MVR&  \emph{US News} &  NRC  &   Eig. C. &   PageRank \\
\hline
SpringRank      & -& 0.96 &     0.80 &   0.72 & 0.84 & 0.57 \\
MVR     & 0.96 & - &     0.81 &   0.73 & 0.80 & 0.48 \\
\emph{US News} & 0.80 & 0.81 &     -&   0.73 & 0.69 & 0.41 \\
NRC  & 0.72 & 0.73 &     0.73 &   - & 0.68 & 0.41 \\
Eig. C.      & 0.84 & 0.80 &     0.69 &   0.68 & - & 0.74 \\
 PageRank      & 0.57 & 0.48 &     0.41 &   0.41 & 0.74 & - \\
\hline
\\
\textbf{Business} &   SpringRank &  MVR&  \emph{US News} &  NRC  &   Eig. C. &   PageRank \\
\hline
SpringRank      & - & 0.98 &     0.74 &   - & 0.92 & 0.75 \\
MVR     & 0.98 & -&     0.72 &   - & 0.92 & 0.69 \\
\emph{US News}  & 0.74 & 0.72 &    - &  -& 0.68 & 0.60 \\
NRC & - & -&     -&   - & - & - \\
 Eig. C.      & 0.92 & 0.92 &     0.68 &   - & - & 0.72 \\
PageRank      & 0.75 & 0.69 &     0.60 &  -& 0.72 & - \\
\hline
\\
\textbf{History} &   SpringRank &  MVR&  \emph{US News} &  NRC  &   Eig. C. &   PageRank \\
\hline
SpringRank      & - & 0.95 &     0.86 &     0.66 & 0.86 & 0.69 \\
MVR     & 0.95 & - &     0.86 &     0.65 & 0.77 & 0.57 \\
\emph{US News}  & 0.86 & 0.86 &     - &     0.66 & 0.72 & 0.51 \\
NRC & 0.66 & 0.65 &     0.66 &   - & 0.59 & 0.44 \\
 Eig. C.   & 0.86 & 0.77 &     0.72 &     0.59 & - & 0.88 \\
PageRank      & 0.69 & 0.57 &     0.51 &     0.44 & 0.88 & - \\
\hline
\end{tabular}
\caption{Pearson correlation coefficients between various rankings of faculty hiring networks. All coefficients are statistically significant ($p<10^{-9}$). SpringRank is most highly correlated with Minimum Violations Ranks across all three faculty hiring networks. Among {\it US News} and NRC rankings, SpringRank is more similar to {\it US News}. Values for {\it US News} and NRC were drawn from Ref. \cite{clauset2015systematic} for comparison to the ranks available at the same time that the faculty hiring data were collected. The NRC does not rank business departments.}
\end{table}

\begin{table}[h]
	\centering
\begin{tabular}{| l| l| l| l| l| l| l| l| l| l| l| l|}
	\hline
	DataSet	&Type	&$N$	&$M$ &$H/M$	&Acc. $\eps_a$	&$\hat{\beta}{_L}$	&$\hat{\beta}_a$	&Viol. (\%) / Bound	&Wt. viol. (per viol.)	&Depth	&$p$-value	\\ 
	\hline 
	Parakeet G1~\cite{hobson2015social}	&Anim. Dom.	&21	&838 &0.174	&0.930	&2.70	&6.03	&76 (9.1\%) / 42	&0.008 (0.089) 	&2.604	&$<10^{-4}$	\\ 
	Parakeet G2~\cite{hobson2015social}	&Anim. Dom.	&19	&961 &0.193	&0.932	&2.78	&18.12	&75 (7.8\%) / 36	&0.011 (0.139) 	&1.879	&$<10^{-4}$	\\ 
	Asian Elephants~\cite{elephant}	&Anim. Dom.	&20	&23 &0.078	&0.923	&2.33	&3.44	&2 (8.7\%) / 0	&0.001 (0.040) 	&3.000	&0.4466	\\ 
	Business~\cite{clauset2015systematic}	&Fac. Hiring	&112 &7353	&0.251	&0.881	&2.04	&3.14	&1171 (15.9\%) / 808	& 0.019 (0.119) 	&2.125	&$<10^{-4}$	\\ 
	Computer Science~\cite{clauset2015systematic}	&Fac. Hiring	&205 & 4033	&0.220	&0.882	&2.23	&8.74	&516 (12.8\%) / 255	&0.013 (0.105) 	&2.423	&$<10^{-4}$	\\ 
	History~\cite{clauset2015systematic}	&Fac. Hiring	&144	 & 3921 &0.186	&0.909	&2.39	&5.74	&397 (10.1\%) / 227	&0.012 (0.119) 	&2.234	&$<10^{-4}$	\\ 
	A\underbar lak\= apuram~\cite{power2017}	&Soc. Support	&415 & 2497	&0.222	&0.867	&1.98	&7.95	&347 (13.9\%) / 120	&0.011 (0.079) 	&3.618	&$<10^{-4}$	\\ 
	Te\underbar npa\d t\d ti~\cite{power2017}	&Soc. Support	&361 & 1809	&0.241	&0.858	&1.89	&8.20	&262 (14.5\%) / 120	&0.012 (0.082) 	&3.749	&$<10^{-4}$	\\
	\hline
\end{tabular}
	\caption{\textbf{Statistics for SpringRank applied to real-world networks.} \normalfont{Column details are as follows: $N$ is the number of nodes; $M$ is the number of edges; $H/m$ is the ground state energy per edge; Accuracy $\eps_a$ refers to accuracy in 5-fold cross-validation tests using temperature $\hat{\beta}_a$; $\hat{\beta}_L$ and $\hat{\beta}_a$ are temperatures optimizing edge prediction accuracies $\eps_L$ and $\eps_a$ respectively; Violations refers to the number of edges that violate the direction of the hierarchy as a number, as a percentage of all edges, with a lower bound provided for reference, computed as the number of unavoidable violations due to reciprocated edges; Weighted violations are the sum of each violation weighted by the difference in ranks between the offending nodes; Depth is $s_{\text{max}}-s_{\text{min}}$; $p$-value refers to the null model described in the Materials and Methods.} Relevant performance statistics for NCAA datasets (53 networks) are reported elsewhere; see Fig.~\ref{SIfig:EDNCAA}. }
	\label{tableresults}
\end{table}

\clearpage
\section{Supplemental Figures}

\begin{figure}[h]
	\includegraphics[width=0.48\linewidth]{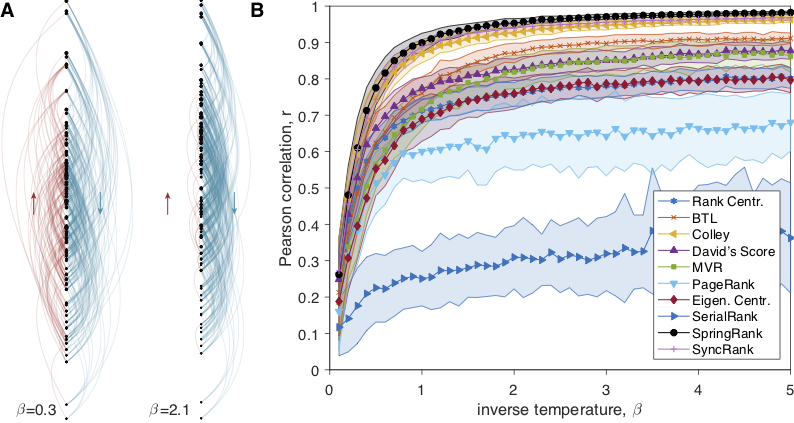}
	\includegraphics[width=0.48\linewidth]{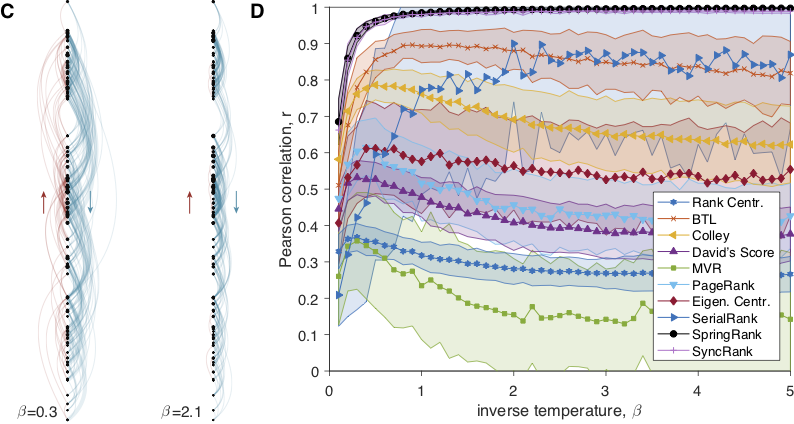}
	\caption{\textbf{Performance (Pearson correlation) on synthetic data.} Tests were performed as in Fig.~\ref{fig:synthetic}, but here performance is measured using Pearson correlation. This favors algorithms like SpringRank and BTL, that produce real-valued ranks, over ordinal ranking schemes like Minimum Violation Ranking which are not expected to recover latent positions.  (A) Linear hierarchy diagrams show latent ranks $s_\text{planted}$ of 100 nodes, drawn from a standard normal distribution, with edges drawn via the generative model Eq.~\eqref{eq:generativemodel} for indicated $\beta$ (noise) values. Blue edges point down the hierarchy and red edges point up, indicated by arrows. (B) Mean accuracies $\pm$ one standard deviation (symbols $\pm$ shading) are measured as the Pearson correlation between method output and $s_\text{planted}$ for 100 replicates. (C, D) Identical to A and B but for hierarchies of $N=102$ nodes divided into three tiers. All plots have mean degree $5$; see Fig.~\ref{fig:synthetic} for performance curves for Spearman correlation $r$. See Materials and Methods for synthetic network generation.}
	\label{sfig:pearson}
\end{figure}

\begin{figure*}[tbhp] 
 \centering
	\subfloat[US HS]{{\includegraphics[width=0.2\linewidth]{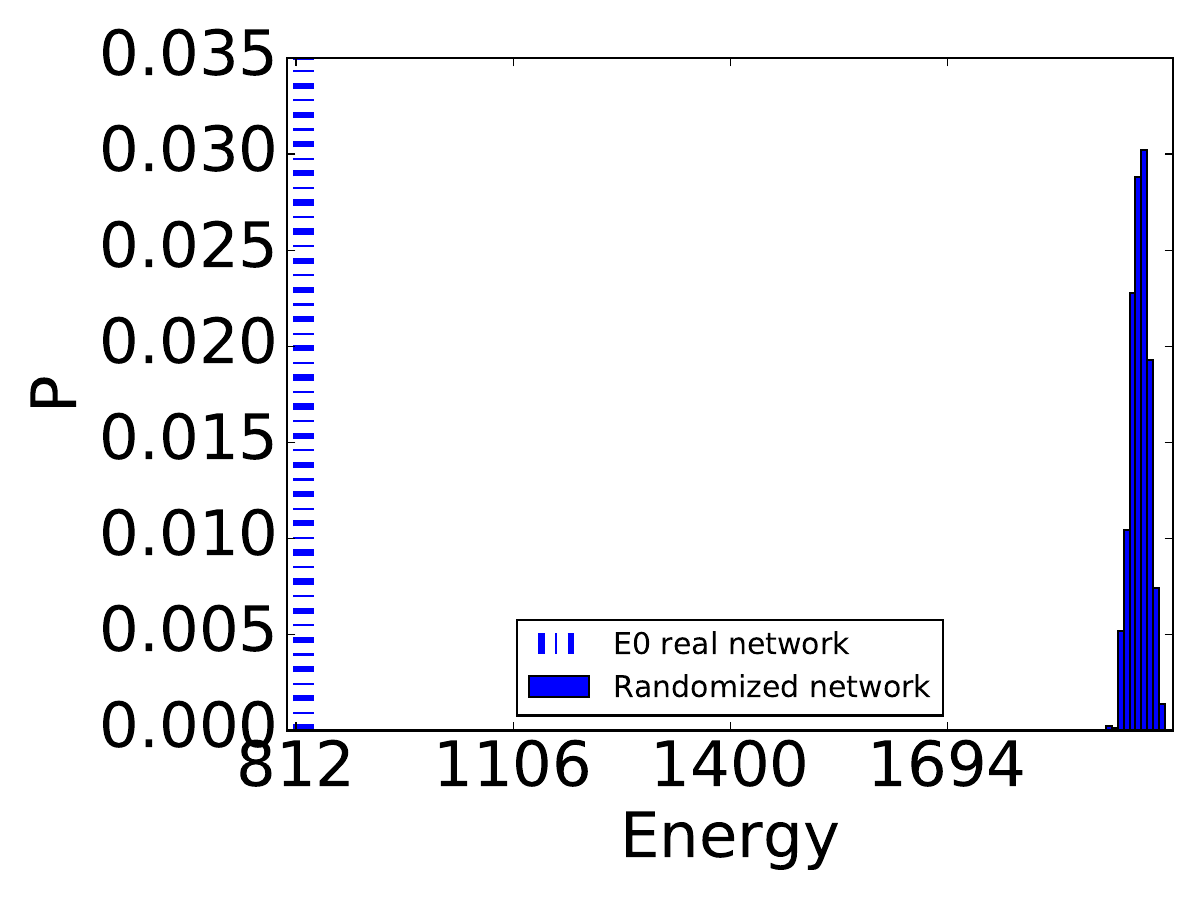} }}
	\subfloat[US BS]{{\includegraphics[width=0.2\linewidth]{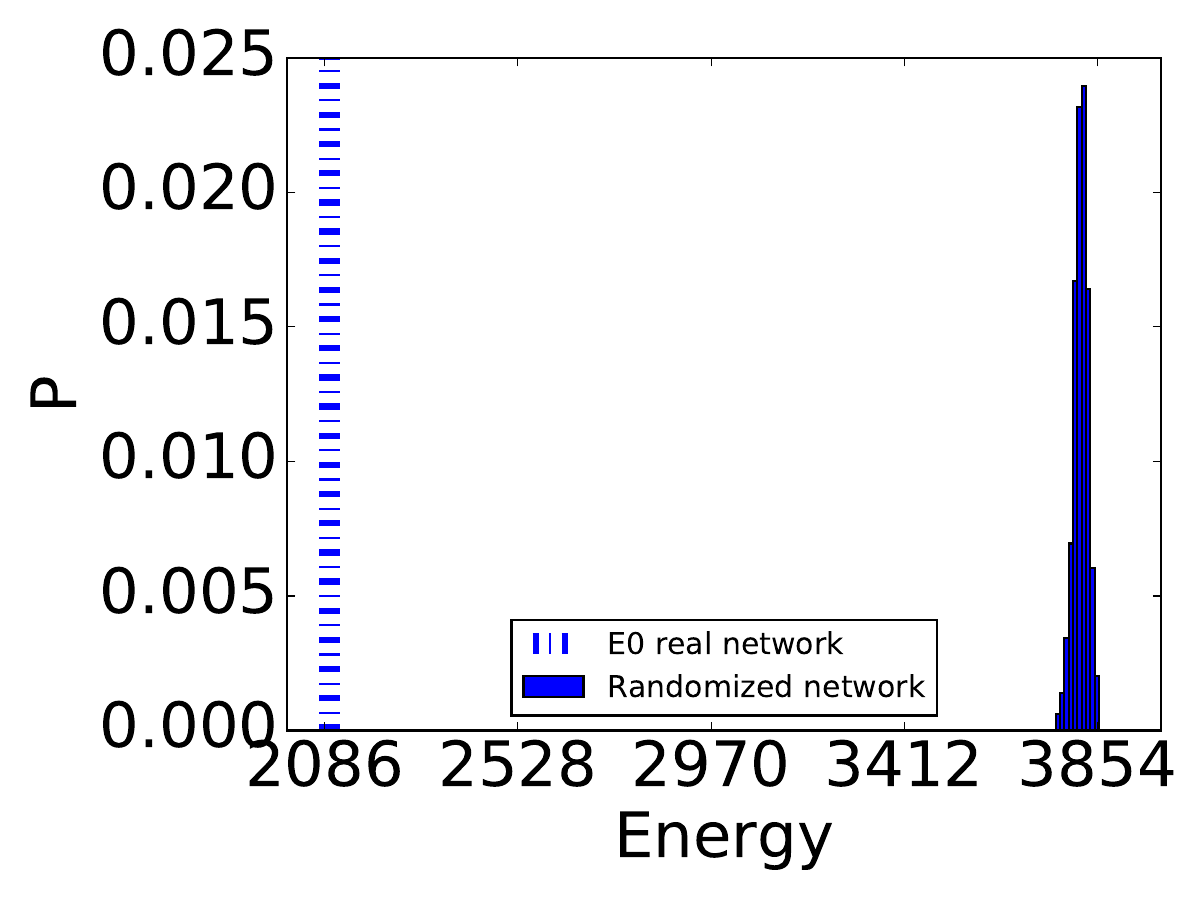} }}
	\subfloat[US CS]{{\includegraphics[width=0.2\linewidth]{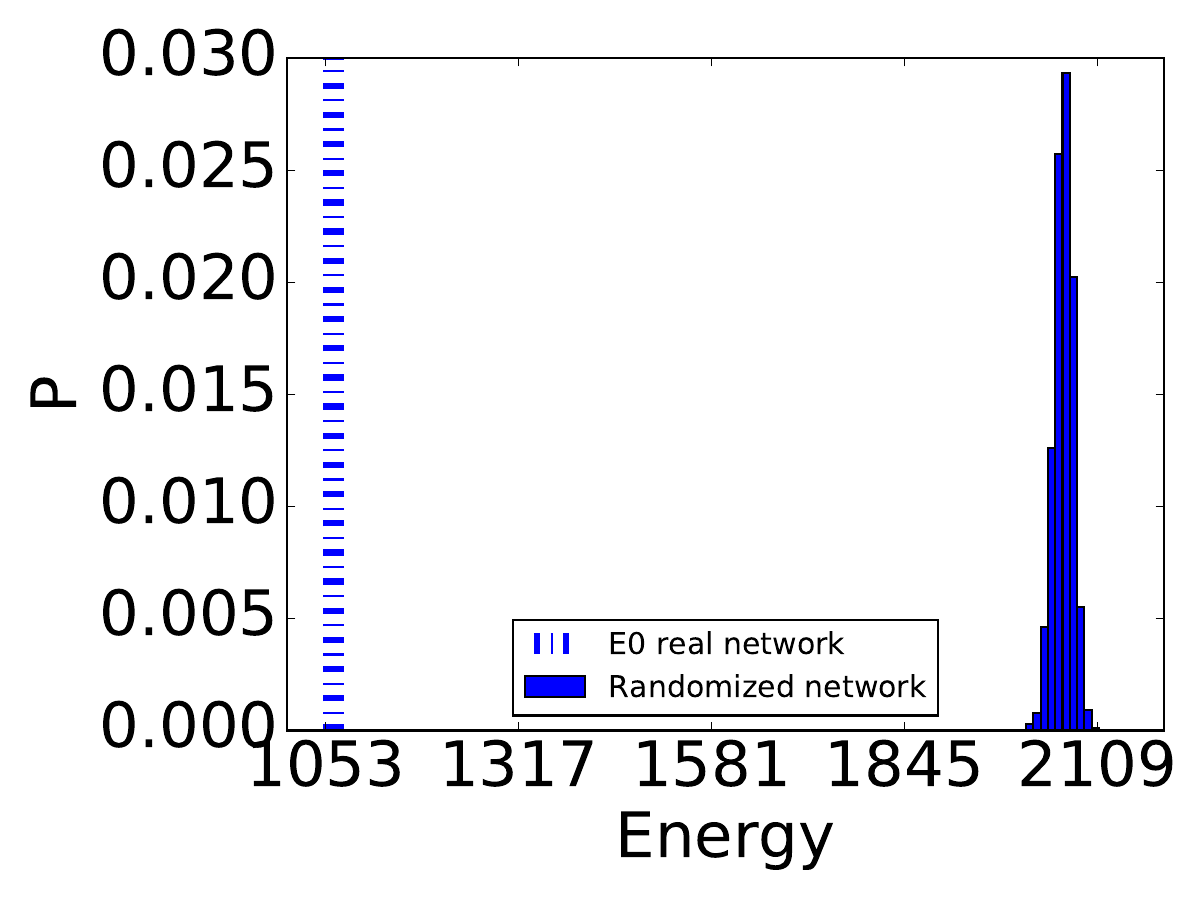} }}
	\subfloat[Te\underbar npa\d t\d ti]{{\includegraphics[width=0.2\linewidth]{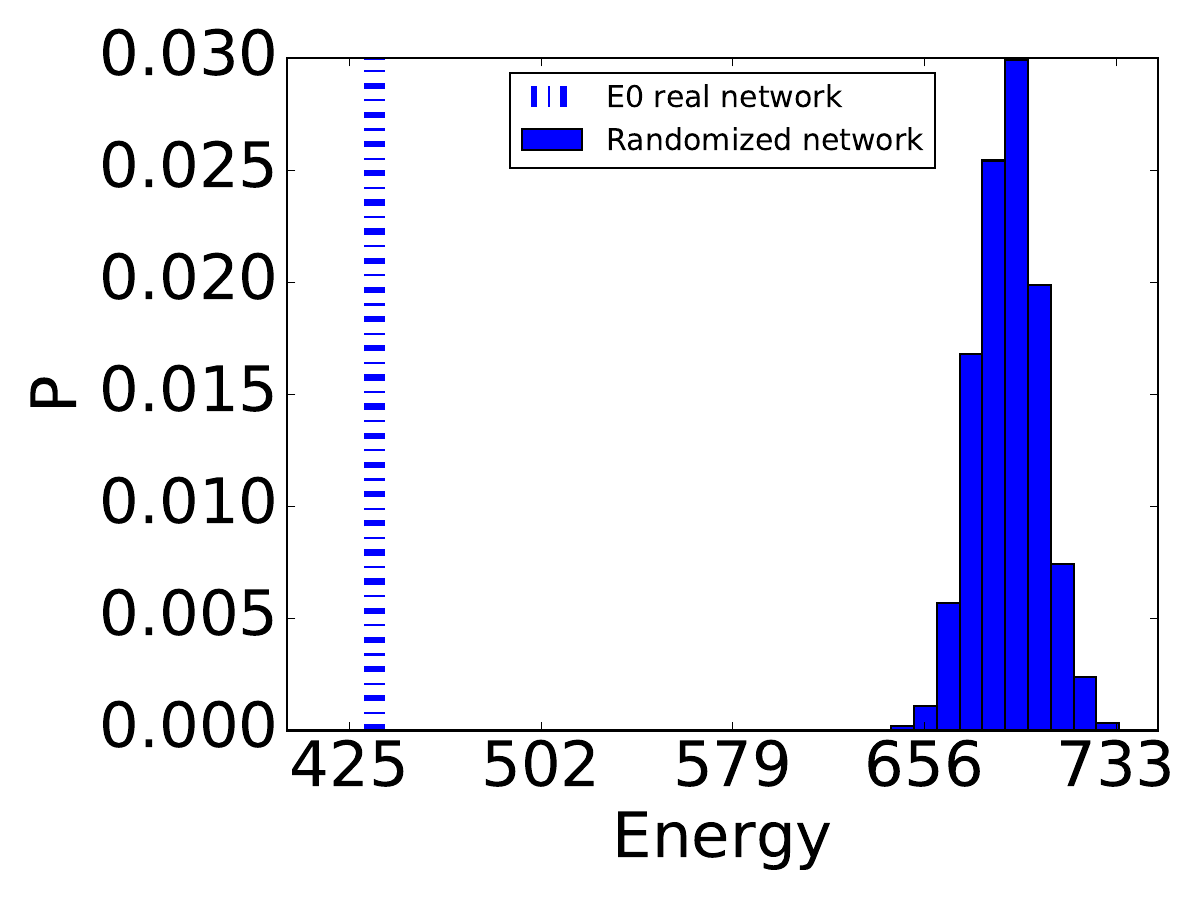} }}
	\subfloat[A\underbar lak\= apuram]{{\includegraphics[width=0.2\linewidth]{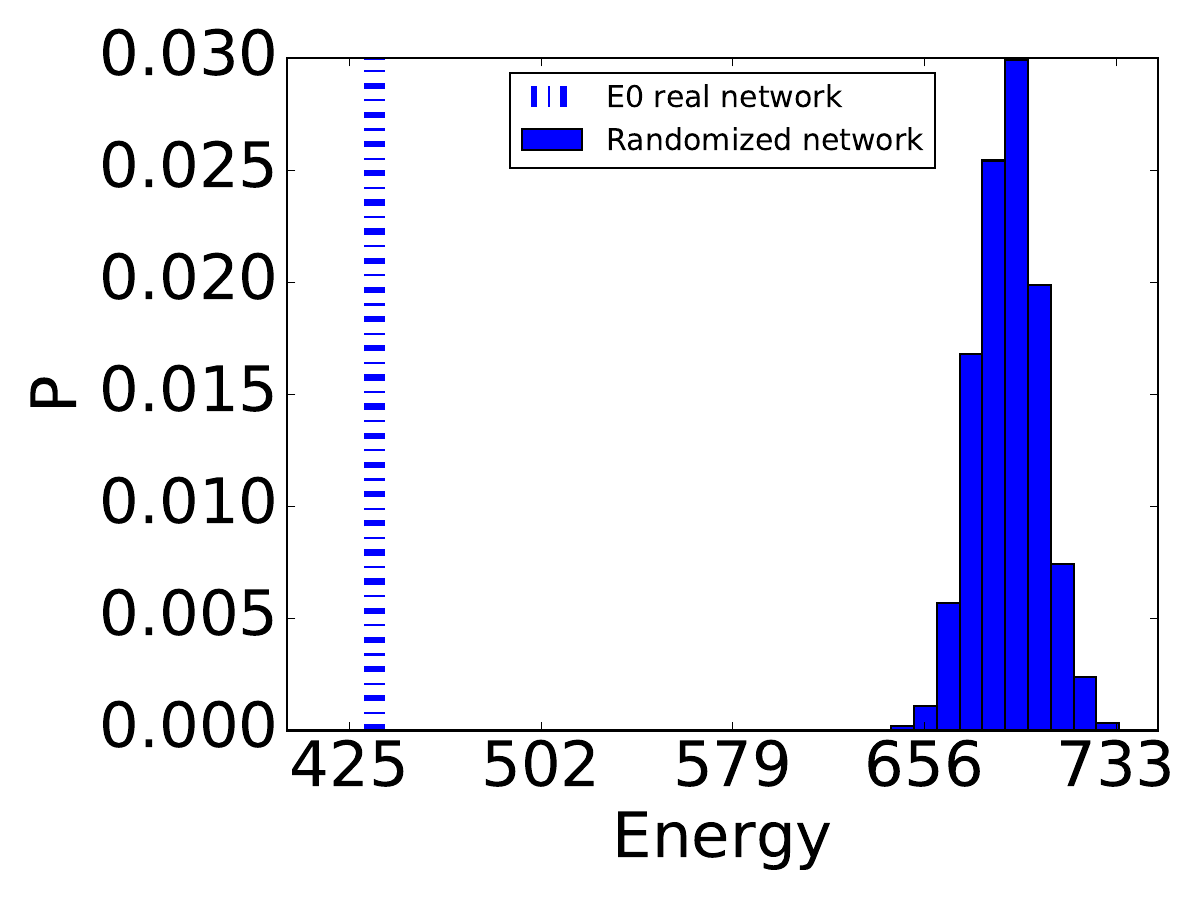} }}\\%
	\subfloat[parakeet G1]{{\includegraphics[width=0.16\linewidth]{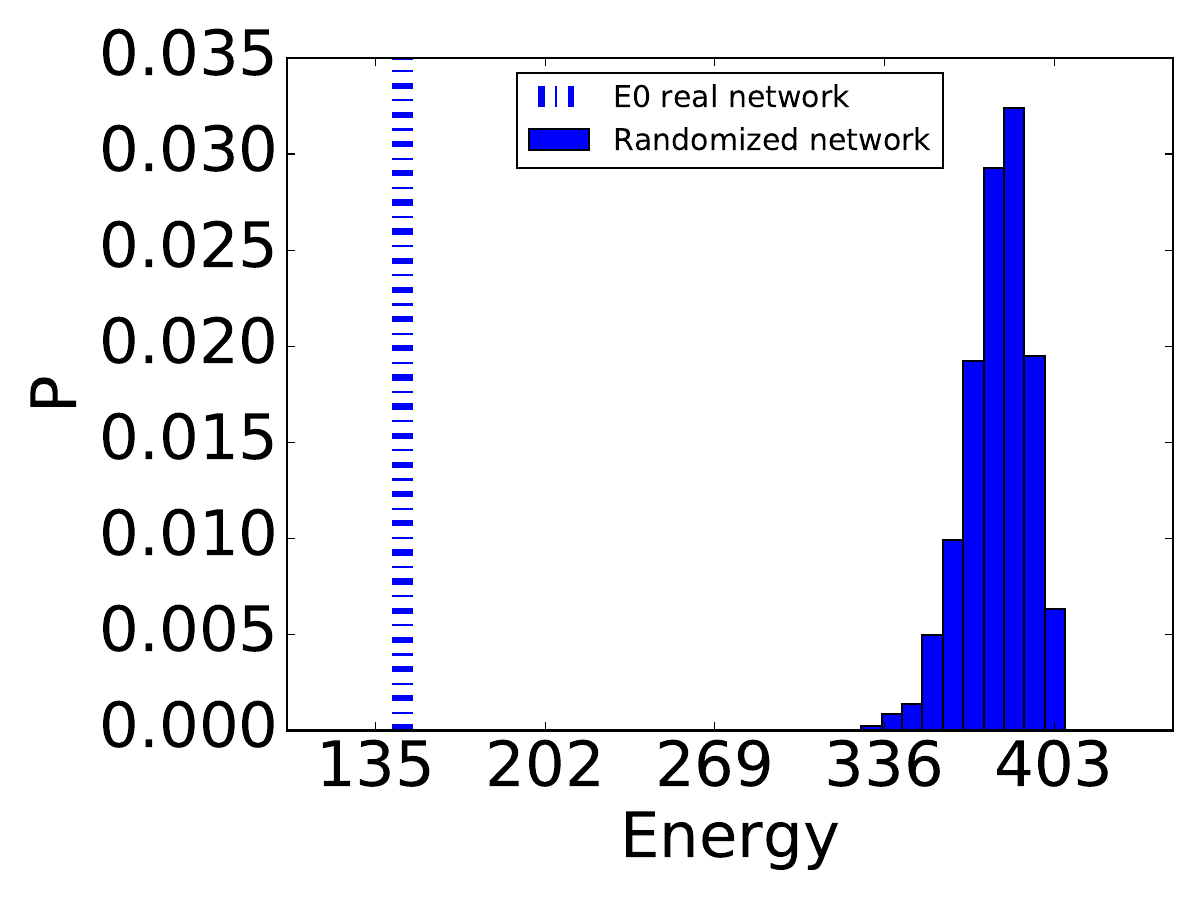} }}
	\subfloat[parakeet G2]{{\includegraphics[width=0.16\linewidth]{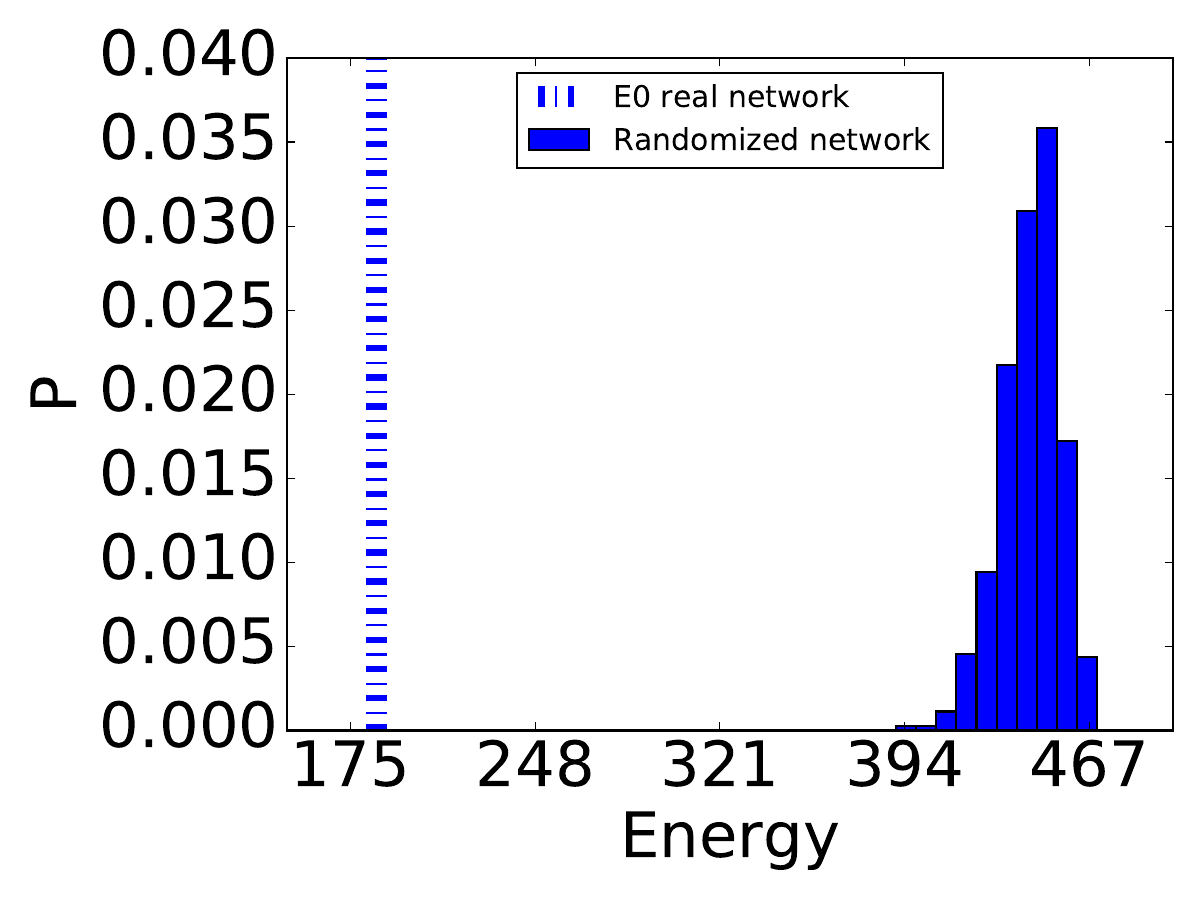} }}
	\subfloat[planted $\beta=5.0$]{{\includegraphics[width=0.16\linewidth]{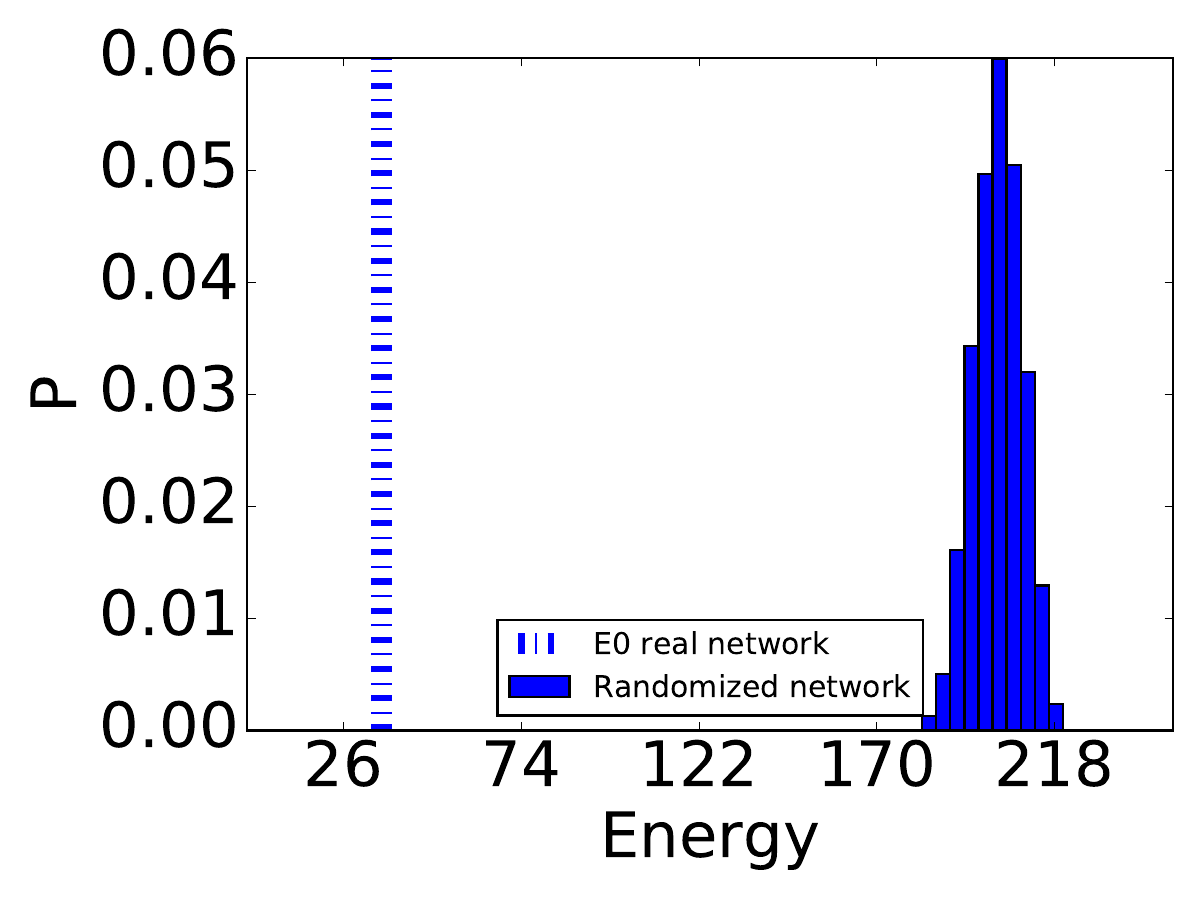} }}
	\subfloat[planted $\beta=0.1$]{{\includegraphics[width=0.16\linewidth]{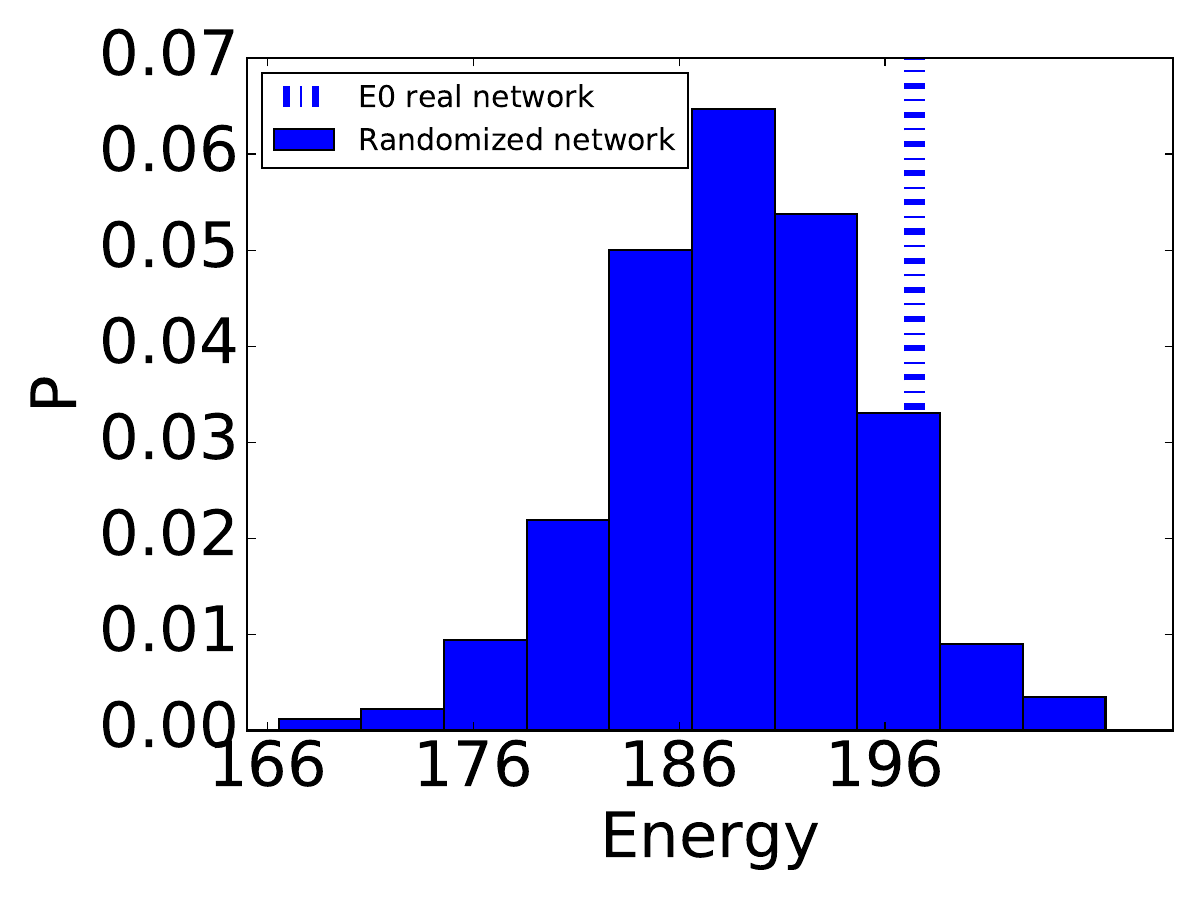} }}
	\subfloat[Asian elephant]{{\includegraphics[width=0.16\linewidth]{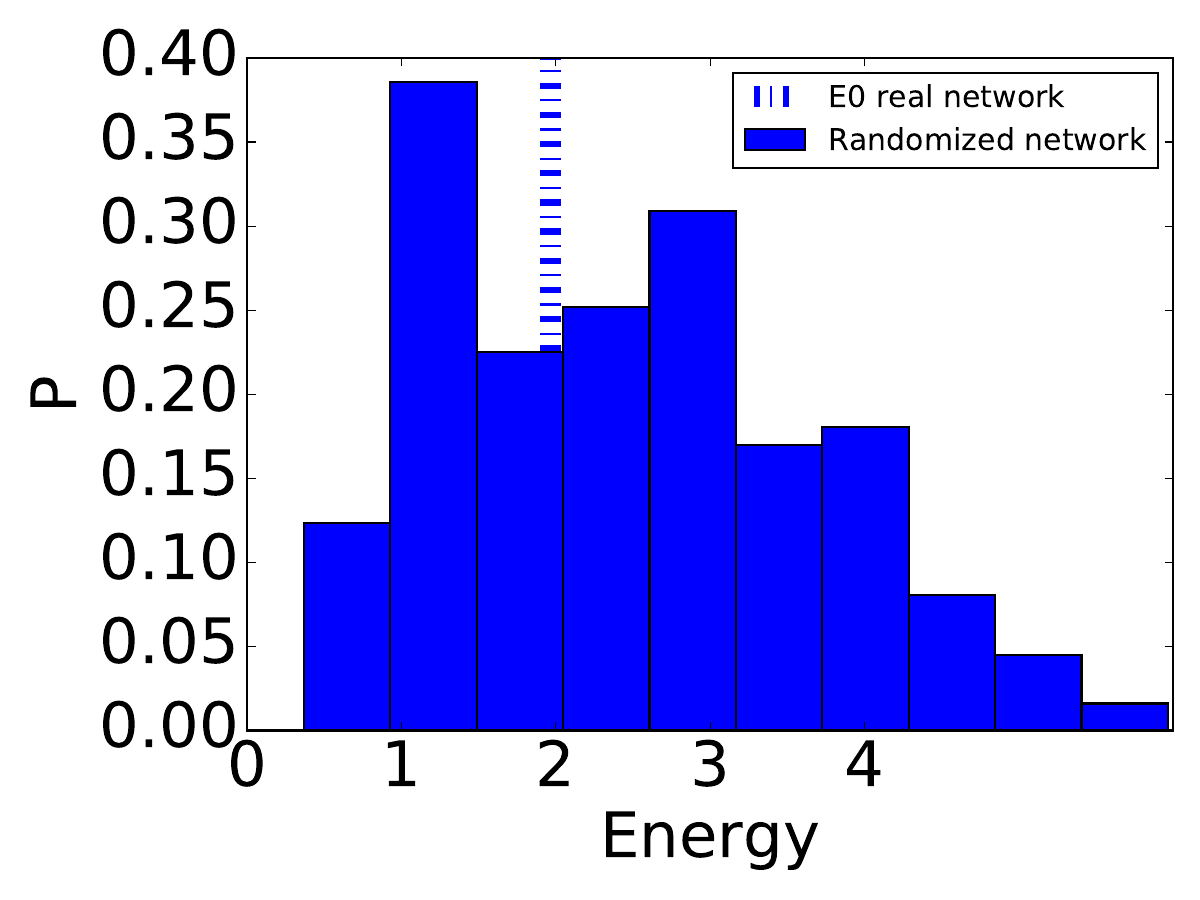} }}
	\subfloat[ER $N=100, \langle k \rangle=3$]{{\includegraphics[width=0.16\linewidth]{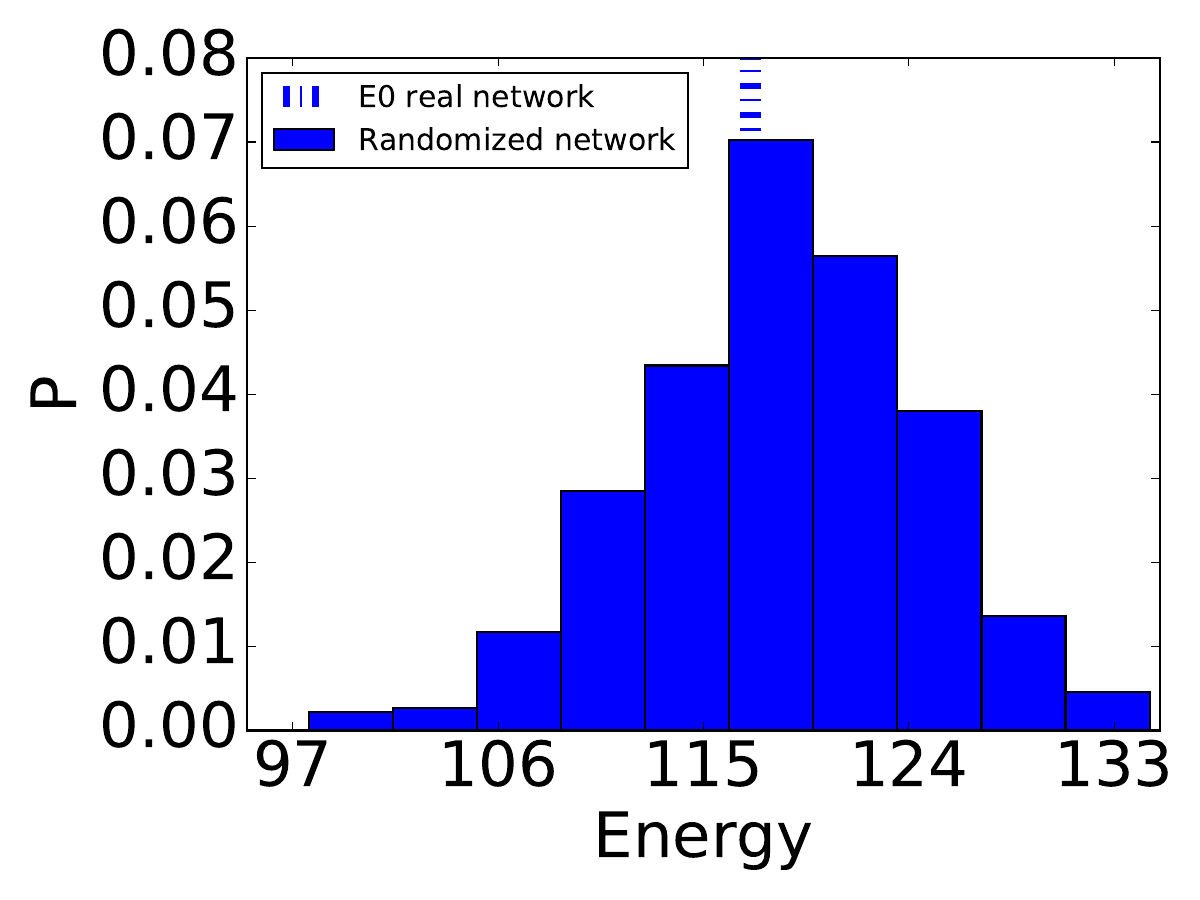} }}
	\caption{\textbf{Statistical significance testing using the null model distribution of energies.} Results are 1000 realizations of the null model where edge directions are randomized while keeping the total number of interactions between each pair fixed, for real and synthetic networks: a-c) US History (HS), Business (BS) and Computer Science (CS) faculty hiring networks~\cite{clauset2015systematic}; d-e) social support networks of two Indian villages~\cite{power2017} considering 5 types of interactions (see main manuscript); f,g) aggression network of parakeet Group 1 and 2 (as in~\cite{hobson2015social}); h,i) planted network using SpringRank generative model with $N=100$ and mean degree $\langle k \rangle=5$, Gaussian prior for the ranks with average $\mu=0.5$ and variance $1$ ($\alpha=1/\beta$) and two noise levels $\beta=5.0$ and $\beta=0.1$; j) dominance network of asian elephants~\cite{elephant}; k) Erd\H{o}s-R\'enyi directed random network with $N=100$ and $\langle k \rangle=3$. The vertical line is the energy obtained on the real network. In all but the last two cases we reject the null hypothesis that edge directions are independent of the ranks, and conclude that the hierarchy is statistically significant.}
	\label{SIfig:nullmodel}
\end{figure*}

\begin{figure*}[t]%
	\centering
	\includegraphics[width=0.48\linewidth]{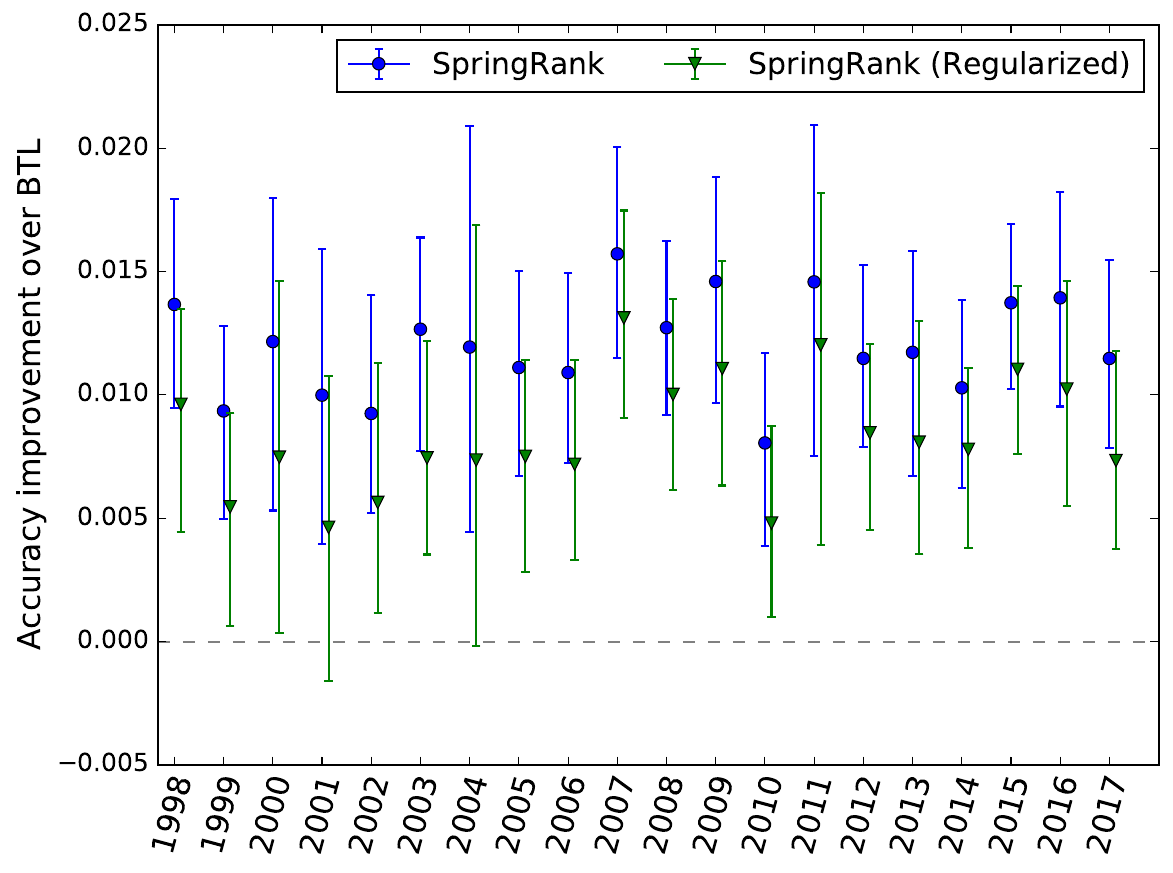}
	\includegraphics[width=0.48\linewidth]{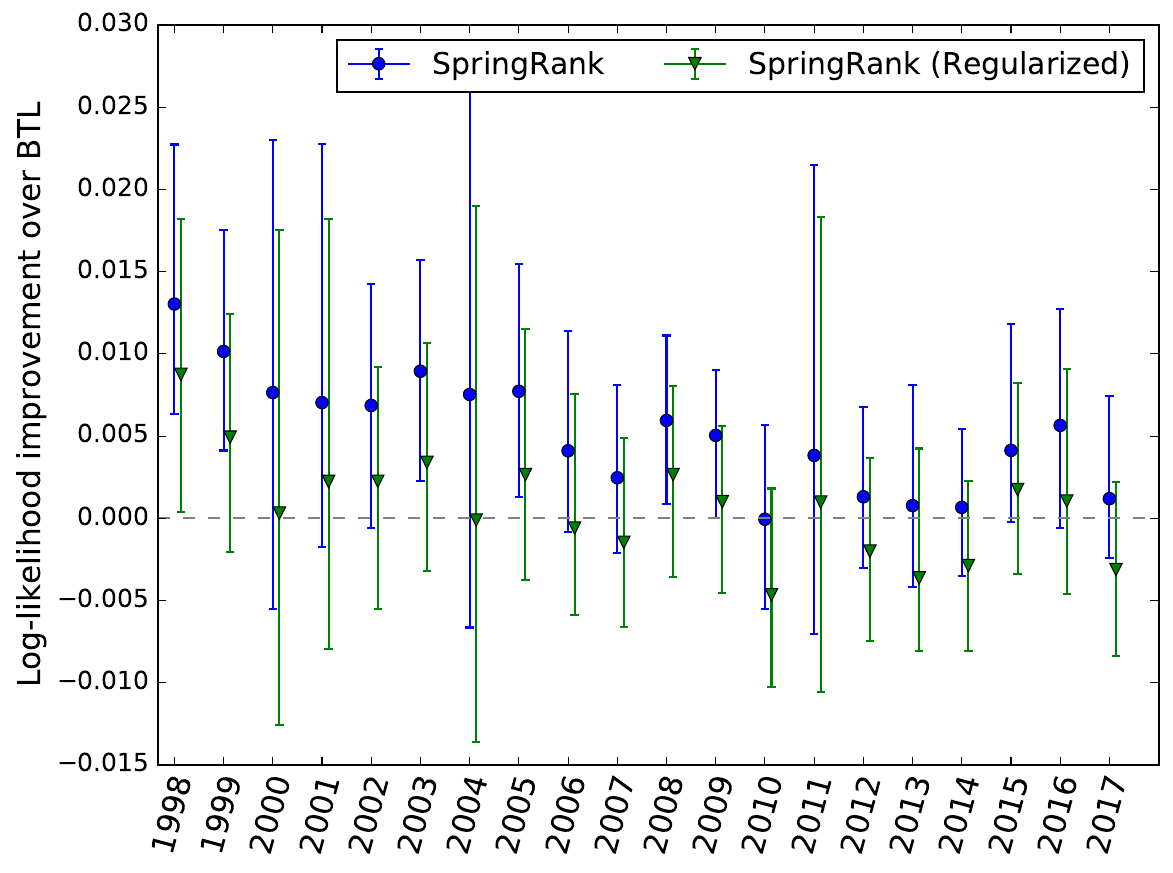}
	\includegraphics[width=0.48\linewidth]{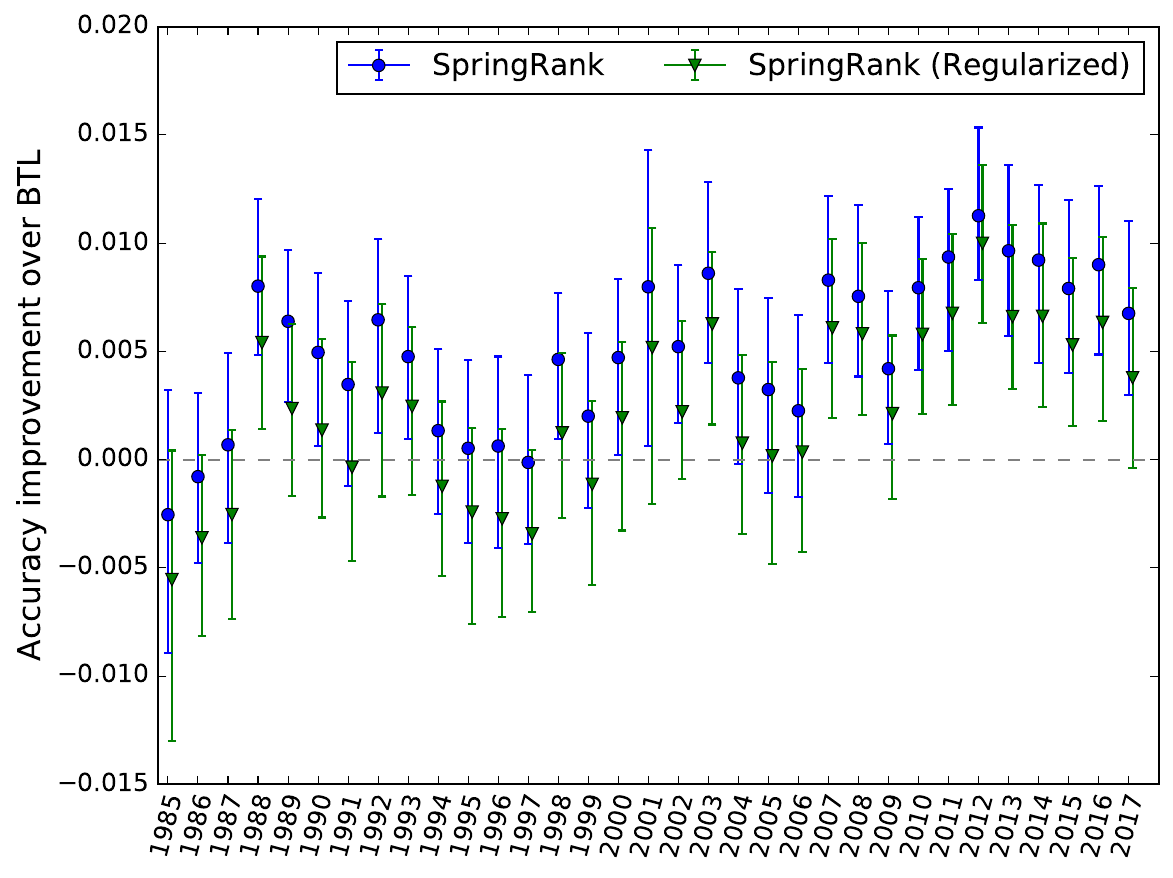}
	\includegraphics[width=0.48\linewidth]{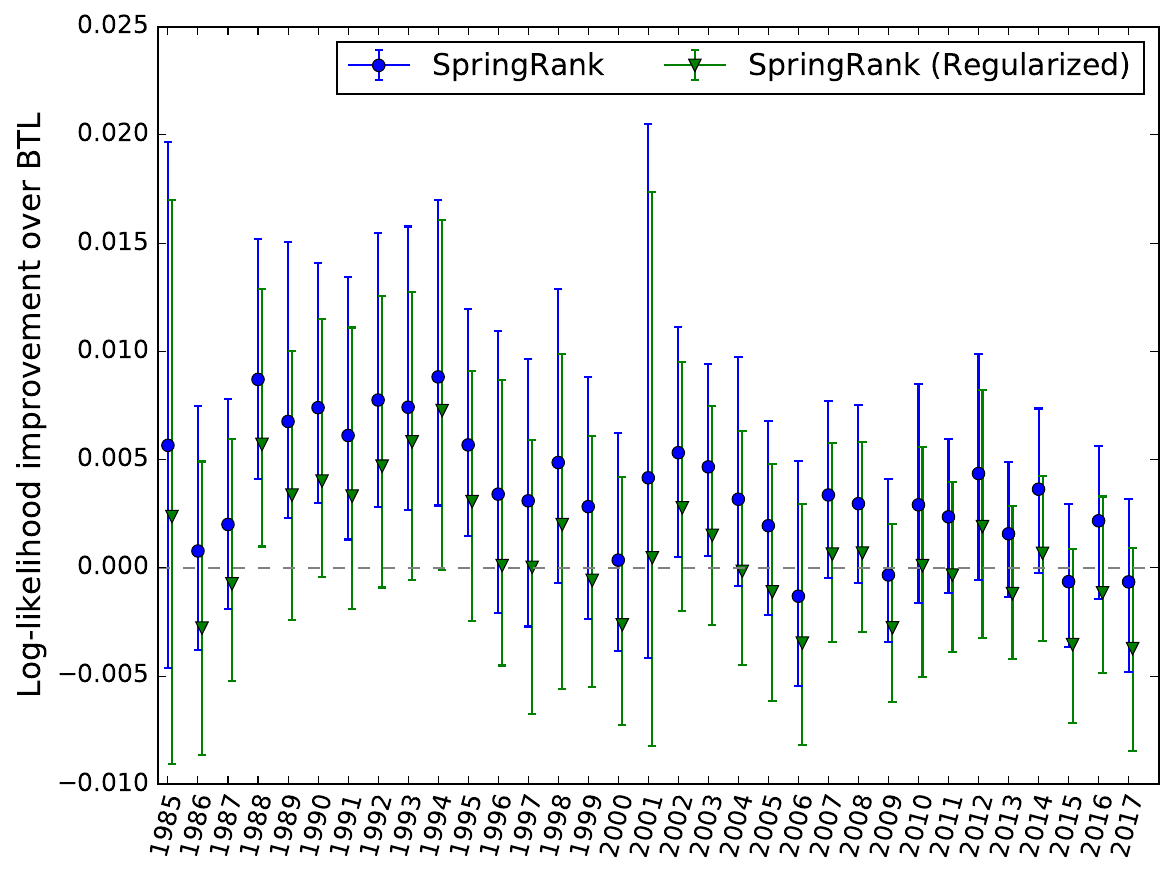}
	\includegraphics[width=0.48\linewidth]{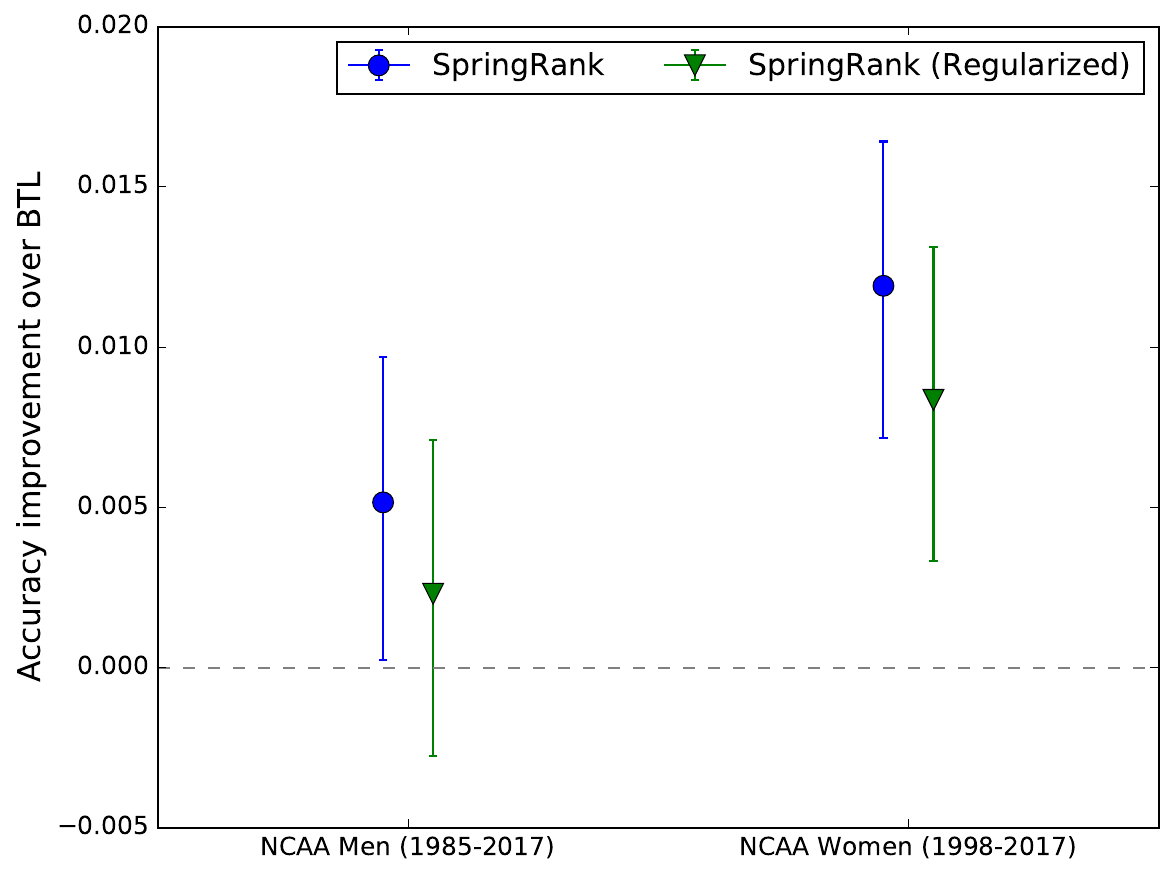}
	\includegraphics[width=0.48\linewidth]{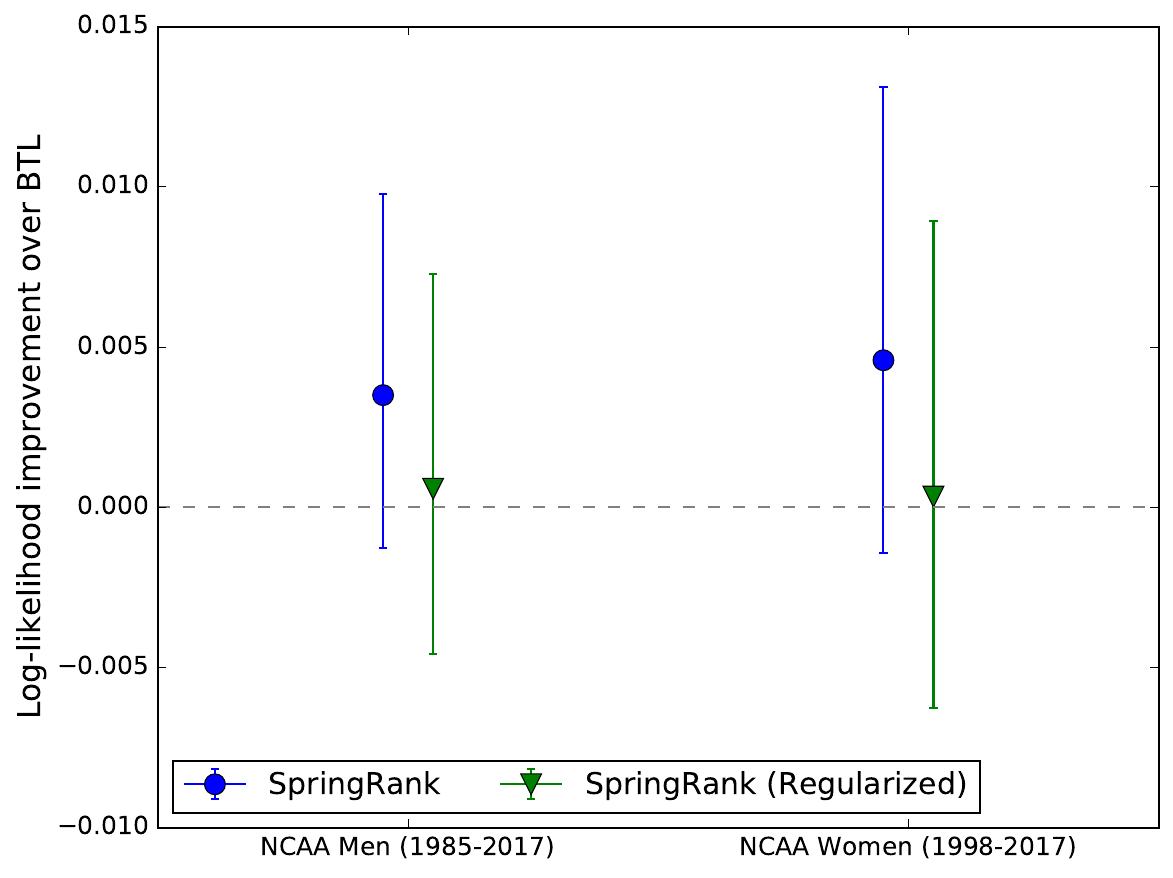}
	\caption{\textbf{Edge prediction accuracy over BTL for NCAA basketball datasets}. Distribution of differences in performance of edge prediction of SpringRank compared to BTL on NCAA College Basketball regular season matches for (top) Women and (middle) Men, defined as (left) the probabilistic edge-prediction accuracy $\eps_a$ Eq.~\eqref{eq:localaccuracy} and (right) the conditional log-likelihood $\eps_L$ Eq.~\eqref{eq:loglikelihood}. Error bars indicate quartiles and markers show medians, corresponding to $50$ independent trials of 5-fold cross-validation, for a total of 250 test sets for each dataset.  The bottom plot is obtained by considering the distributions over all the seasons together. In terms of number of correctly predicted outcomes, SpringRank correctly predicts on average 8 to 16 more outcomes than BTL for each of the 20 Women NCAA seasons and up to 12 more outcomes for each of the 33 Men NCAA seasons; for the latter dataset, BTL has an average better prediction in 3 out of the 33 seasons. The number of matches played per season in the test set varies from the past to the most recents years from 747 to 1079. } 
	\label{SIfig:EDNCAA}
\end{figure*}

\begin{figure*}[htbp]
 \centering
	\includegraphics[width=0.9\linewidth]{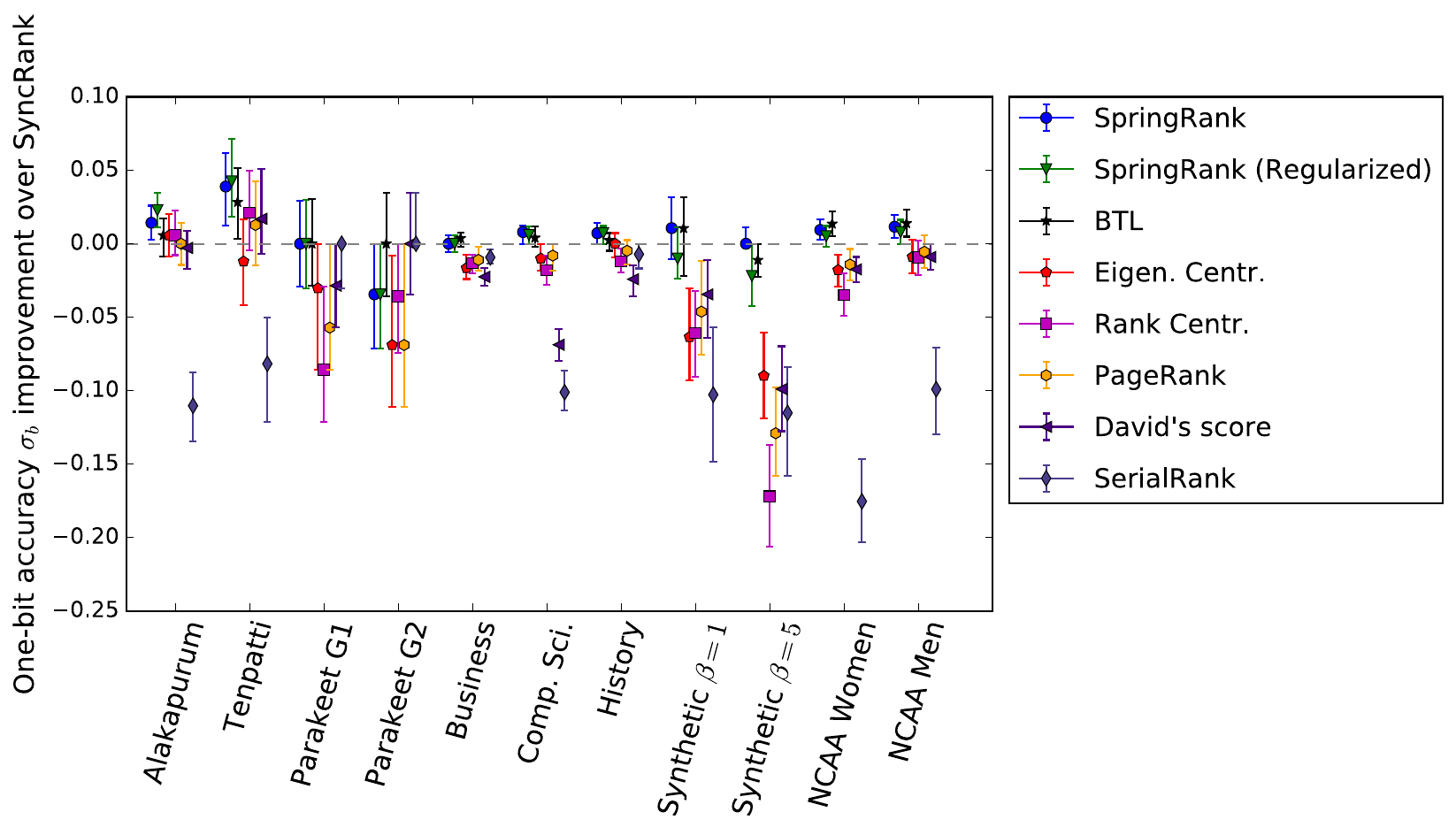}
	\caption{\textbf{Bitwise edge direction prediction}. Symbols show medians of bitwise edge prediction accuracies Eq.~\eqref{maxfractionED} over 50 realization of 5-fold cross-validation (for a total of 250 trials) compared with the median accuracy for SyncRank; error bars indicate quartiles. Thus, points above the dashed line at zero indicate better predictions than SyncRank, while values below indicate that SyncRank performed better.}
 \label{SIfig:ED}
\end{figure*}

\begin{figure*}[t]%
	\centering
	\includegraphics[width=0.48\linewidth]{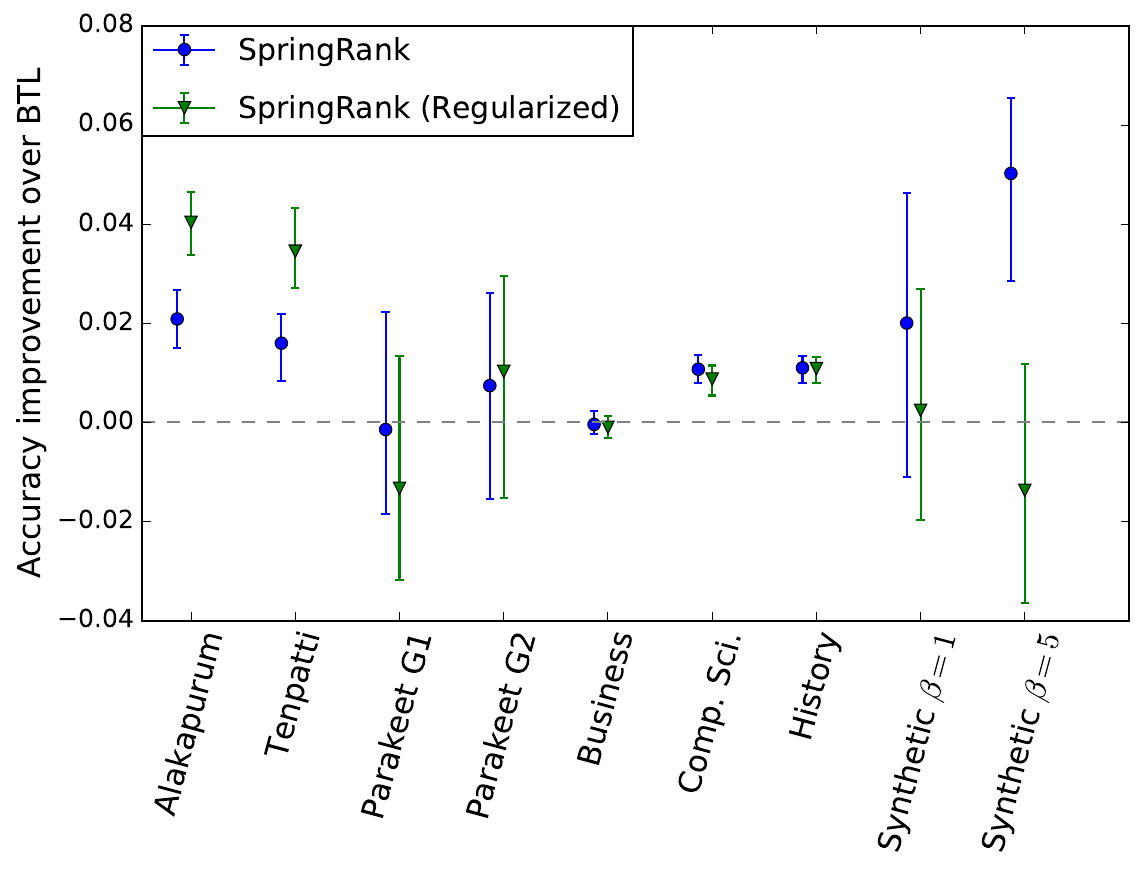}
	\includegraphics[width=0.48\linewidth]{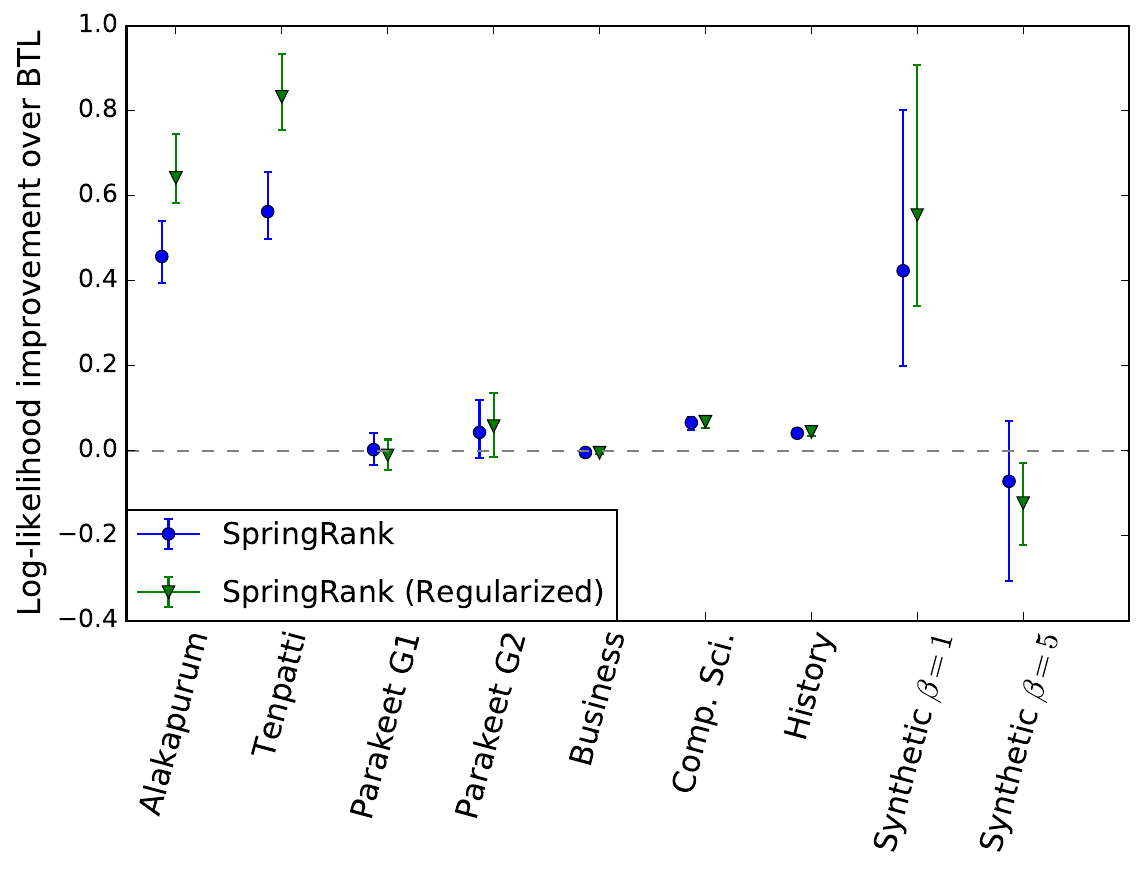}
	\includegraphics[width=0.96\linewidth]{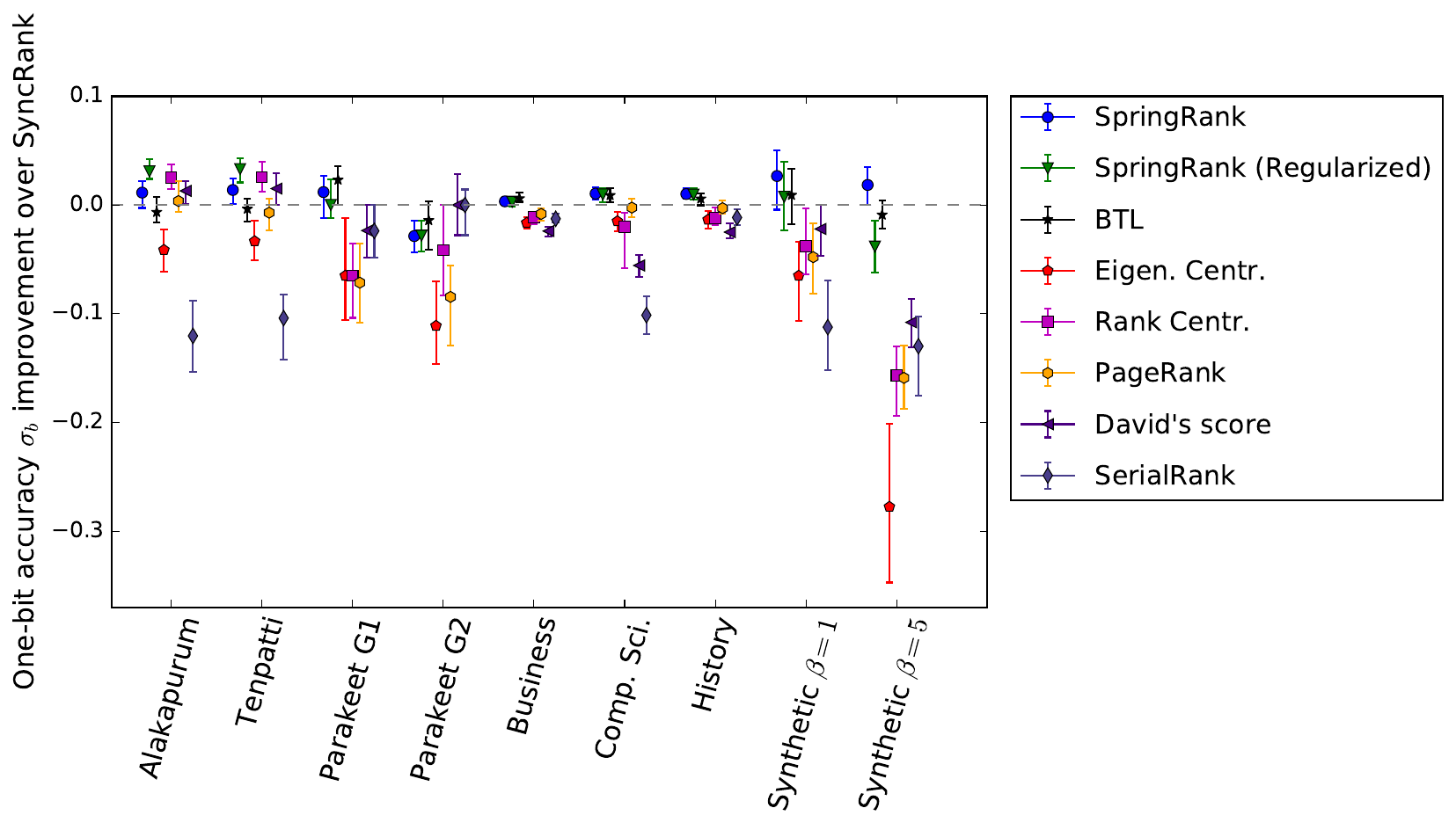}
	\caption{\textbf{Edge prediction accuracy with 2-fold cross-validation}.  Top: the accuracy of probabilistic edge prediction of SpringRank compared to the median accuracy of BTL on real and synthetic networks defined as (top left) edge-prediction accuracy $\eps_a$ Eq.~\eqref{eq:localaccuracy} and (top right) the conditional log-likelihood $\eps_L$ Eq.~\eqref{eq:loglikelihood}; (bottom) bitwise edge prediction accuracies $\eps_b$ Eq.~\eqref{maxfractionED} of SpringRank and other algorithms compared with the median accuracy of SyncRank.   Error bars indicate quartiles and markers show medians, corresponding to $50$ independent trials of 2-fold cross-validation, for a total of 100 test sets for each network. The two synthetic networks are generated with $N=100$, average degree $5$, and Gaussian-distributed ranks as in Fig.~\ref{fig:synthetic}A, with inverse temperatures $\beta=1$ and $\beta=5$.  Notice that these results are similar those of Fig. \ref{fig:ED}, obtained using 5-fold cross-validation. }
	\label{fig:ED2}%
\end{figure*}

\clearpage
\section*{Computer Science}
\vspace{-0.5cm}
\begin{figure}[ht!]
	\centering
	\begin{tabular}{m{4.25cm}  m{3.8cm} | m{8cm}}
		\includegraphics[height=12cm]{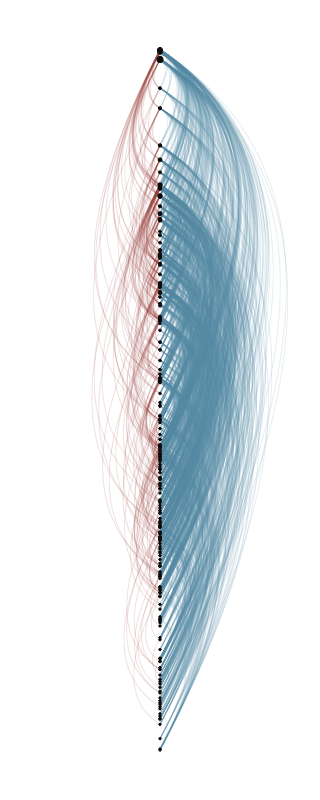}\vspace{15pt}
		& \includegraphics[height=13cm]{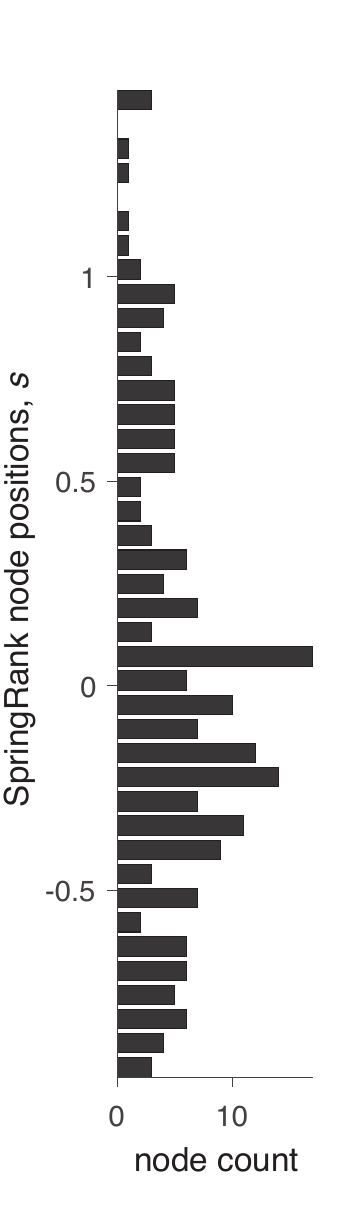}
		&\begin{tabular}{c}
			\includegraphics[width=8cm]{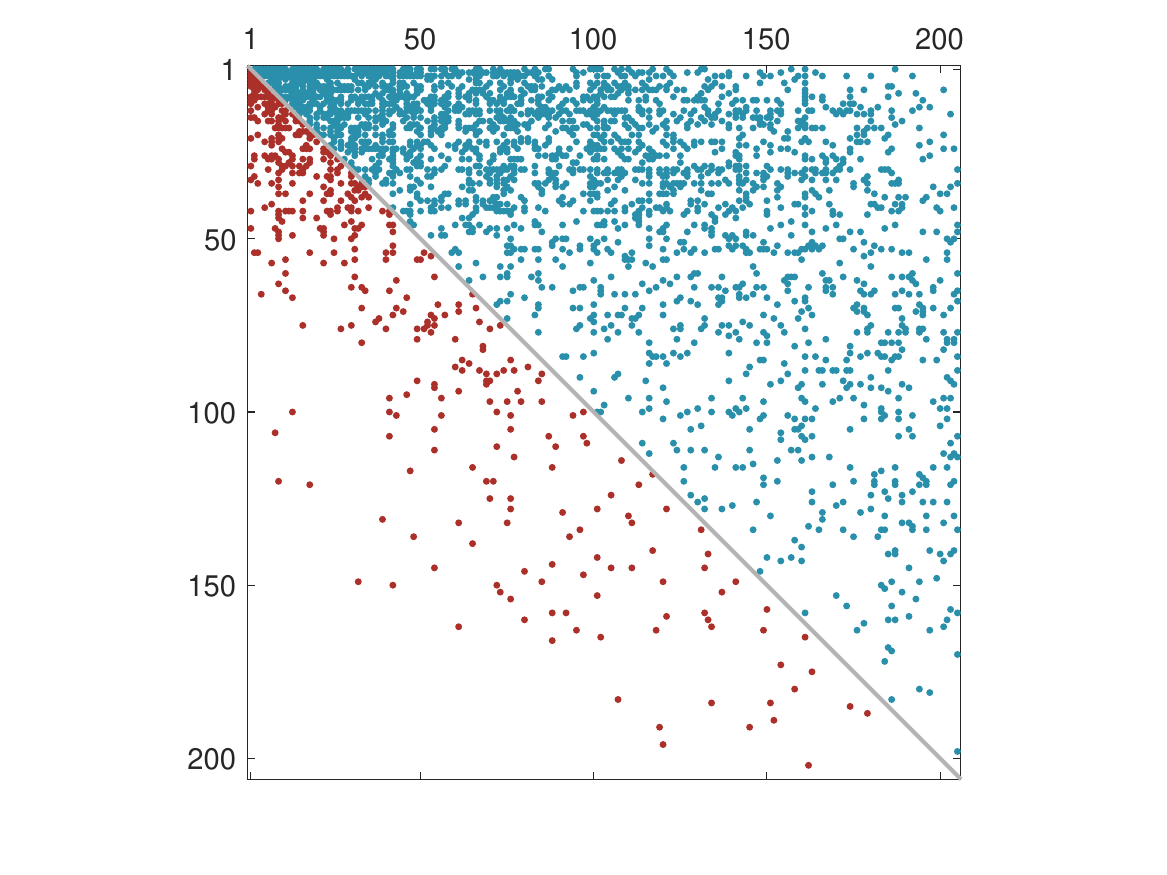}\\
			\includegraphics[width=7cm]{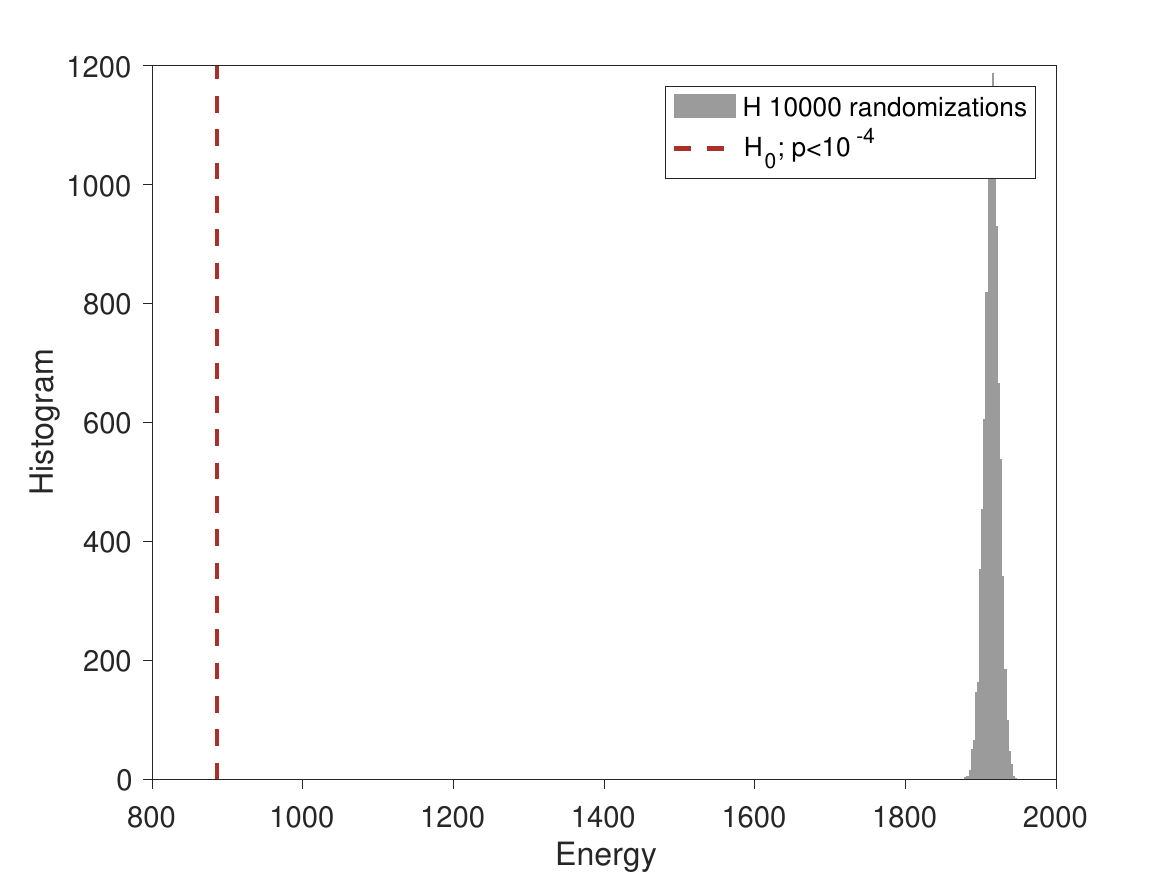}
		\end{tabular}
	\end{tabular}
	\includegraphics[width=18cm]{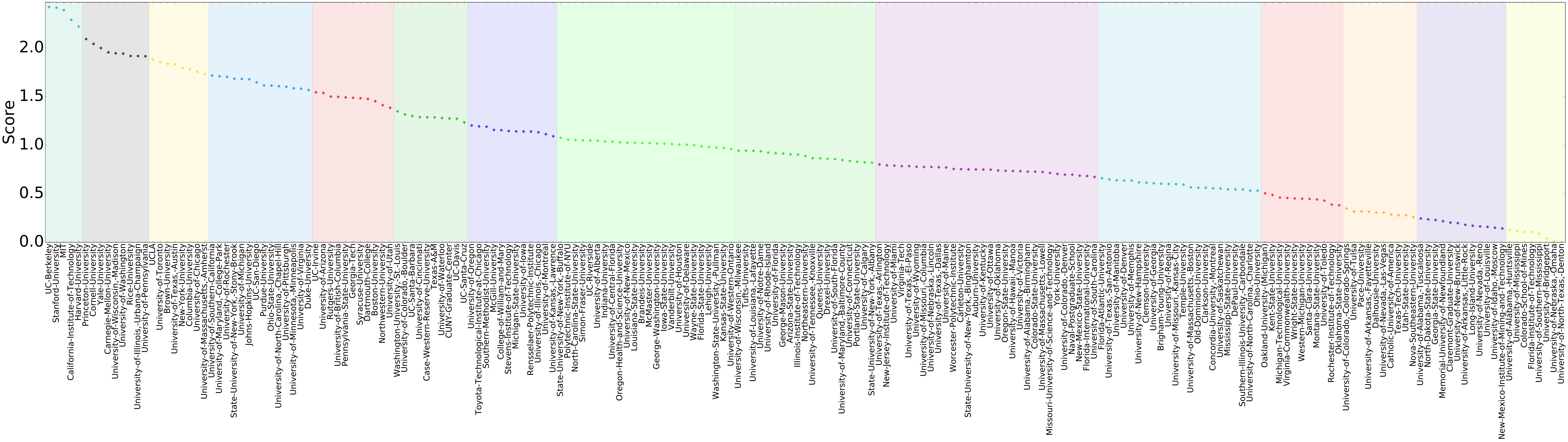}
	\caption{\textbf{Summary of SpringRank applied to Computer Science faculty hiring network \cite{clauset2015systematic}.} \reportcaption \ \reportcaptionb}
	\label{SI:CS}
\end{figure}

\clearpage
\section*{History}
\vspace{-0.5cm}
\begin{figure}[ht!]
	\centering
	\begin{tabular}{m{4.25cm}  m{3.8cm} | m{8cm}}
		\includegraphics[height=12cm]{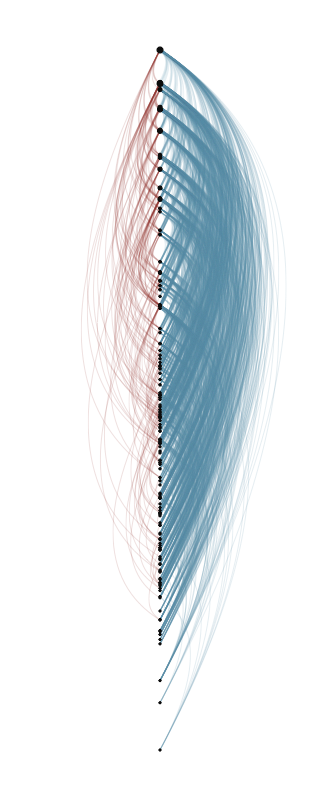}\vspace{15pt}
		& \includegraphics[height=13cm]{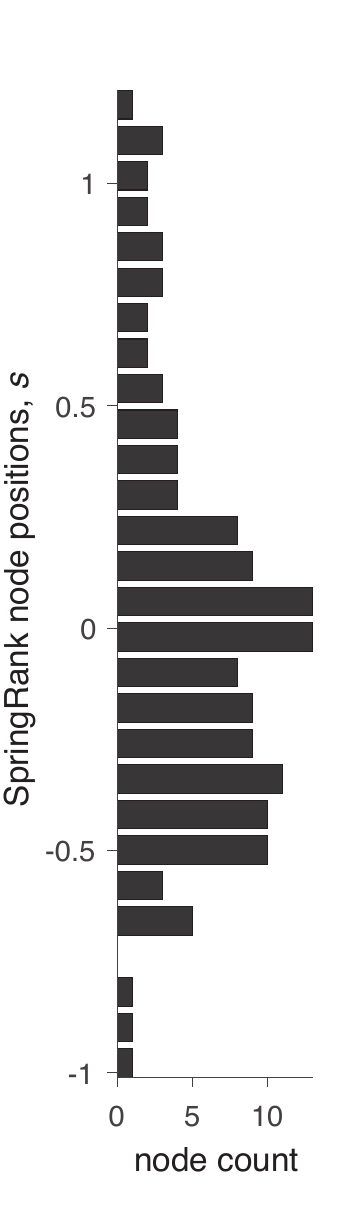}
		&\begin{tabular}{c}
			\includegraphics[width=8cm]{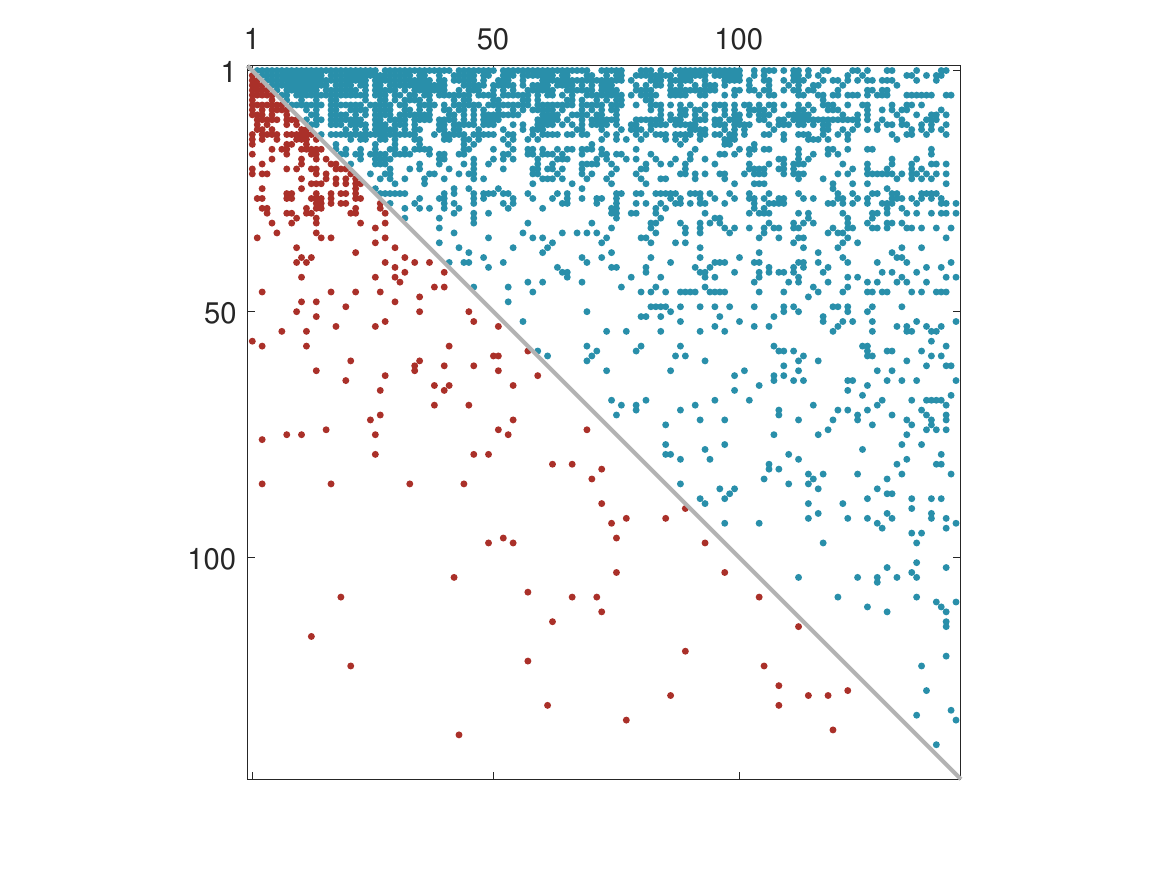}\\
			\includegraphics[width=7cm]{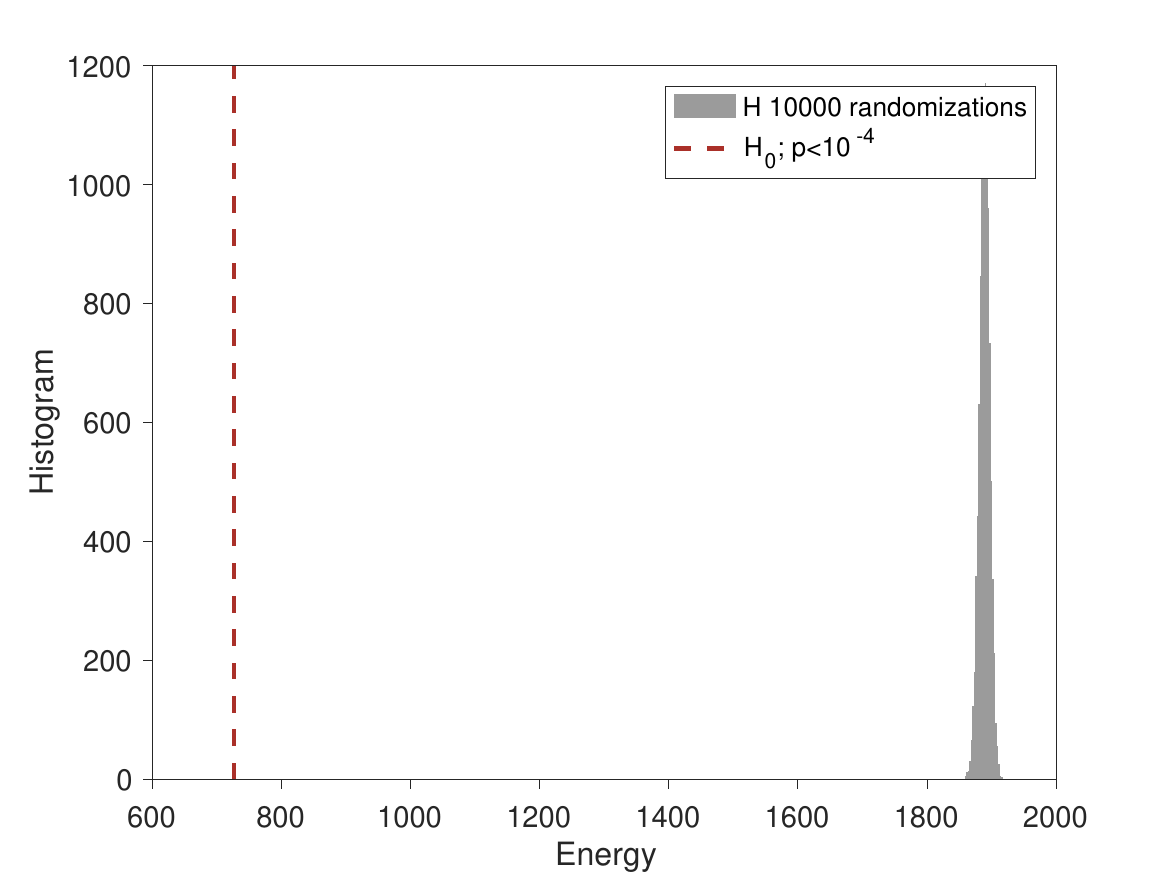}
		\end{tabular}
	\end{tabular}
	\includegraphics[width=18cm]{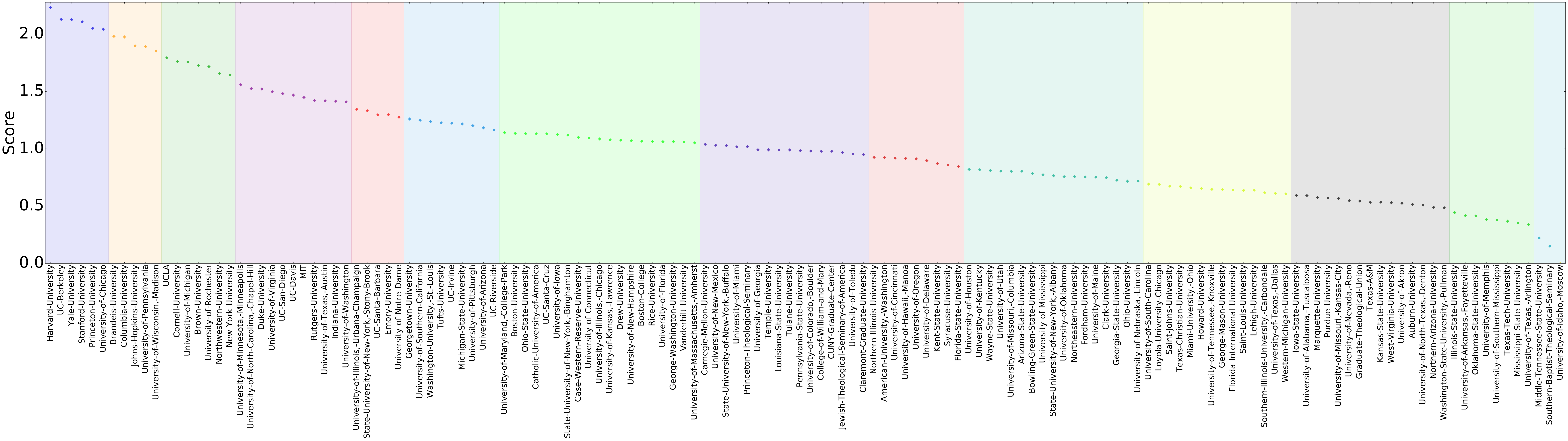}
	\caption{\textbf{Summary of SpringRank applied to History faculty hiring network \cite{clauset2015systematic}.} \reportcaption\ \reportcaptionb}
	\label{SI:HS}
\end{figure}

\clearpage
\section*{Business}
\vspace{-0.5cm}
\begin{figure}[ht!]
	\centering
	\begin{tabular}{m{4.25cm}  m{3.8cm} | m{8cm}}
		\includegraphics[height=12cm]{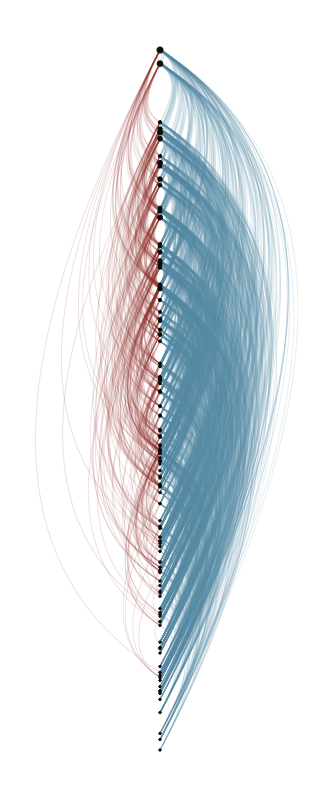}\vspace{15pt}
		& \includegraphics[height=13cm]{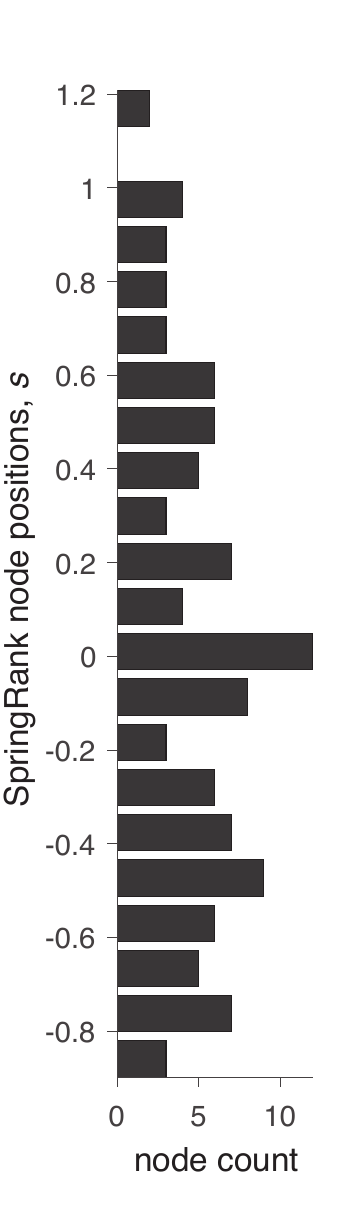}
		&\begin{tabular}{c}
			\includegraphics[width=8cm]{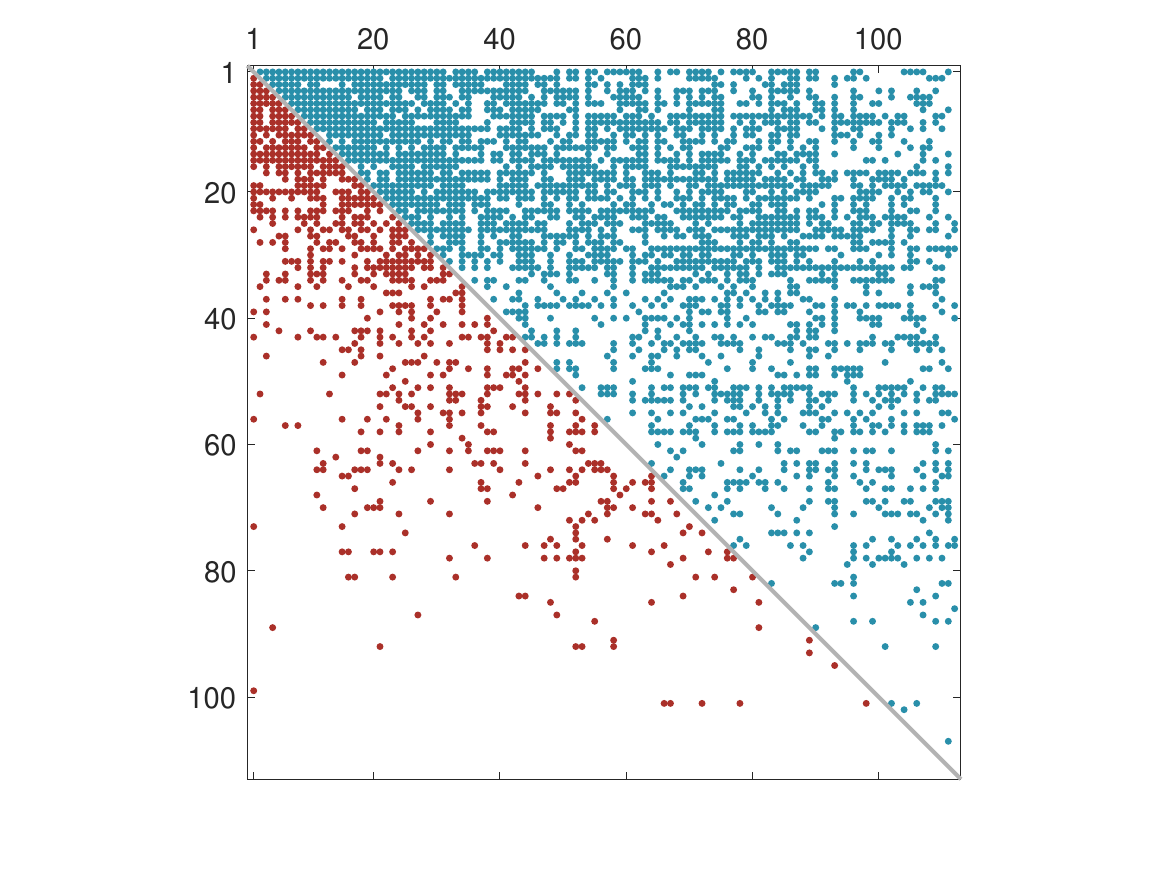}\\
			\includegraphics[width=7cm]{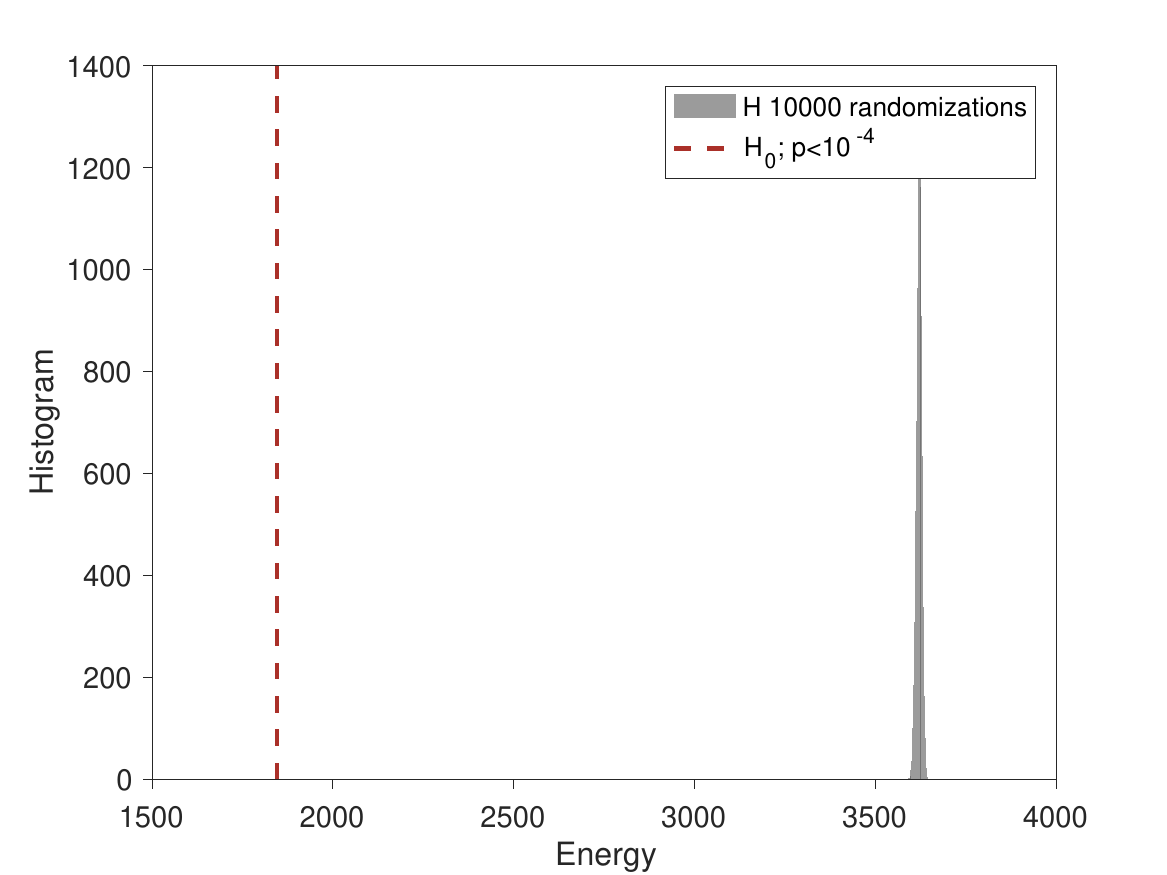}
		\end{tabular}
	\end{tabular}
	\includegraphics[width=18cm]{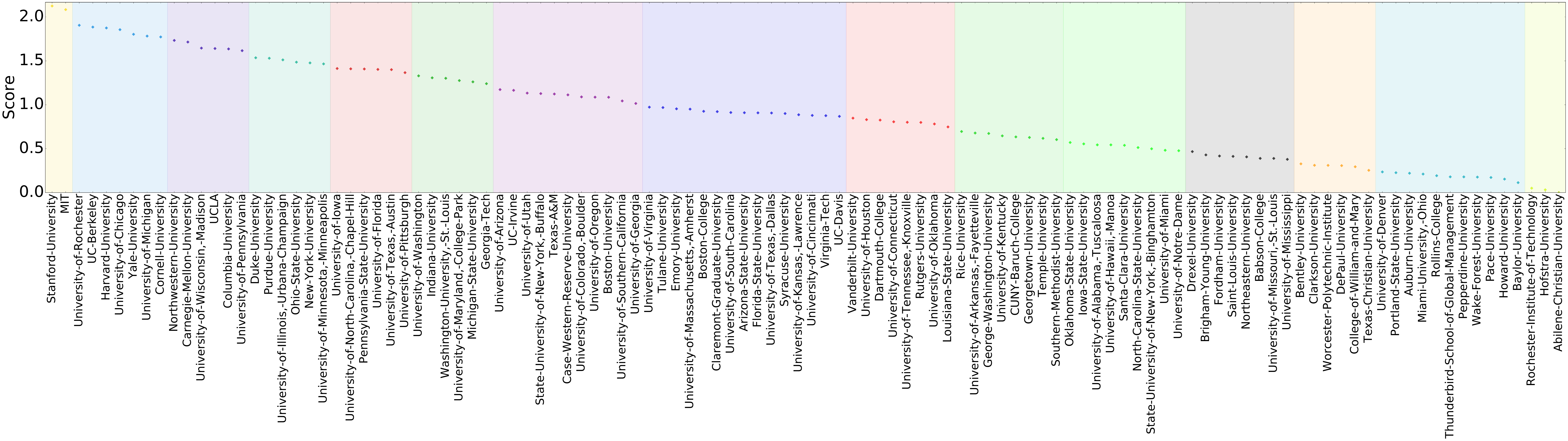}
	\caption{\textbf{Summary of SpringRank applied to Business faculty hiring network \cite{clauset2015systematic}.} \reportcaption\ \reportcaptionb}
	\label{SI:BS}
\end{figure}

\clearpage
\section*{Asian Elephants}
\vspace{-0.5cm}
\begin{figure}[ht!]
	\centering
	\begin{tabular}{m{4.25cm}  m{3.8cm} | m{8cm}}
		\includegraphics[height=355pt]{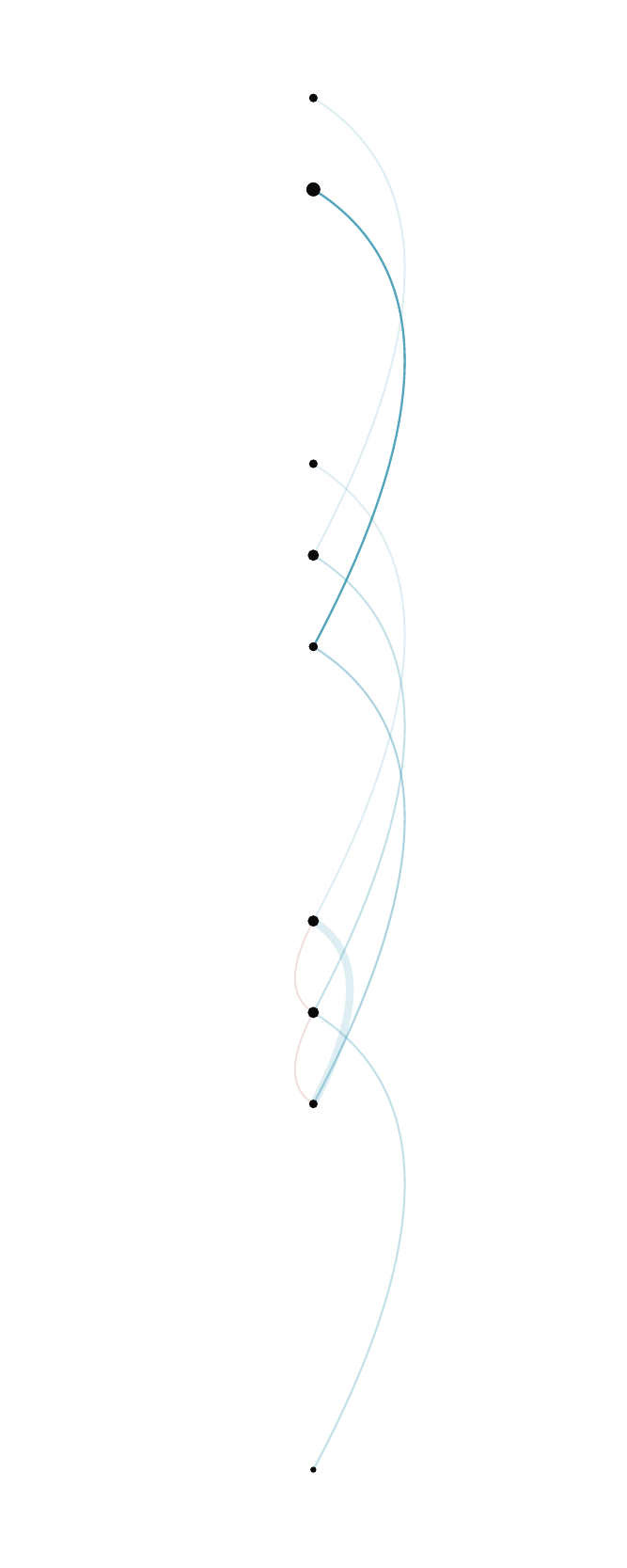}\vspace{15pt}
		& \includegraphics[height=436pt]{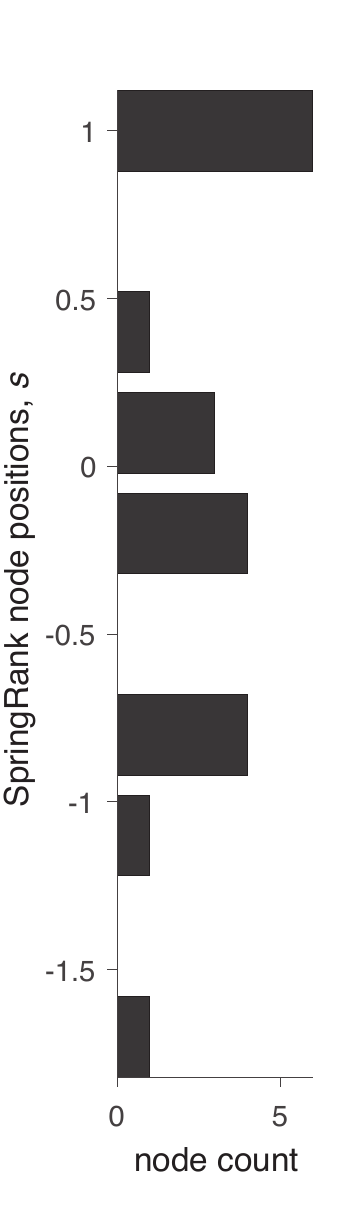}
		&\begin{tabular}{c}
			\includegraphics[width=8cm]{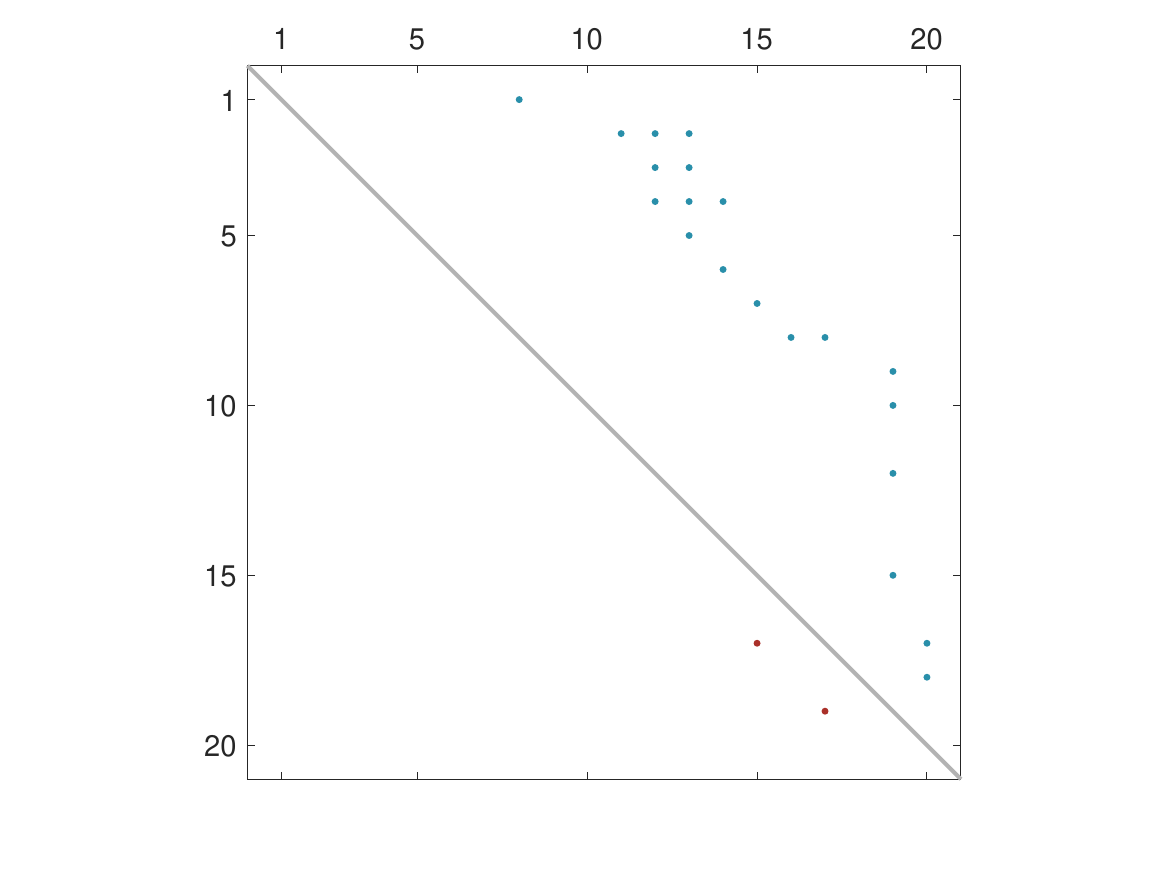}\\
			\includegraphics[width=8cm]{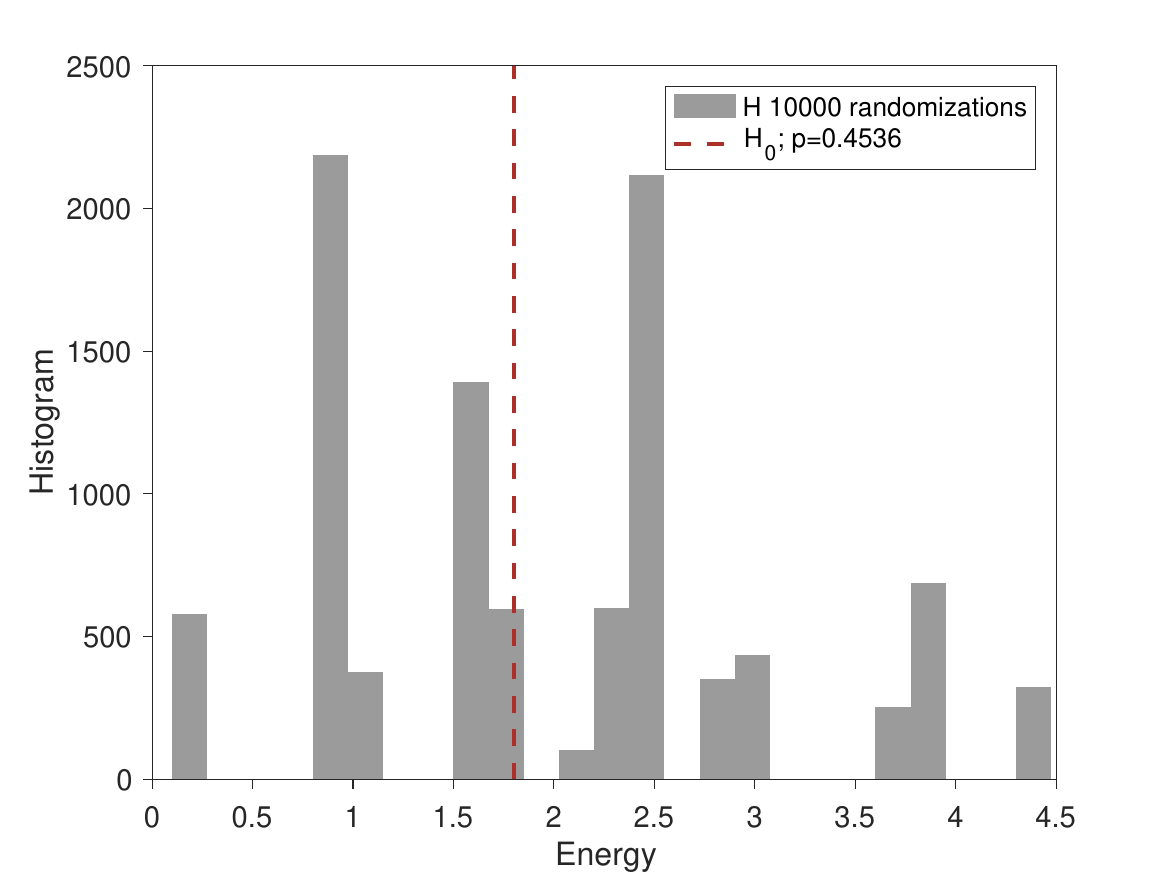}
		\end{tabular}
	\end{tabular}
	\caption{\textbf{Summary of SpringRank applied to Asian Elephants network \cite{elephant}.} \reportcaption}
	\label{SFelephants}
\end{figure}
\clearpage
\section*{Parakeet G1}
\vspace{-0.5cm}
\begin{figure}[ht!]
	\centering
	\begin{tabular}{m{4.25cm}  m{3.8cm} | m{8cm}}
		\includegraphics[height=12cm]{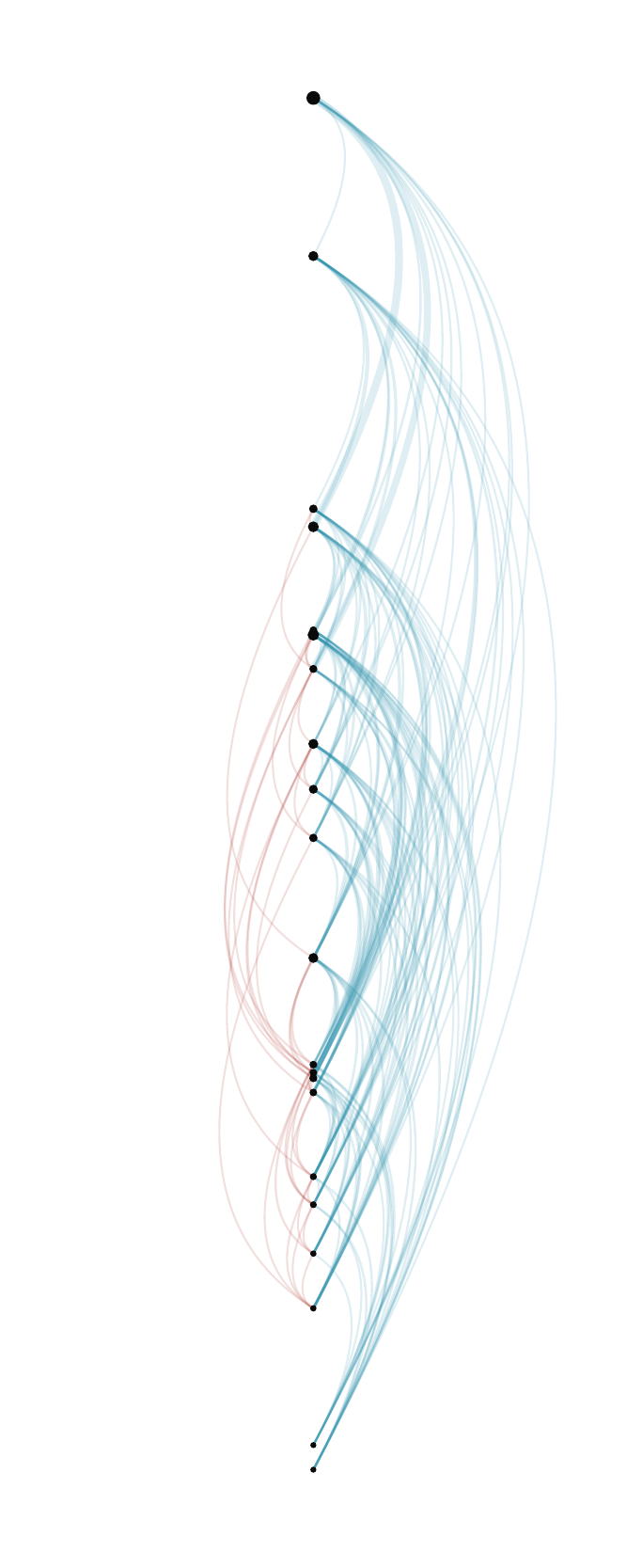}\vspace{15pt}
		& \includegraphics[height=13cm]{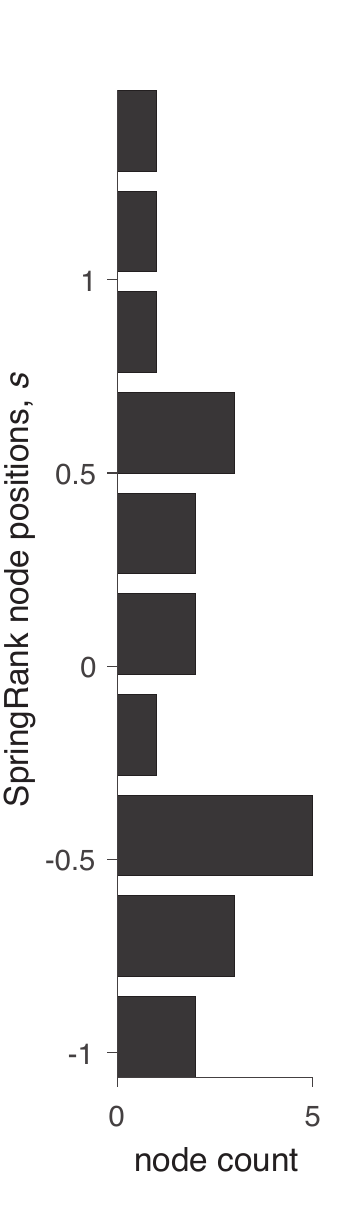}
		&\begin{tabular}{c}
			\includegraphics[width=8cm]{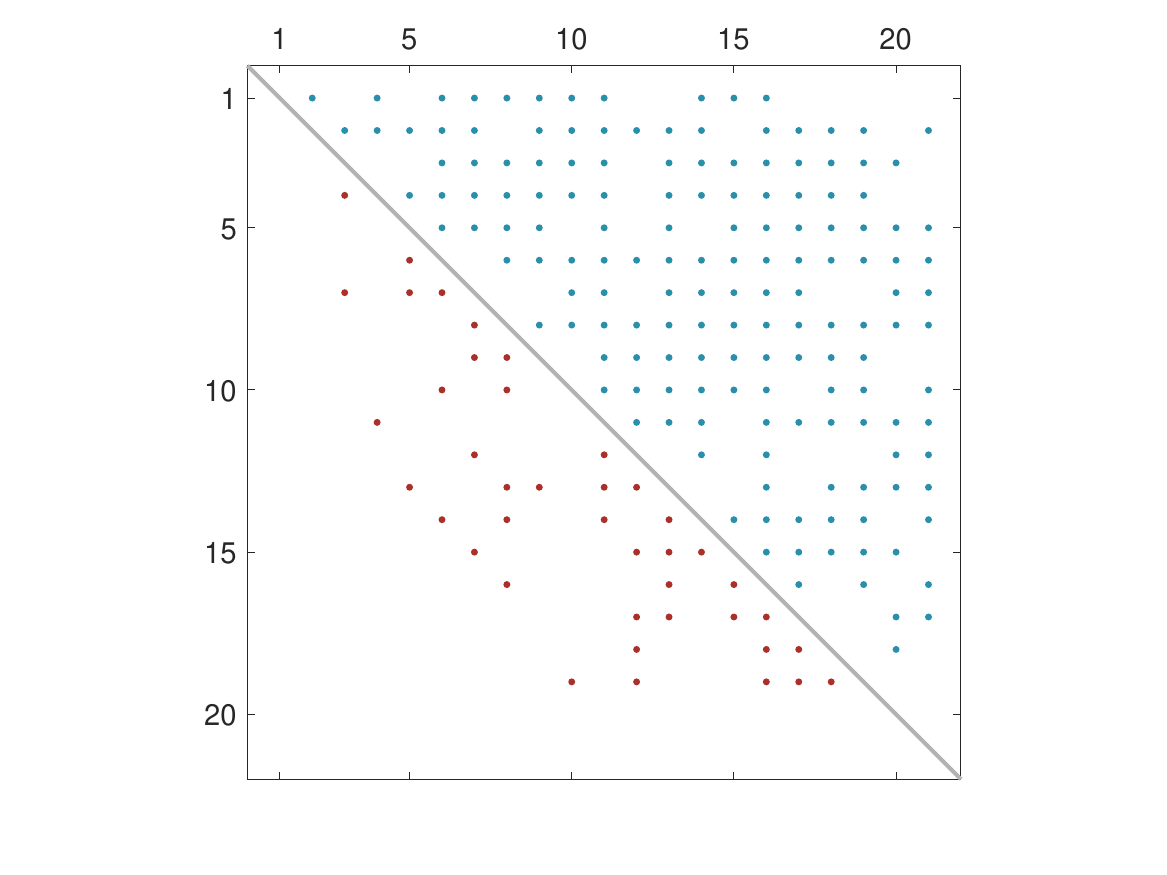}\\
			\includegraphics[width=7cm]{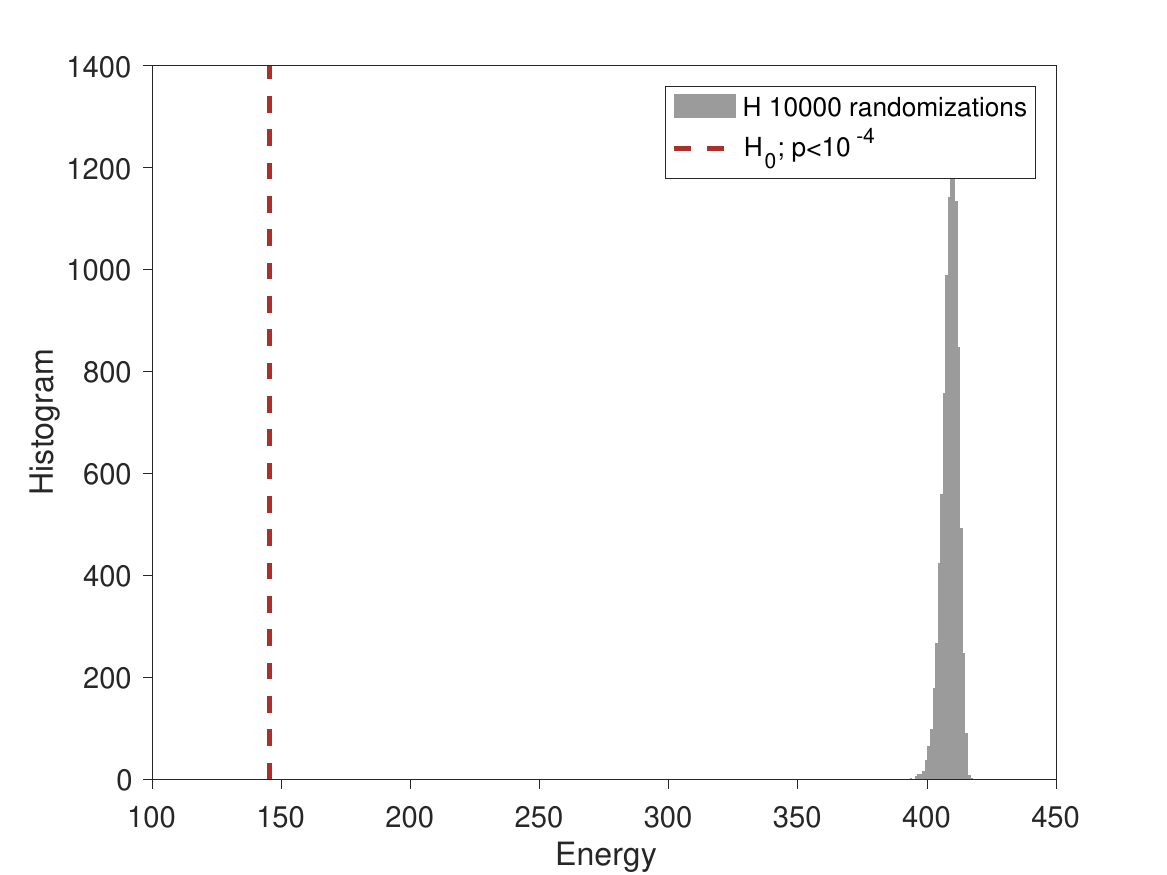}
		\end{tabular}
	\end{tabular}
		\includegraphics[width=8cm]{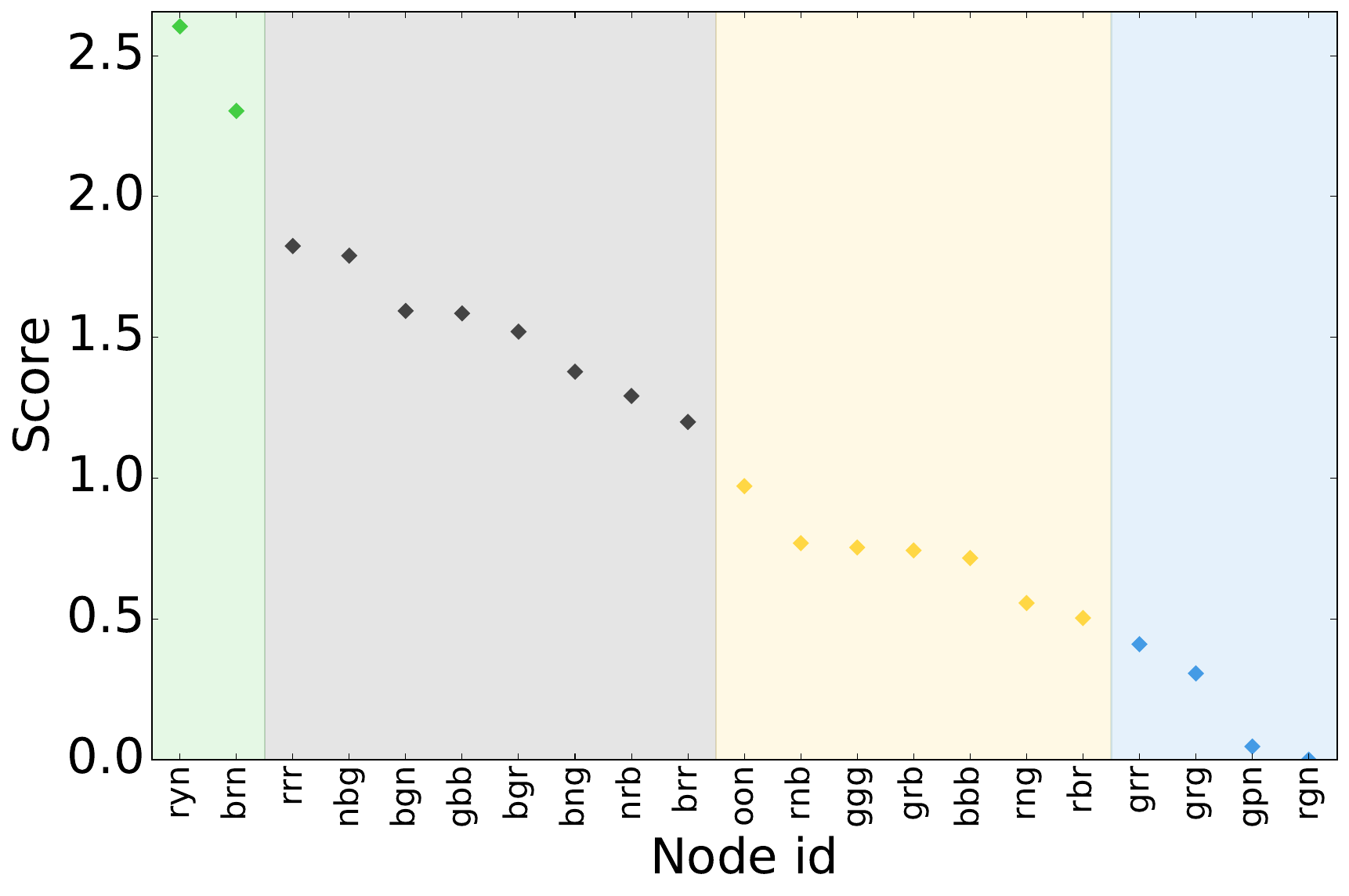}
	\caption{\textbf{Summary of SpringRank applied to Parakeet G1 network \cite{hobson2015social}.} \reportcaption \ \reportcaptionb}
	\label{SI:PK1}
\end{figure}

\clearpage
\section*{Parakeet G2}
\vspace{-0.5cm}
\begin{figure}[ht!]
	\centering
	\begin{tabular}{m{4.25cm}  m{3.8cm} | m{8cm}}
		\includegraphics[height=12cm]{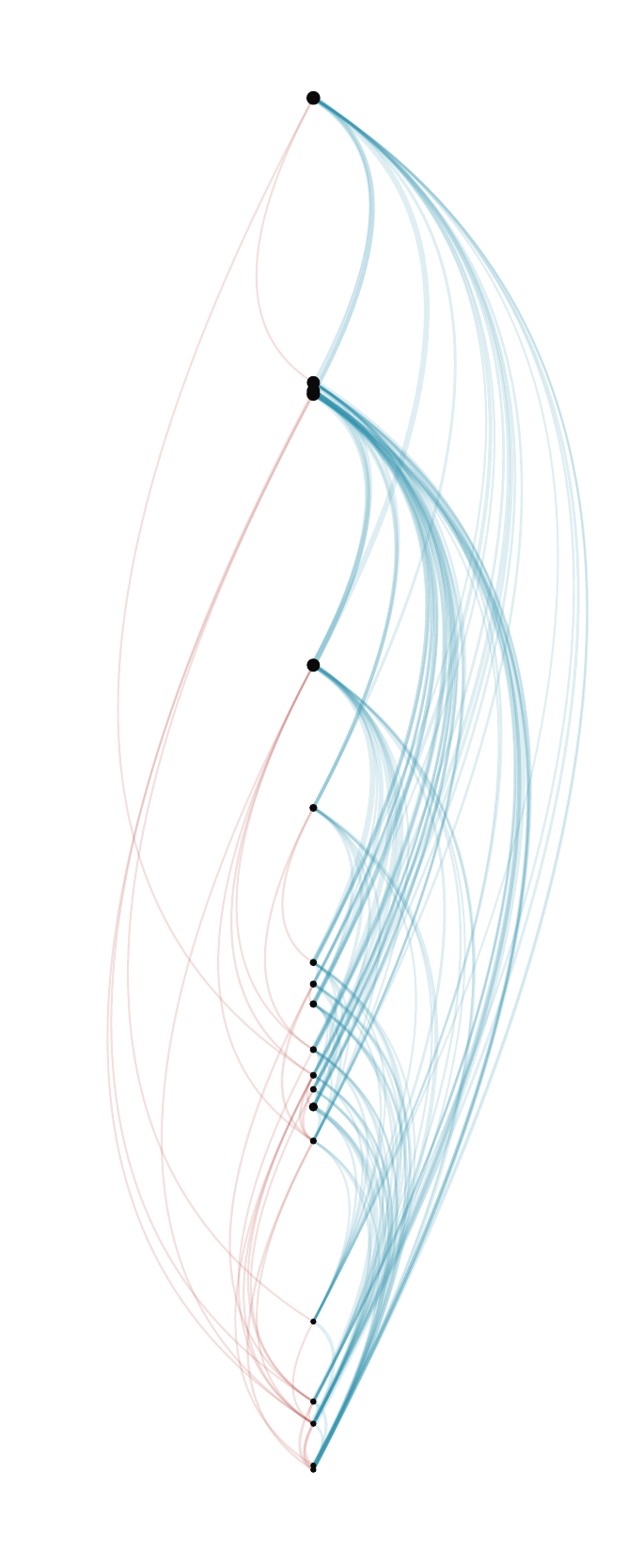}\vspace{15pt}
		& \includegraphics[height=13cm]{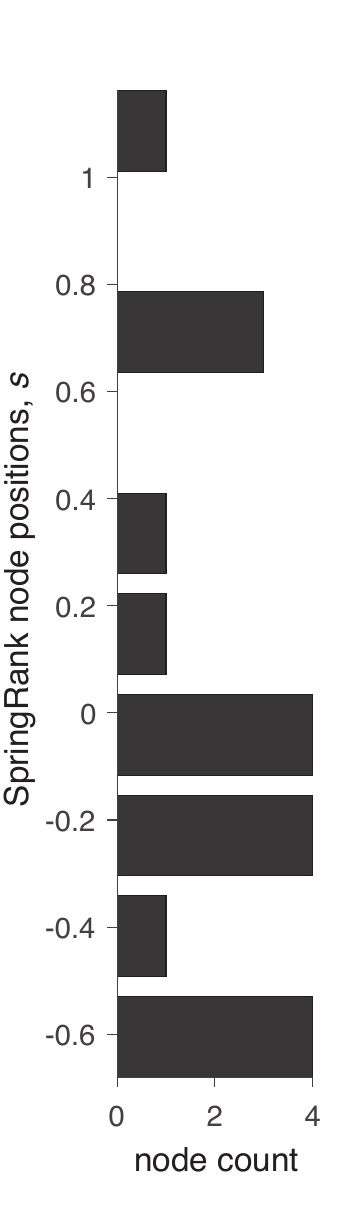}
		&\begin{tabular}{c}
			\includegraphics[width=8cm]{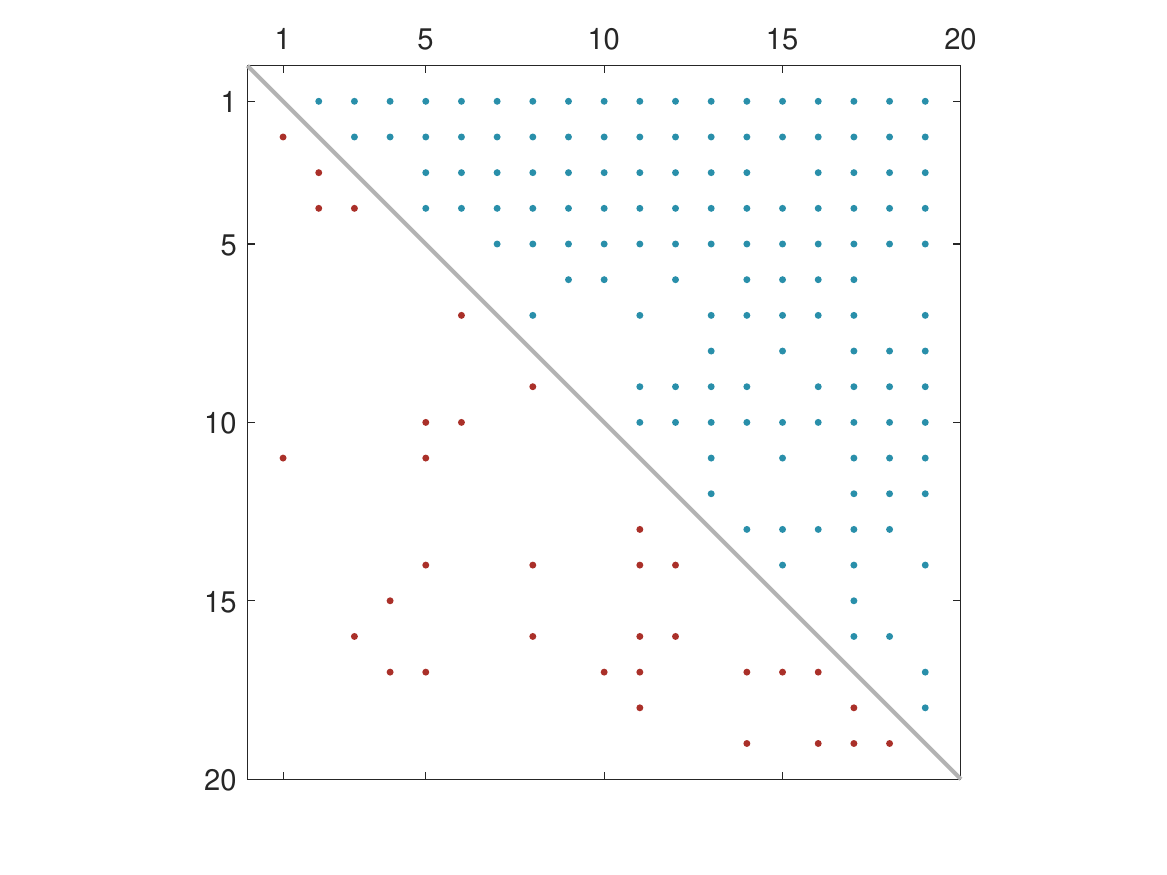}\\
			\includegraphics[width=7cm]{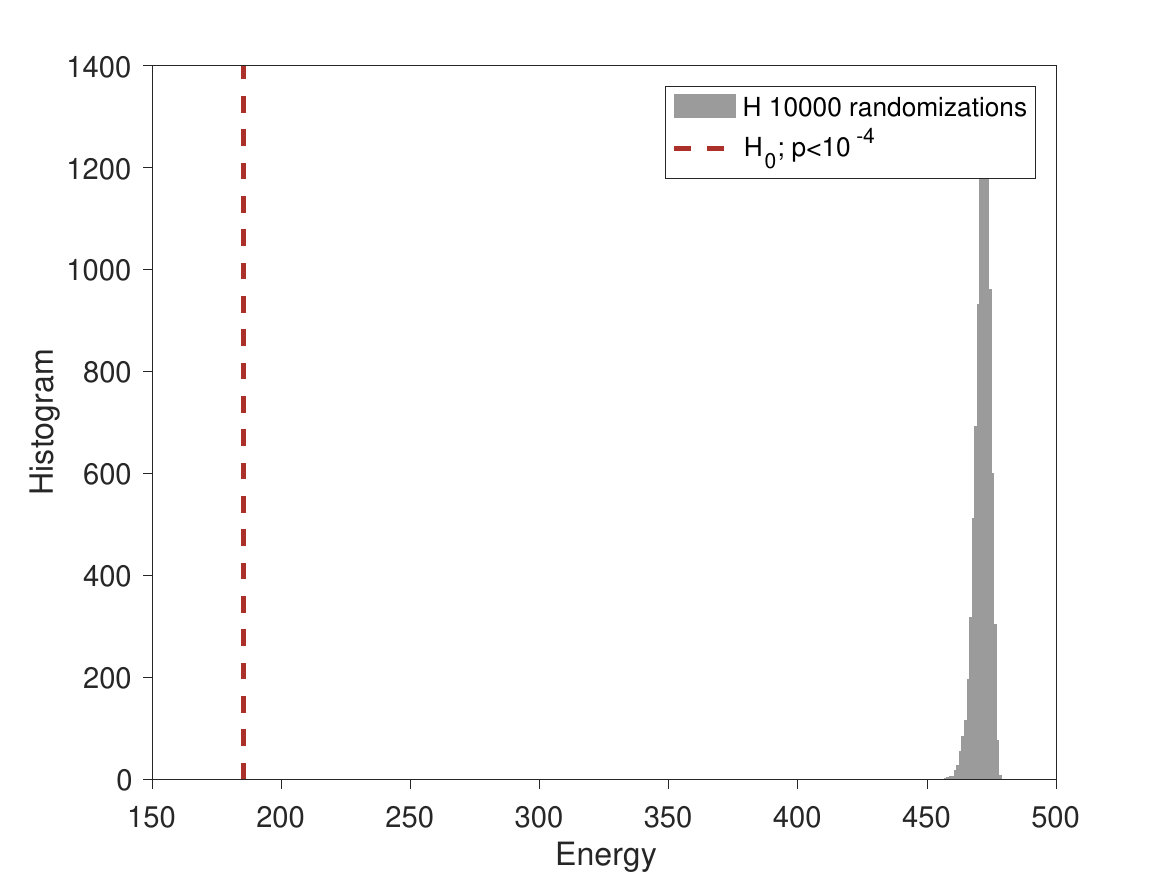}
		\end{tabular}
	\end{tabular}
		\includegraphics[width=8cm]{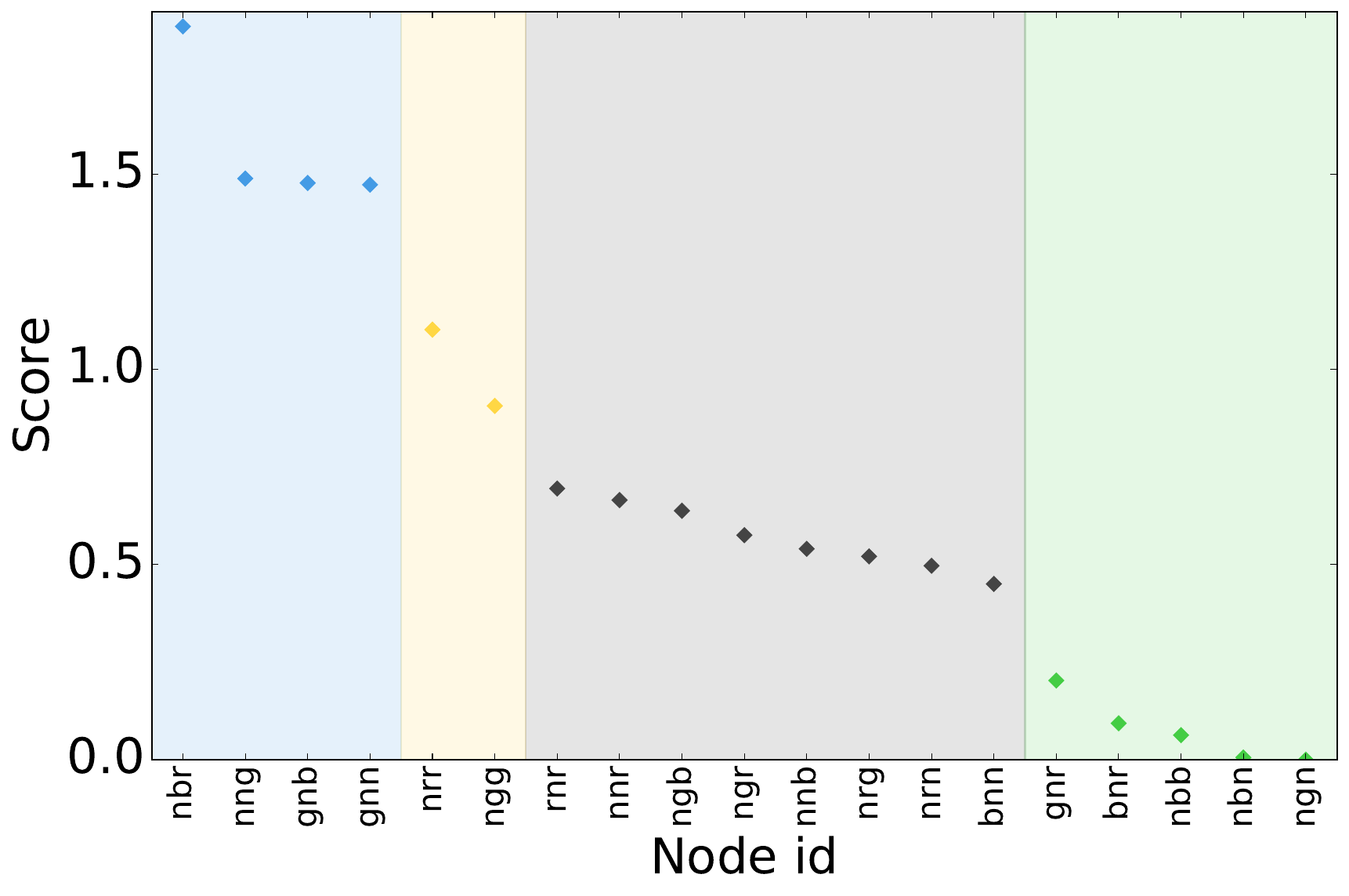}
	\caption{\textbf{Summary of SpringRank applied to Parakeet G2 network \cite{hobson2015social}.} \reportcaption\ \reportcaptionb}
	\label{SI:PK2}
\end{figure}

\clearpage
\section*{Te\underbar npa\d t\d ti}
\vspace{-0.5cm}
\begin{figure}[ht!]
	\centering
	\begin{tabular}{m{4.25cm}  m{3.8cm} | m{8cm}}
		\includegraphics[height=12cm]{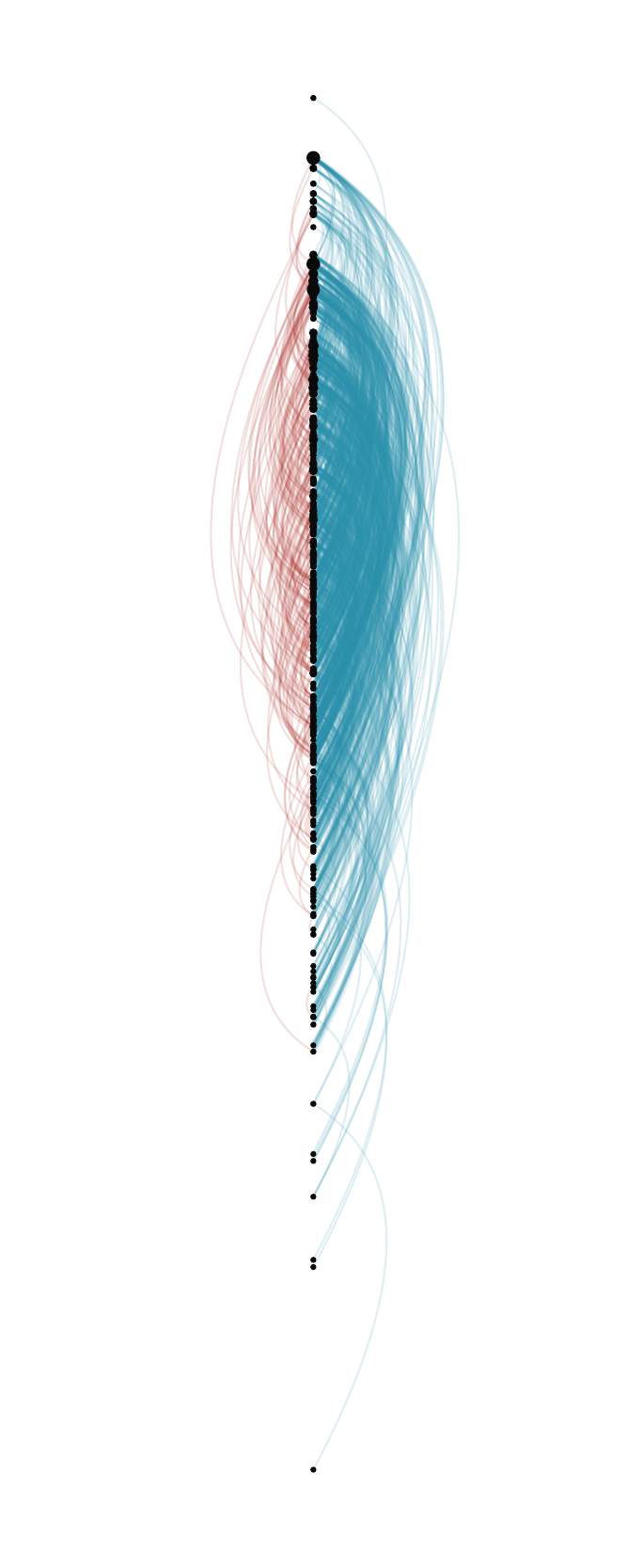}\vspace{15pt}
		& \includegraphics[height=13cm]{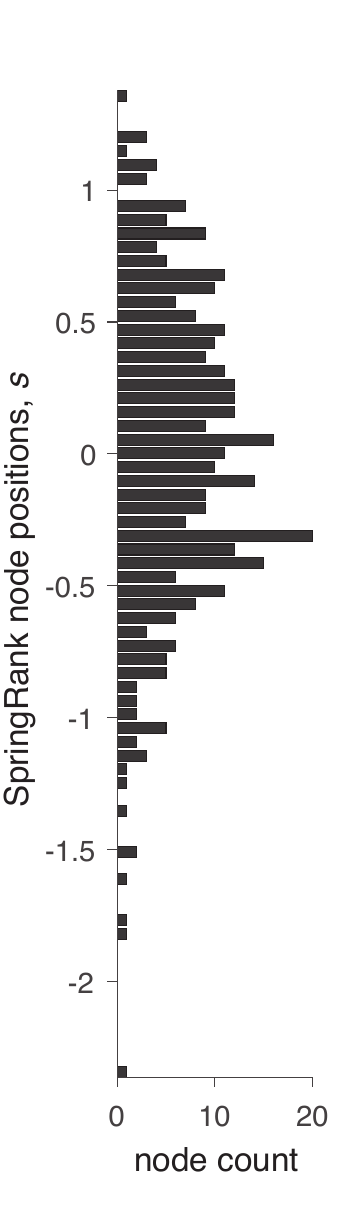}
		&\begin{tabular}{c}
			\includegraphics[width=8cm]{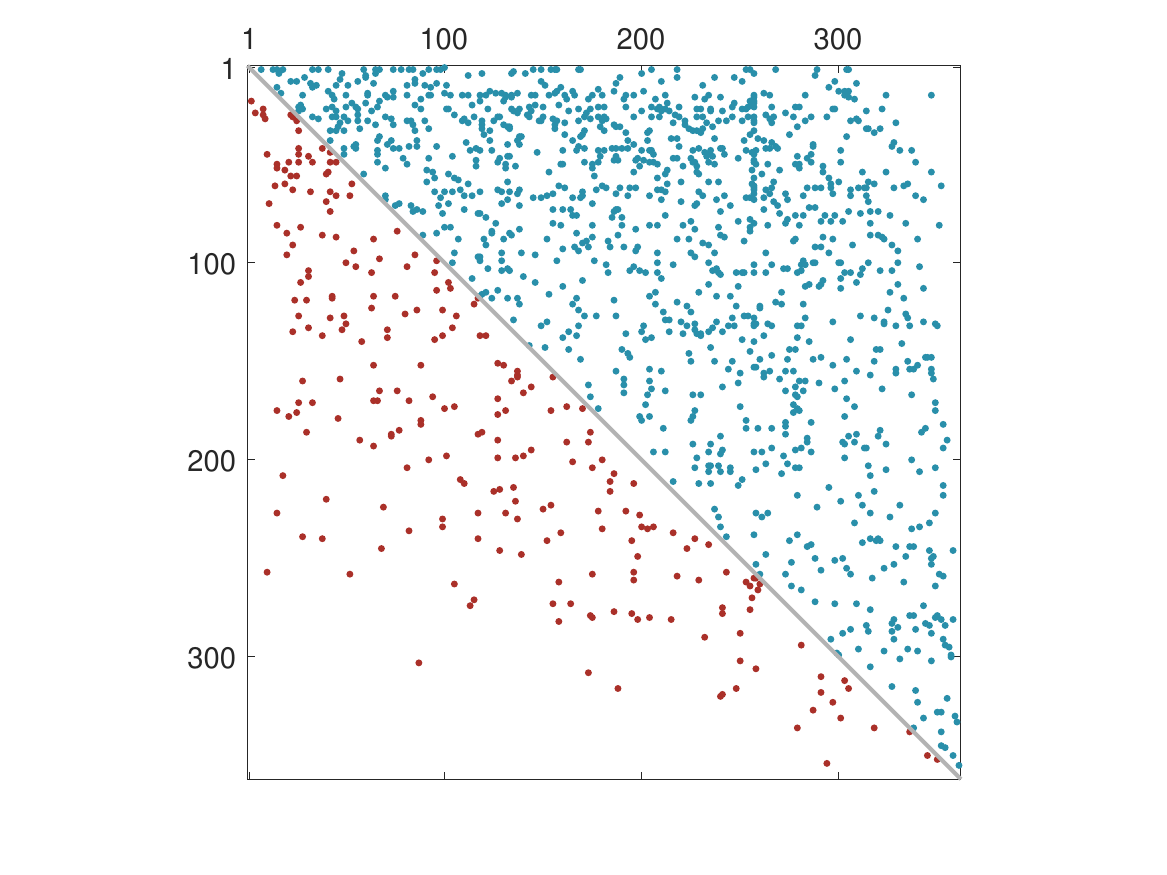}\\
			\includegraphics[width=7cm]{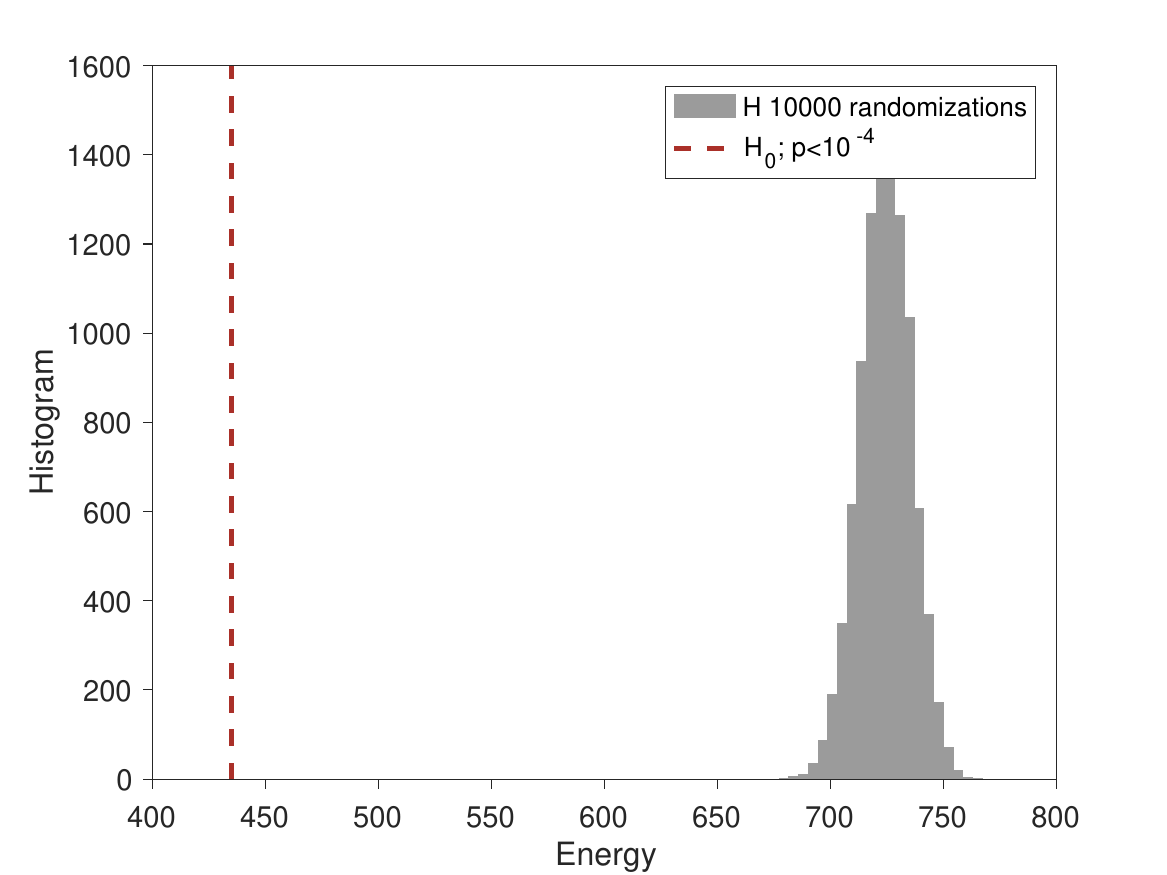}
		\end{tabular}
	\end{tabular}
		\includegraphics[width=18cm]{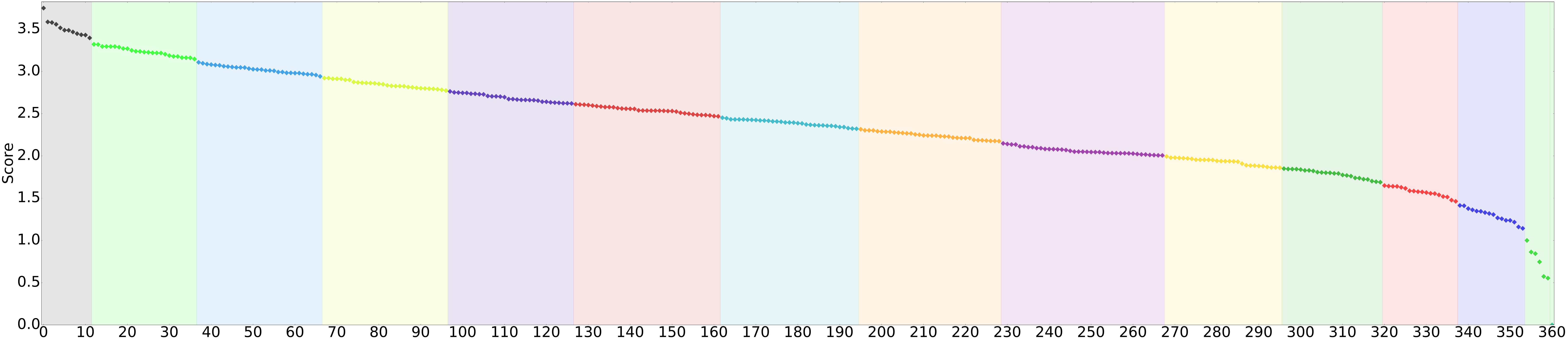}
	\caption{\textbf{Summary of SpringRank applied to Te\underbar npa\d t\d ti social support network \cite{power2017}.} \reportcaption \ \reportcaptionb}
	\label{SI:Ten}
\end{figure}

\clearpage
\section*{A\underbar lak\= apuram}
\vspace{-0.5cm}
\begin{figure}[h!]
	\centering
	\begin{tabular}{m{4.25cm}  m{3.8cm} | m{8cm}}
		\includegraphics[height=12cm]{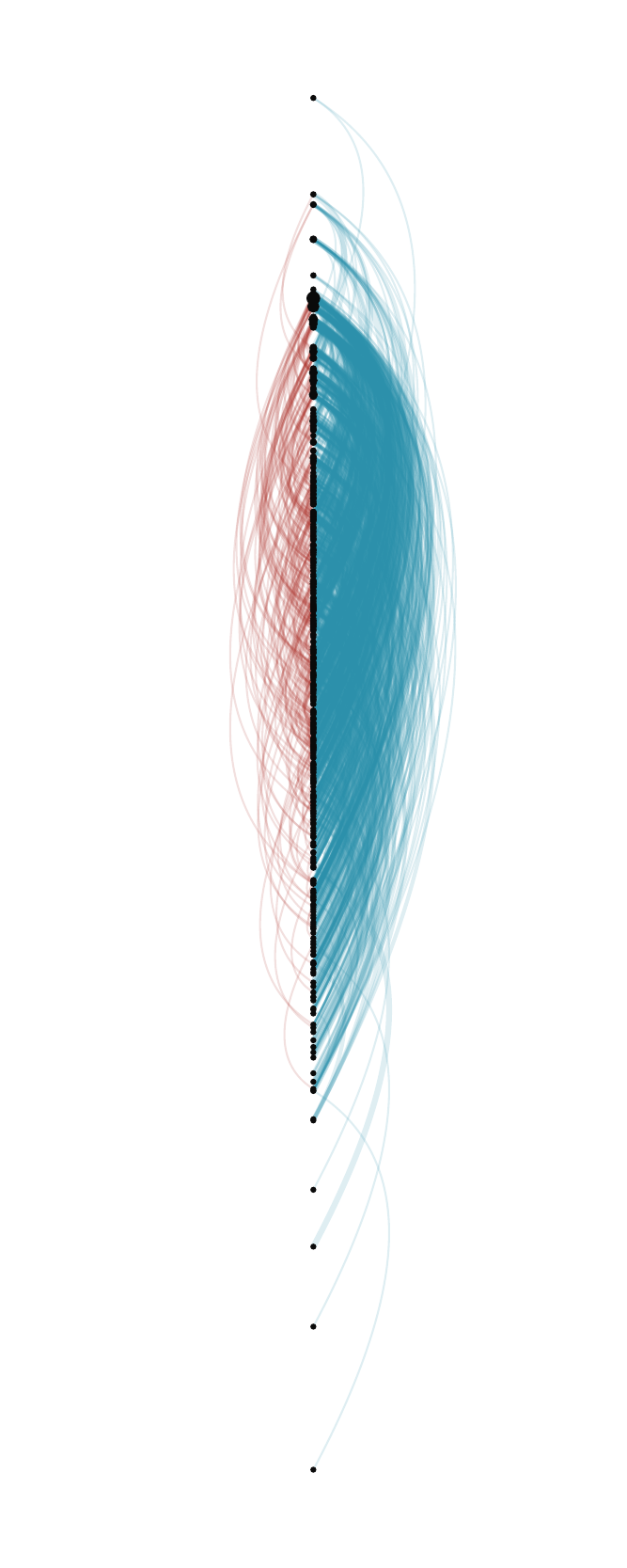}\vspace{15pt}
		& \includegraphics[height=13cm]{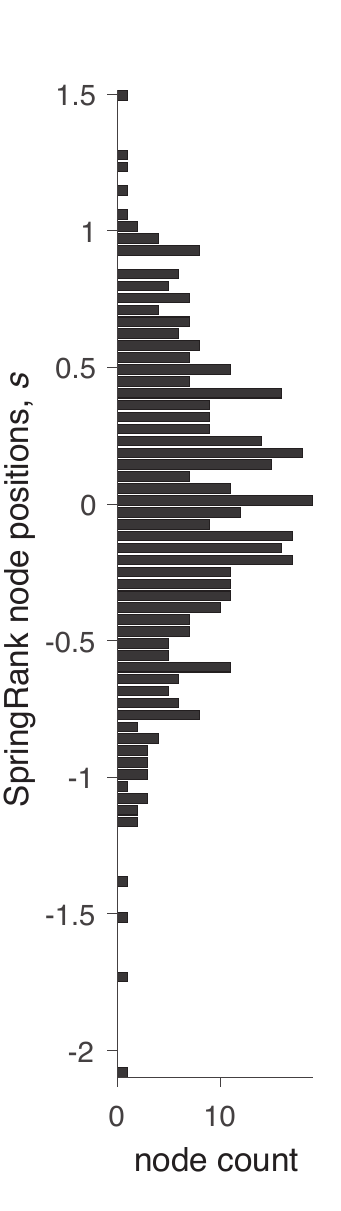}
		&\begin{tabular}{c}
			\includegraphics[width=8cm]{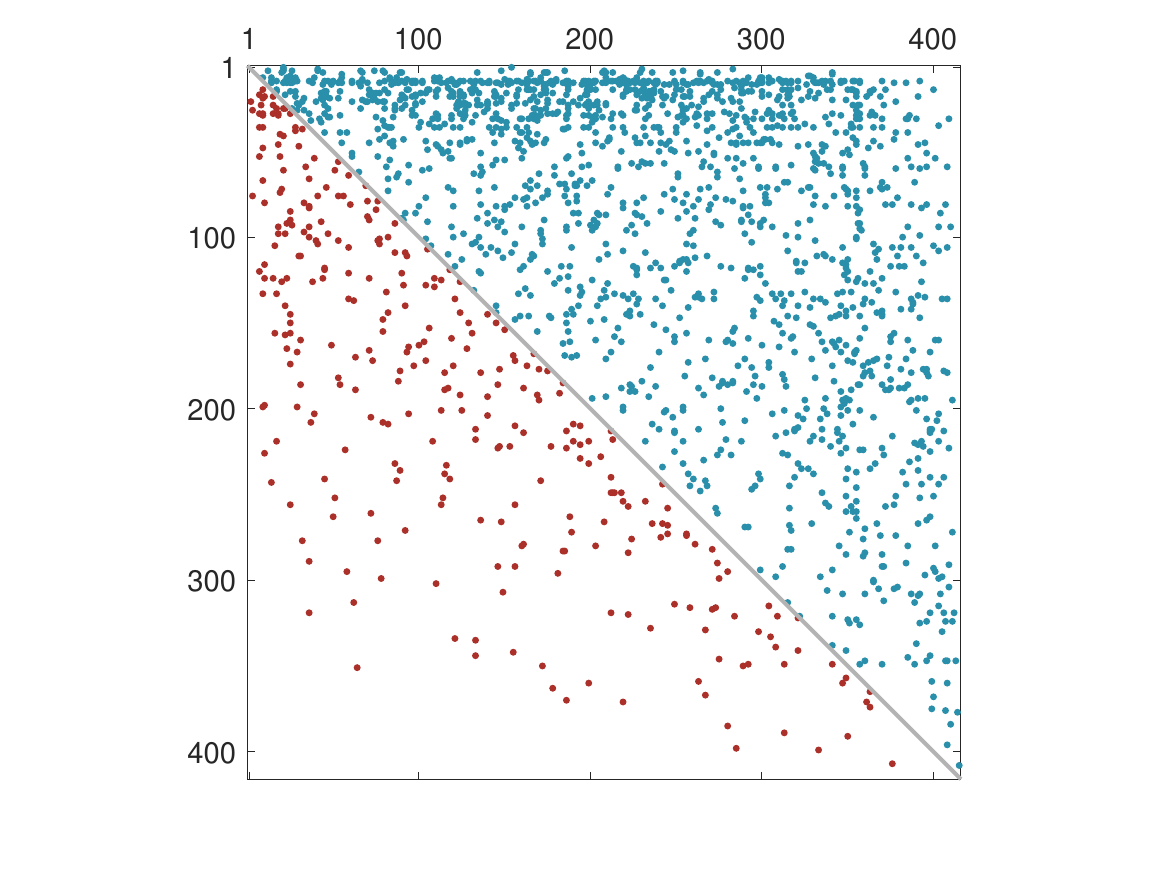}\\
			\includegraphics[width=7cm]{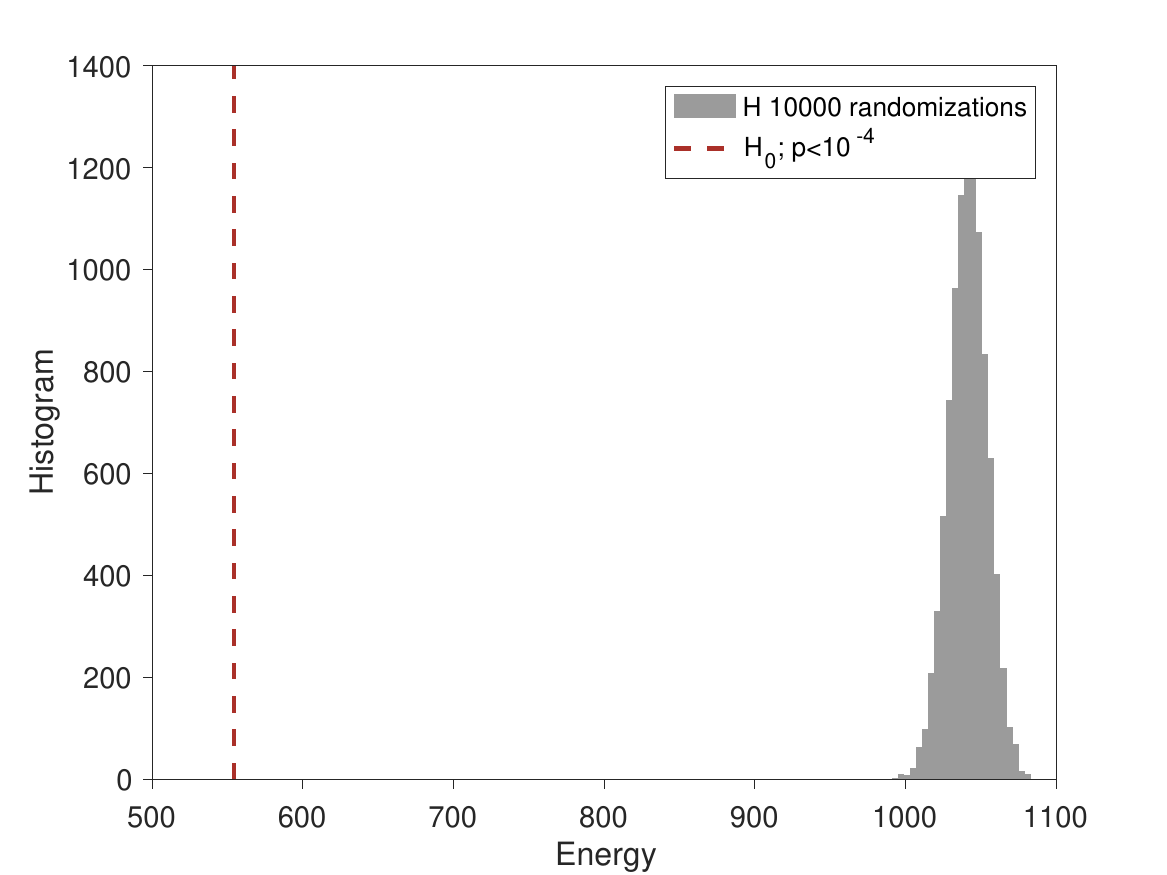}
		\end{tabular}
	\end{tabular}
		\includegraphics[width=18cm]{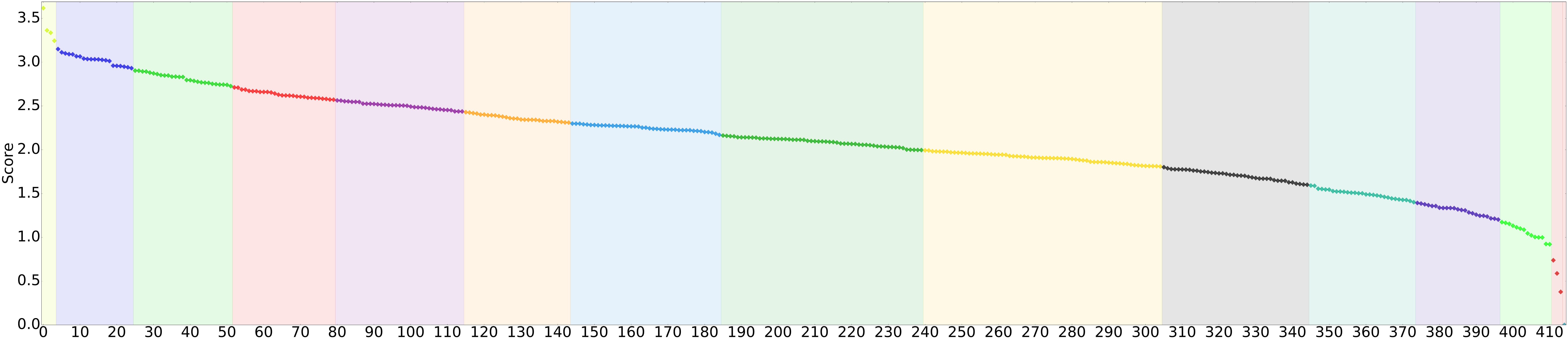}
	\caption{\textbf{Summary of SpringRank applied to A\underbar lak\= apuram social support network \cite{power2017}.} \reportcaption  \ \reportcaptionb}
	\label{SI:Ala}
\end{figure}

\end{widetext}

\end{document}